\documentclass[11pt]{article}

\usepackage[margin=1in]{geometry}
\usepackage[T1]{fontenc}
\usepackage{lmodern}
\usepackage{graphicx}
\graphicspath{{./Figures/}}
\usepackage{authblk}

\usepackage{amssymb}
\usepackage{multirow}
\usepackage[x11names]{xcolor}
\usepackage{mathtools}
\usepackage{esint}
\newcommand*{\dd}{\mathop{}\!\mathrm{d}}

\usepackage{amsmath}
\allowdisplaybreaks
\numberwithin{equation}{section}
\usepackage{bm}

\usepackage{dsfont}
\usepackage{float, subcaption}
\usepackage[super]{nth}

\usepackage{tikz}
\usetikzlibrary{positioning}

\usepackage[backend=biber,style=phys,sorting=none,url=false,hyperref=true,eprint=true,maxcitenames=2,maxbibnames=10,abbreviate=false]{biblatex}
\usepackage{hyperref}
\hypersetup{colorlinks=true, linkcolor=blue!60!black, citecolor=blue!60!black, urlcolor=blue!60!black}
\usepackage{cleveref}

\usepackage{algorithm}
\usepackage{algorithmic}

\usepackage{etoolbox}
\AtBeginEnvironment{algorithmic}{\small}

\title{Learning Interaction Kernels for Nonlocal Bacterial Growth Models}

\author[1]{Jacob W. Murri\thanks{Corresponding author: \href{mailto:jwmurri@math.ucla.edu}{jwmurri@math.ucla.edu}}}
\author[2,3]{Cameron Dudiak}
\author[3,4,5]{James N. Wilking}
\author[2,3]{Scott G. McCalla}
\author[1,6]{Hayden Schaeffer}
\affil[1]{Department of Mathematics, University of California, Los Angeles, Los Angeles, CA 90095}
\affil[2]{Department of Mathematical Sciences, Montana State University, Bozeman, MT 59717}
\affil[3]{Center for Biofilm Engineering, Montana State University, Bozeman, MT 59717}
\affil[4]{Chemical and Biological Engineering Department, Montana State University, Bozeman, MT 59717}
\affil[5]{Department of Physiology \& Biomedical Engineering, Mayo Clinic, Rochester, MN 55905}
\affil[6]{California NanoSystems Institute (CNSI), University of California, Los Angeles, Los Angeles, CA 90095}
\date{}

\begin{document}

\maketitle

\begin{abstract}

A previously developed system of geometric nonlocal differential equations models the growth of \emph{Paenibacillus dendritiformis} bacterial colonies by tracking their boundaries using a Lagrangian approach. This model contains self-interaction and cross-interaction kernels which are difficult to derive from first principles. We consider the problem of using data to learn these interaction kernels. Image processing and segmentation techniques are employed to extract contours from videos of growing bacterial colonies, and interaction kernels are modeled using simple parametrized functions. Model parameters are trained using a constrained optimization approach with the Nelder--Mead simplex method and L-BFGS-B to minimize the error between simulated and target contours. The fitted interaction kernels accurately predict the evolution of \emph{P. dendritiformis} colonies and reproduce the ``stationary fronts'' phenomenon whereby two nearby colonies inhibit each other's growth. We compare models using shared interaction kernels and colony growth rates with those using distinct self- and cross-interaction kernels and varying per-colony growth rates. We also introduce a nonlocal transport mechanism directed along the intercolony center-of-mass axis.%
\end{abstract}


\noindent\textbf{ORCID iDs:} JWM, 0009-0007-9273-4309; CD, 0009-0006-2307-6376; JNW, 0000-0002-7255-7209; SGM, 0000-0003-0093-4658; HS, 0009-0002-9015-1357

\section{Introduction}

Biofilms are spatially structured microbial communities embedded within an extracellular matrix whose growth and organization arise from the interaction of cellular proliferation, nutrient transport, mechanical forces, and biochemical signaling across multiple spatial and temporal scales \cite{costerton1999bacterial,hall2004bacterial,flemming2016biofilms,wilking2011biofilms}. These systems play an important role in a broad range of biological and engineering settings, including chronic infections, antibiotic tolerance, and industrial biofouling \cite{costerton1999bacterial,stewart2001antibiotic,flemming2010biofouling}. Despite extensive experimental investigation, the development of predictive mathematical models capable of reproducing the full spatiotemporal evolution of biofilm and bacterial colony morphologies remains a significant challenge.

A defining feature of many microbial growth systems is the emergence of coherent interfaces separating occupied and unoccupied regions. The evolution of these interfaces reflects the collective influence of local growth processes, geometric regularization, competition for resources, and long-range communication mechanisms such as quorum sensing and chemically mediated inhibition. Consequently, interfacial dynamics provide a natural reduced description of colony-scale behavior, allowing complex biological interactions to be encoded through the motion of evolving geometric objects. Such reduced descriptions are particularly attractive when many relevant variables, including chemical concentrations and intracellular signaling states, are either unobserved or only partially observed experimentally; this is typical for biofilm imaging studies.

A wide range of mathematical frameworks have been developed to study microbial growth and biological pattern formation. Agent-based models resolve individual cells and their interactions, providing detailed descriptions of collective phenomena such as differentiation, quorum sensing, and branching morphologies \cite{warren2019spatiotemporal,benjacob2000cooperative}. Such models are most informative when the relevant cellular mechanisms are known, but they become computationally expensive for colonies of millions of cells and contain many parameters that are difficult to measure. At the continuum level, systems of ordinary and partial differential equations have been used to model population densities, nutrient fields, signaling molecules, and mechanical effects, often incorporating reaction--diffusion dynamics, fluid transport, elasticity, and free-boundary evolution \cite{klapper2002viscoelastic,dockery2001finger,klapper2010mathematical}. More sophisticated formulations employ immersed-boundary, phase-field, and level-set methodologies to capture evolving interfaces and mechanical interactions \cite{peskin2002immersed,osher1988fronts,chen2002phase}. While these approaches have yielded substantial biological insight, they often involve large numbers of parameters and latent variables, making quantitative studies difficult in the low-data and high-noise regimes typical of biological experiments.

An alternative perspective is provided by geometric evolution equations, in which the state of each biofilm colony is represented directly by one or more evolving interfaces. Such formulations connect naturally to the extensive mathematical literature on curvature-driven flows, front propagation, and free-boundary problems \cite{evans1992phase,giga2006surface,sethian1999level}. Classical examples include mean curvature flow and its variants, where interfacial motion is governed by local geometric quantities.  Curvature-driven flows of this kind have long been used in materials science and metallurgy, for example to describe the motion of grain boundaries and the stability of growing interfaces \cite{mullins1999,mullins_sekerka1963}, and they arise as sharp-interface limits of phase-field models of phase separation and coarsening such as the Allen--Cahn and Cahn--Hilliard equations \cite{allen_cahn_original,cahn_hilliard_original,pego1989,dai_promislow_2013}. Interfacial motion of this type also emerges from reaction--diffusion systems with fast reaction and slow diffusion \cite{rubinstein1989fast}. However, many multiscale and biological systems exhibit interactions mediated by signaling molecules, mechanical stresses, or other processes acting over finite spatial scales. These mechanisms naturally motivate the incorporation of nonlocal effects into geometric evolution laws.  In multiscale reaction--diffusion systems, for example, interfaces interact through slowly varying chemical fields, and eliminating these fields yields effective nonlocal interactions between fronts; in singularly perturbed three-component systems, such front interactions can be derived rigorously in one spatial dimension \cite{van2010front} and observed in numerical simulations in two \cite{van2012coexistence}.

Nonlocal geometric partial differential equations (PDEs) have emerged in a variety of contexts, including materials science, aggregation dynamics, phase separation, image processing, and collective behavior \cite{bertozzi2007aggregation,carrillo2010asymptotic,kolokolnikov2011stability,slepcev2019coarsening,du2012analysis}. In these models, interface motion depends both on local geometric information and on spatially distributed interactions encoded through integral operators or interaction kernels. Such formulations provide a mathematically natural framework for describing quorum sensing, toxin-mediated inhibition, competition between neighboring colonies, and mechanically mediated communication through the extracellular matrix. Moreover, nonlocal models can often be interpreted as effective descriptions of unresolved or latent dynamics, a perspective emphasized in the work of Du and collaborators and developed further in recent approaches to learning nonlocal operators from partially observed systems \cite{du2019nonlocal,mccalla2018consistent,smithjanus,lyu2026mvnnmeasurevaluedneuralnetwork}.

Our motivating application is the growth of colonies of the bacterium \emph{Paenibacillus dendritiformis}. When two colonies of the same strain grow on the same agar plate, they do not merge: their fronts stop advancing at a finite distance from each other, leaving a stable gap between the colonies \cite{be2009deadly}. This mutual inhibition is mediated by a protein, sibling lethal factor (SLF), which the colonies secrete in response to competition \cite{be2010lethal}. Such stationary, or arrested, fronts are a natural target for interfacial models with nonlocal interactions, since the colonies inhibit each other's growth before they come into contact.

The model considered in this work builds upon the nonlocal interfacial framework introduced in \cite{Scott2016Fronts,vonbrecht2023frontsarrestiianalytical}, in which colony boundaries evolve under the combined influence of growth, curvature regularization, and nonlocal inhibitory interactions. Some previous numerical work fit a simplified model with shared colony growth rates and interaction kernels \cite{dud22,Scott2016Fronts}, but we find that using a more general form of the model introduced in \cite{Scott2016Fronts,vonbrecht2023frontsarrestiianalytical}, with varying per-colony growth rates and distinct self- and cross-interaction kernels, improves model fits. This extension greatly enlarges the class of admissible dynamics while remaining amenable to both analysis and numerical computation. 

We also consider an extension of the model: a nonlocal transport mechanism which permits attractive or repulsive colony-scale drift induced by interactions between neighboring colonies. The resulting dynamics combine curvature-driven regularization, nonlocal normal forcing, and global transport effects within a unified geometric framework. However, we find that the effect of the nonlocal transport mechanism is minimal at best (losses improve by only $0.1$--$2.8\%$), indicating that the standard model from \cite{Scott2016Fronts,vonbrecht2023frontsarrestiianalytical} already captures the separation between colonies well. 

A secondary contribution is a comparison of two parametrized classes of interaction kernels: Gaussian kernels, used in earlier work, and a new class of Ramp kernels. Learning the kernels nonparametrically, as has been done for systems of interacting particles and agents \cite{lu2019nonparametric,lu2021learning1,lu2021learning2,li2021identifiability,liu2023random}, is a natural direction for future work, although the limited contour data make it challenging.

A central objective of this work is to infer the parameters governing this nonlocal geometric PDE directly from experimental observations. This problem lies at the intersection of inverse problems, scientific machine learning, and data-driven model discovery. Recent advances have produced powerful techniques for learning dynamical systems from data, including sparse equation-discovery methods \cite{schmidt2009distilling,brunton2016discovering,schaeffer2017sparse}, neural differential equations \cite{chen2018neural}, weak-form PDE identification approaches \cite{messenger2021weak}, and operator-learning frameworks such as DeepONets and Fourier Neural Operators \cite{lu2021learning,li2020fno}, as well as foundation models trained across many differential equations \cite{LIU2024106707,yang2023incontext,liu2025bcatblockcausaltransformer}. While highly successful in many settings, these approaches are generally designed for function-valued states evolving on fixed spatial domains and often rely on large volumes of high-quality observations.

The setting considered here presents several challenges that fall outside the assumptions underlying many existing learning frameworks. First, the state variable is not a function on a fixed domain but rather a collection of evolving interfaces represented as closed curves. Second, the available data are limited, noisy, and only partially observed, with important physical variables remaining latent. Third, the governing dynamics involve geometric constraints and nonlocal interactions that are difficult to encode within generic neural architectures. As a result, purely data-driven methods frequently suffer from poor interpretability, limited extrapolation capability, and difficulties in enforcing geometric consistency.

These considerations motivate the use of structured mechanistic models that incorporate nonlocal interactions explicitly. By restricting attention to a parametrized family of biologically motivated geometric evolution laws, one obtains a model class that remains sufficiently expressive to capture experimentally observed phenomena while retaining interpretability and enabling parameter estimation from limited data. In this sense, the present work contributes to a growing body of research demonstrating that carefully structured nonlocal models can provide a powerful alternative to purely black-box learning approaches in partially observed scientific systems.

In addition to model development, we address a major practical bottleneck in quantitative biological modeling: the extraction of reliable geometric information from experimental imagery. We develop an improved segmentation and contour-processing pipeline for \emph{P. dendritiformis} colony growth, producing smooth and temporally consistent interface representations suitable for downstream analysis. These data enable the formulation of a PDE-constrained optimization problem in which model parameters are inferred by minimizing discrepancies between observed and simulated interface evolution.

Our contributions are therefore as follows. First, we develop a robust image-processing pipeline that produces high-quality geometric representations of experimental growth dynamics. Second, we fit a nonlocal geometric model of bacterial colony evolution to experimental data, using its full generality by incorporating distinct per-colony growth rates and self- and cross-interaction kernels. Third, we compare this model with simpler parametrizations of earlier work by testing how sharing colony growth rates and interaction kernels affects model fit. Fourth, we demonstrate the robustness of the model by fitting an additional nonlocal transport mechanism and showing that the improvement is minimal at best. Fifth, we produce a comparison of two parametrized classes of interaction kernels: Gaussian kernels, used in earlier work, and a new class of Ramp kernels. Overall, our work demonstrates that the resulting model is identifiable, trainable, and quantitatively predictive in a low-data, high-noise regime, providing evidence that nonlocal geometric models constitute an effective framework for scientific machine learning in systems with substantial latent structure.

The remainder of the paper is organized as follows. Section~\ref{sec:mp} introduces the mathematical formulation of the model and associated representations of planar contours. Section~\ref{sec:mt} describes the parameter estimation framework and training methodology. Section~\ref{sec:r} presents numerical results and comparisons with experimental data. Finally, Section~\ref{sec:c} discusses the implications of the model, its limitations, and directions for future work.

\section{Mathematical Preliminaries} \label{sec:mp}

\subsection{Notation}\label{sec:notation}

This section will clarify and emphasize some of the notation used in this work. All of the contours will be functions defined on $\mathbb{T}^1 = [-\pi, \pi]$, where $\mathbb{T}^1$ is used instead of $[-\pi, \pi]$ to connote periodic boundary conditions. Variables in $\mathbb{T}^1$ will typically be denoted $u$, $v$, or $w$.

Vector-valued quantities will be bolded (for example, $\bm{\tau}$ for a tangent vector or $\mathbf{n}$ for a normal vector). The quantities in this work often depend on several indices. Vector indices will always be denoted with a subscript, while the contour index uses a superscript with parentheses (typically using the letter $i$ as a variable).  Likewise, spatial indices will typically use the letter $j$ and temporal indices will use the letter $k$. These latter two indices can appear as subscripts if referring to discrete points. For example, 
\[
(\bm{\gamma}_{j,k}^{(i)})_2 \quad \text{ or } \quad \bm{\gamma}^{(i)}(u_j, t_k)_2
\]
refers to the 2nd vector component of the $i$th contour at the $j$th spatial location and the $k$th time. The latter usage emphasizes evaluating a function at a point, while the former usage connotes that the contour is being represented by a discrete set of points. The notation $[n]$, where $n \in \mathbb{Z}$, will be used to refer to the set $\{1, \ldots, n\}$.

Partial derivatives are represented in a few ways. They can be explicitly written out (e.g. $\tfrac{\partial \bm{\gamma}}{\partial u}(u, t)$) when clarity is required and the function arguments are emphasized. In this case the variable in the partial derivative refers to the slot of the function which is being differentiated with respect to. To compress notation, $\partial_u \bm{\gamma}(u, t)$ or $\bm{\gamma}_u$ may be used (the latter only if the function arguments are also suppressed).

The notation $\arctan(y, x)$ refers to the two-argument arctangent function which returns the angle $\theta$ between the positive $x$-axis and the ray from the origin to the point $(x, y)$, confined to $(-\pi, \pi]$.

\subsection{Modeling using Planar Contours}

In \cite{Scott2016Fronts}, the authors studied a class of nonlocal partial differential equations which model the interfacial dynamics related to growth of \emph{P. dendritiformis} colonies. That model, in which every force acting on a colony boundary is directed normal to the boundary, is the model whose interaction kernels we learn in this work. An extension which also allows a rigid, non-normal drift of each colony is described in Section \ref{sec:transport_force}; it is compared against the standard model in Section \ref{sec:results_transport}. 

Suppose the colony boundaries are modeled by a collection of $m$ planar counterclockwise-oriented contours denoted as $\mathcal{C} = {\{\bm{\gamma}^{(1)},\ldots,\bm{\gamma}^{(m)} \}}$. These parametrized curves are defined as functions on $\mathbb{T}^1 \times [0, T]$ where $[0, T]$ is a certain interval of time:
\[ 
 \bm{\gamma}^{(i)} (u,t) = \begin{bmatrix}
    \text{   } (\bm{\gamma}^{(i)}(u,t))_1 \text{   } \\
     \text{   } (\bm{\gamma}^{(i)}(u,t))_2 \text{   }  \\
 \end{bmatrix}, \qquad 
 \bm{\gamma}^{(i)}:\mathbb{T}^1 \times [0, T] \to\mathbb{R}^2,\]
 where $\mathbb{T}^1$ is as defined in Section \ref{sec:notation}. Each $\bm{\gamma}^{(i)}$ is assumed to be a simple closed contour with counterclockwise orientation. We also assume that the parametrizations are smooth and regular in the sense that $|\partial_u{\bm{\gamma}^{(i)}} (u,t)| \neq 0$ for any $(u, t) \in \mathbb{T}^1 \times [0, T]$ and $i \in [m]$.

 Let $\bm{\tau}^{(i)}(u,t)$ and $\mathbf{n}^{(i)}(u,t)$ represent the unit tangent and outward normal vectors to the curve $\bm{\gamma}^{(i)}$ at the point $u \in\mathbb{T}^1$ at time $t$, i.e. 
 \begin{equation}\label{eq:unit_tangent}
\bm{\tau}^{(i)}(u,t) = \frac{\partial_u{\bm{\gamma}^{(i)}} (u,t)}{|\partial_u{\bm{\gamma}^{(i)}} (u,t)|} 
 \end{equation}
 \begin{equation}\label{eq:unit_normal}
 \mathbf{n}^{(i)}(u, t)
 = \begin{bmatrix}
    \text{   } (\bm{\tau}^{(i)}(u,t))_2\text{   } \\
     \text{   } -(\bm{\tau}^{(i)}(u,t))_1 \text{   }  \\
 \end{bmatrix}
 \end{equation}
 Defining $(a, b)^\perp := (b, -a)$, we have that $\mathbf{n}^{(i)} ={\bm{\tau}^{(i)}}^\perp$ and $\bm{\tau}^{(i)} = -{\mathbf{n}^{(i)}}^\perp$. Furthermore, let $\kappa_{\bm{\gamma}^{(i)}}(u, t)$ represent the signed curvature of the contour $\bm{\gamma}^{(i)}$ at a point $u \in \mathbb{T}^1$ at time $t$:
 \[
\kappa_{\bm{\gamma}^{(i)}}(u, t) =\frac{\partial_u\bm{\tau}^{(i)}(u, t)}{|\partial_u\bm{\gamma}^{(i)}(u,t)|} \cdot \mathbf{n}_{\gamma^{(i)}}(u, t),
\]
so that the identity
\begin{equation}\label{eq:curvature}
\kappa_{\bm{\gamma}^{(i)}}(u, t) \mathbf{n}^{(i)}(u, t) = \frac{\partial_u\bm{\tau}^{(i)}(u, t)}{|\partial_u\bm{\gamma}^{(i)}(u,t)|}
\end{equation}
holds.

The evolution of these planar contours is given by
\begin{equation}\label{eq:vague_evolution}
\partial_t  \bm{\gamma}^{(i)} (u,t) = \left( \mu^{}\kappa_{\bm{\gamma}^{(i)}}(u,t)+c^{(i)} -v_{\text{NL}}^{(i)}(u, t) \right)\mathbf{n}^{(i)}(u, t),
\end{equation}
where $\mu$ and $\{c^{(1)}, \ldots, c^{(m)}\}$ are constants which describe (respectively) the dependence of colony growth on signed curvature, and a per-colony constant growth or death rate. Per-colony constant growth rates as in \cite{Scott2016Fronts}, rather than the single shared growth rate of \cite{dud22}, are used to address varying initial size or vigor of distinct colonies. Nonlocal forces are described by $v_{\text{NL}}^{(i)}$, the normal-direction nonlocal force modeled by interaction kernels.

\textbf{Nonlocal Force from Interaction Kernels. } The nonlocal force $v_{\text{NL}}^{(i)}(u, t)$ modeling interactions between contours is described by interaction kernels $f$ and $g$: 
\begin{align*}
&v^{(i)}_{\text{NL}}(u,t) \\
&:= \int_{\mathbb{T}^1} f\left(\frac{1}{2} \left\lvert \bm{\gamma}^{(i)}(u,t) - \bm{\gamma}^{(i)}(v,t) \right\rvert^2\right) | \partial_u \bm{\gamma}^{(i)}(v,t) | \dd v \\
&+ \sum_{j \neq i} \int_{\mathbb{T}^1} g\left(\frac{1}{2} \left\lvert \bm{\gamma}^{(i)}(u,t) - \bm{\gamma}^{(j)}(v,t) \right\rvert^2\right) | \partial_u\bm{\gamma}^{(j)}(v,t)| \dd v. 
\end{align*} 
To align with the framework in \cite{Scott2016Fronts}, the inputs to interaction kernels are one-half of the squared distance between two points. For $f$, this pair of points lies on a single contour; thus $f$ describes self-interaction. The input to $g$ is one-half of the squared distance between a pair of points lying on two separate contours; hence $g$ describes the interaction between colonies. We note that while \cite{Scott2016Fronts} allow $f$ and $g$ to be distinct, the numerical study of \cite{dud22} trains a model with $f = g$. We take the former approach, and in fact find that the best-fitting interaction kernels for self- and cross-interactions are very different, and requiring $f = g$ raises the best loss by $34$--$323$\% (see Section \ref{sec:results_prior} for more details). Within the context of a \emph{P. dendritiformis} bacterial growth model, $f$ and $g$ could describe the effect of a toxic protein, known as sibling lethal factor (SLF), which colonies release and which inhibits growth. 

The interaction kernels are required to have several properties:
\begin{enumerate}
    \item Interaction kernels should be functions from $\mathbb{R}^+$ to $\mathbb{R}^+$, i.e. only take nonnegative values. We assume that the interactions only serve to inhibit growth, not increase it (as in the example of SLF protein inhibiting growth of bacterial colonies).
    \item Interaction kernels are the same for each pair of colonies (excepting self-interaction).
    \item Interaction kernels should be continuous, monotone decreasing functions (the strength of interaction should be continuous in the distance, as well as decrease with increasing distance).
    \item Interaction kernels should tend to zero as their input tends to infinity (colonies which are far enough away should have arbitrarily small interaction).
\end{enumerate}

Substituting the nonlocal force from interaction kernels into \eqref{eq:vague_evolution} yields the full model

\begin{equation}\begin{split}\label{eq:full_model}
&\frac{\partial}{\partial t} \bm{\gamma}^{(i)} (u, t)
= \,\mathbf{n}^{(i)}(u,t) \left[ \mu \, \kappa_{\bm{\gamma}^{(i)}}(u,t)+c^{(i)} - \right. \\
&\int_{\mathbb{T}^1} f\left(\frac{1}{2} \left\lvert \bm{\gamma}^{(i)}(u,t) - \bm{\gamma}^{(i)}(v,t) \right\rvert^2\right) \left\lvert \partial_u \bm{\gamma}^{(i)}(v,t) \right\rvert \dd v  \\
&\left. - \sum_{j \neq i} \int_{\mathbb{T}^1} g\left(\frac{1}{2} \left\lvert \bm{\gamma}^{(i)}(u,t) - \bm{\gamma}^{(j)}(v,t) \right\rvert^2\right) \left\lvert \partial_u \bm{\gamma}^{(j)}(v,t)\right\rvert \dd v\right], \\
& \qquad \qquad \qquad \qquad \qquad \qquad \qquad \qquad \qquad \qquad \qquad i \in [m].
\end{split}\end{equation}

\subsection{Nondimensionalization}\label{sec:nondim}

By introducing a time scale $\tau > 0$ and nondimensional time $t' := t/\tau$, the model equations can be rewritten to eliminate the curvature coefficient $\mu$, at the cost of introducing $\tau$ itself as a new model parameter:
\begin{equation}\begin{split}
&\frac{\partial}{\partial t'} \bm{\gamma}^{(i)} (u, t') 
= \,\mathbf{n}^{(i)}(u,t') \left[ \tau \mu \, \kappa_{\bm{\gamma}^{(i)}}(u,t')+\tau c^{(i)} - \right. \\
&\tau \int_{\mathbb{T}^1} f\left(\frac{1}{2} \left\lvert \bm{\gamma}^{(i)}(u,t') - \bm{\gamma}^{(i)}(v,t') \right\rvert^2\right) \left\lvert \partial_u \bm{\gamma}^{(i)}(v,t') \right\rvert \dd v - \\
&\left.  \sum_{j \neq i} \tau \int_{\mathbb{T}^1} g\left(\frac{1}{2} \left\lvert \bm{\gamma}^{(i)}(u,t') - \bm{\gamma}^{(j)}(v,t') \right\rvert^2\right) \left\lvert \partial_u \bm{\gamma}^{(j)}(v,t')\right\rvert \dd v\right], \\
& \qquad \qquad \qquad \qquad \qquad \qquad \qquad \qquad \qquad \qquad \qquad i \in [m].
\end{split}\end{equation}
After renaming $t'$ to $t$, setting $\tau = 1/\mu$, and renaming $c^{(i)\prime} = \tau c^{(i)}$ to $c^{(i)}$, the coefficient on the curvature term is eliminated:
\begin{equation}\begin{split}\label{eq:full_model_nondim}
&\frac{\partial}{\partial t} \bm{\gamma}^{(i)} (u, t)
= \,\mathbf{n}^{(i)}(u,t) \left[\kappa_{\bm{\gamma}^{(i)}}(u,t)+c^{(i)} - \right. \\
& \int_{\mathbb{T}^1} \tau f\left(\frac{1}{2} \left\lvert \bm{\gamma}^{(i)}(u,t) - \bm{\gamma}^{(i)}(v,t) \right\rvert^2\right) \left\lvert \partial_u \bm{\gamma}^{(i)}(v,t) \right\rvert \dd v - \\
&\left. \sum_{j \neq i}\int_{\mathbb{T}^1} \tau g\left(\frac{1}{2} \left\lvert \bm{\gamma}^{(i)}(u,t) - \bm{\gamma}^{(j)}(v,t) \right\rvert^2\right) \left\lvert \partial_u \bm{\gamma}^{(j)}(v,t)\right\rvert \dd v\right], \\
& \qquad \qquad \qquad \qquad \qquad \qquad \qquad \qquad \qquad \qquad \qquad i \in [m].
\end{split}\end{equation}
Even though the total number of parameters stays the same, this representation is easier to work with numerically because the governing equation's diffusion term (and the associated time step restriction) does not depend on a model parameter. The disadvantage is that the interval between images, measured in nondimensional time, now depends on $\tau$; Section \ref{sec:optimization_problems} explains how the time step is chosen so that the discretized model remains smooth in $\tau$. The interaction kernels now appear only through the rescaled kernels $\tau f$ and $\tau g$, which we train directly (Section \ref{sec:parameters}).

\textbf{Interpretation of $\tau$.} Throughout this work, physical time is measured in units of the interval between successive images, so consecutive images are separated by $1/\tau$ units of nondimensional time. Since the time scale that sets the curvature coefficient to one is $\tau = 1/\mu$, the parameter $\tau$ is the reciprocal of the physical curvature coefficient, and in physical units the growth rate of colony $i$ is $c^{(i)}/\tau$. A smaller value of $\tau$ therefore means that curvature-driven smoothing acts more strongly over each interval between images. Also, notably, curvature-driven smoothing is inversely proportional to radius for circular colonies, so it affects smaller colonies more heavily and may help to differentiate growth rates between colonies. The optimization finds a large range of values of $\tau$ appropriate for different settings and datasets; Section \ref{sec:results_standard} shows that the best fits with the standard model have $\tau \approx 0.88$--$0.89$ (Dataset 1), $0.19$--$0.40$ (Dataset 2) and $0.028$--$0.095$ (Dataset 3). 

\subsection{Constant-Speed and Tangent-Angle Representations}\label{sec:representations}

Rather than evolving the parametrized contours $\bm{\gamma}^{(i)}$ directly, we follow \cite{vonbrecht2023frontsarrestiianalytical} and represent each contour by its average speed, its centroid, and its tangent-angle function. This leads to a form of the governing equations that is well suited to stable numerical computation (Section \ref{sec:dynamics_main}). This section summarizes the definitions; Supplementary Section~\ref{sec:derivation} gives the details and derives the governing equations.

Given a regular planar contour $\bm{\gamma}: \mathbb{T}^1 \to \mathbb{R}^2$, its average speed
\[
\sigma = \frac{1}{2\pi} \int_{\mathbb{T}^1} |\bm{\gamma}'(u)| \dd u
\]
is its arc length divided by $2\pi$. Reparametrizing by the solution $\xi: \mathbb{T}^1 \to \mathbb{T}^1$ of the initial value problem
\[
\xi'(u) = \frac{\sigma}{|\bm{\gamma}'(\xi(u))|}, \qquad \xi(-\pi) = -\pi
\]
gives the \emph{constant-speed representation} $\bm{\psi} := \bm{\gamma} \circ \xi$ of the same curve, which satisfies $|\bm{\psi}'(u)| = \sigma$ for all $u \in \mathbb{T}^1$ (Algorithm \ref{alg:reparametrize} describes the numerical implementation). Its \emph{shape} $\bm{\phi} := \bm{\psi}/\sigma$ is a unit-speed curve, and the pair $(\bm{\phi}, \sigma)$ is called the \emph{shape-scaling representation}. In \cite{Scott2016Fronts} the scale is described instead by $\varepsilon := 1/\sigma^2$.

The unit tangent $\bm{\tau} = \bm{\phi}' = \bm{\psi}'/\sigma$ and the outward normal $\mathbf{n} = \bm{\tau}^\perp$ can be written as
\[
\bm{\tau}(u) = \begin{bmatrix} - \sin \theta(u) \\ \,\,\,\,\,\cos\theta(u) \end{bmatrix}, \qquad \mathbf{n}(u) = \begin{bmatrix} \cos \theta(u) \\ \sin\theta(u) \end{bmatrix},
\]
which defines the \emph{tangent-angle function} $\theta(u) := \arctan \left( -\bm{\tau}(u)_1, \bm{\tau}(u)_2 \right)$, chosen to be continuous in $u$. Since $\bm{\tau}$ is periodic, $\theta(\pi) - \theta(-\pi) = 2 \pi k$ for an integer $k$, the winding number. The tangent-angle function $\theta$, the average speed $\sigma$, and the centroid $\bm{\mu} = \frac{1}{2\pi}\int_{\mathbb{T}^1} \bm{\psi}(u) \dd u$ together form the \emph{tangent-angle representation} $(\theta, \sigma, \bm{\mu})$ of the contour.

For an integrable function $f$ on $\mathbb{T}^1$, let $\bar{f} = \mu_f = \fint_{\mathbb{T}^1} f(u) \dd u = \frac{1}{2\pi} \int_{\mathbb{T}^1} f(u) \dd u$ denote its mean. We use two primitive operators: the mean-zero primitive
\[
P_c [f](u) := \int_{-\pi}^u f(v) \dd v + \frac{1}{2\pi} \int_{\mathbb{T}^1}(v-\pi) f(v) \dd v,
\]
which satisfies $\overline{P_c [f]} = 0$ (and is a periodic function on $\mathbb{T}^1$ only if $\mu_f = 0$), and the zero-Dirichlet primitive of $f - \mu_f$,
\[
P_0[f](u) := \int_{-\pi}^u f(v)\dd v - (u + \pi) \mu_f.
\]
The constant-speed contour is recovered from its tangent-angle representation by
\begin{equation}\label{eq:tangent_angle_recovery}
\bm{\psi}(u) = \bm{\mu} + \sigma P_c [\bm{\tau}] (u), \qquad \bm{\tau}(u) = \begin{bmatrix} - \sin \theta(u) \\ \,\,\,\,\,\cos\theta(u) \end{bmatrix}.
\end{equation}

\subsection{Governing Equations in Tangent-Angle Representation}\label{sec:dynamics_main}

In the tangent-angle representation, the nondimensionalized governing equations \eqref{eq:full_model_nondim} become (see Supplementary Section~\ref{sec:derivation} for the derivation)
\begin{equation}\label{eq:tangent-angle-governing-equations}
\begin{split}
\theta_t^{(i)} 
&= 
\frac{1}{{\sigma^{(i)}}^2} \left[ \theta^{(i)}_{uu}  + P_0[a^{(i)}] \theta^{(i)}_u 
- \mathcal{F}^{(i)}_u\right] \\
\partial_t \left[ {\sigma^{(i)}}^2 \right] &= -2\mu_{a^{(i)}} \\
\partial_t \bm{\mu}^{(i)} &= \frac{1}{2 \pi \sigma^{(i)}} \int_{\mathbb{T}^1} P_0[a^{(i)}] \bm{\tau}^{(i)} + \mathcal{F}^{(i)} \mathbf{n}^{(i)} \dd v \\
a^{(i)} &= (\theta^{(i)}_u)^2 - \mathcal{F}^{(i)} \theta^{(i)}_u \\
\mathcal{F}^{(i)}(u, t) &= \sigma^{(i)} \left( c^{(i)} -\right. \\&\tau\int_{\mathbb{T}^1} f\left(\frac{1}{2} \left\lvert \bm{\psi}^{(i)}(u,t) - \bm{\psi}^{(i)}(v,t) \right\rvert^2\right) \sigma^{(i)} \dd v  \\
&\left.- \tau\sum_{j \neq i} \int_{\mathbb{T}^1} g\left(\frac{1}{2} \left\lvert \bm{\psi}^{(i)}(u,t) - \bm{\psi}^{(j)}(v,t) \right\rvert^2\right) \sigma^{(j)} \dd v \right) \\
\bm{\psi}^{(i)} &= \bm{\mu}^{(i)} + \sigma^{(i)} P_c[\bm{\tau}^{(i)}],
\end{split}
\end{equation}
along with the winding number and periodicity constraints for the tangent-angle function:
\[
\theta^{(i)}(\pi, t) - \theta^{(i)}(-\pi, t) = 2 \pi k^{(i)}, \qquad k^{(i)} \in \mathbb{Z}
\]
\[
\theta^{(i)}_u(\pi, t) = \theta^{(i)}_u(-\pi, t).
\]
Note that the centroid equation in \eqref{eq:tangent-angle-governing-equations} carries the factor $1/\sigma^{(i)} = \varepsilon^{(i)\,1/2}$ (in the notation $\varepsilon = 1/\sigma^2$ of \cite{Scott2016Fronts}), whereas the equation for $\theta^{(i)}$ carries $1/{\sigma^{(i)}}^2 = \varepsilon^{(i)}$; the numerical scheme (Algorithm \ref{alg:timestep}) must respect this difference.
Here $\mathcal{F}^{(i)}$ is the normal forcing (growth minus nonlocal inhibition) scaled by $\sigma^{(i)}$, and $P_0[a^{(i)}] \theta^{(i)}_u$ is a nonlocal tangential (advection) term, which only reparametrizes the contour so as to keep its speed constant; it should not be confused with the nonlocal transport mechanism of Section \ref{sec:transport_force}. This form of the equations is well suited to stable numerical methods: its highest-order term is a linear diffusion term whose coefficient $1/{\sigma^{(i)}}^2$ depends only on time, and which can therefore be integrated exactly (Algorithm \ref{alg:propagate2}), while the nonlinear terms involve at most first derivatives of $\theta^{(i)}$.

\subsection{Model Extension: Nonlocal Transport Mechanism}\label{sec:transport_force}

In this subsection we introduce a nonlocal transport mechanism which permits direct attractive or repulsive colony-scale drift induced by interactions between neighboring colonies when intercolony distances decrease below a specified threshold. While the standard model \eqref{eq:full_model} is already capable of expressing colony-scale transport via the nonlocal interaction force, the nonlocal transport mechanism introduced here directly parametrizes this type of transport, thus decoupling it from the self- and cross-interactions of the nonlocal interaction force. Additionally, it can capture differences in growth rate between the front and back edges of colonies which the interaction kernels may not capture easily: an attractive transport mechanism acts like an increased growth rate on the ``front'' edge of a colony (facing its neighbors) and a decreased growth rate on its ``back'' edge, regardless of the precise distance of each edge from the other colonies. Section \ref{sec:results_transport} describes the performance of this extension quantitatively.

Denote the nonlocal transport force by $\mathbf{h}^{(i)}(t) \in \mathbb{R}^2$, $i \in [m]$. We apply this nonlocal transport force to the whole of each colony $i$ as a rigid translation, and take $\mathbf{h}^{(i)}(t)$ to be the sum of pairwise forces acting in the direction of the displacement between centroids of the contours. Let $\bm{\mu}^{(i)}(t) \in \mathbb{R}^2$ be the location of the centroid of contour $i$ at time $t$, which for a constant-speed parametrization may be calculated using the formula
\[\bm{\mu}^{(i)}(t) = \frac{1}{2\pi} \int_{\mathbb{T}^1} \bm{\gamma}^{(i)}(u,t)\dd u.
\]
Then the nonlocal transport mechanism is given by a sum of pairwise forces
\begin{equation}\label{eq:transport_force}
\begin{split}
    \mathbf{h}^{(i)}(t) 
    &= \alpha\sum_{j\neq i} \frac{\bm{\mu}^{(i)}(t) - \bm{\mu}^{(j)}(t)}{|\bm{\mu}^{(i)}(t) - \bm{\mu}^{(j)}(t)|}\, \chi_w\!\left(s^{(ij)}(t) - \beta\right), \\
    s^{(ij)}(t) &:= \min\limits_{u, v\in \mathbb{T}^1}\left\lvert\bm{\gamma}^{(i)}(u, t) - \bm{\gamma}^{(j)}(v, t)\right\rvert,
\end{split}
\end{equation}
where $s^{(ij)}(t)$ is the minimum distance between colonies $i$ and $j$, and the gate $\chi_w : \mathbb{R} \to [0, 1]$ is the smoothstep function of width $w > 0$
\begin{equation}\label{eq:smoothstep}
\chi_w(r) := \begin{cases}
1, & r \leq -w/2, \\
1 - 3\left( \tfrac{1}{2} + \tfrac{r}{w} \right)^2 + 2\left(\tfrac{1}{2} + \tfrac{r}{w} \right)^3 , & |r| < w/2, \\
0, & r \geq w/2.
\end{cases}
\end{equation}
Thus the pairwise force between colonies $i$ and $j$ has full strength $|\alpha|$ when the colonies are closer than $\beta - w/2$, vanishes when they are farther apart than $\beta + w/2$, and has half strength when $s^{(ij)} = \beta$; the threshold $\beta$ is the distance at which the force switches on. The gate $\chi_w$ is continuously differentiable, and as $w \to 0$ it converges to the indicator function $\mathbf{1}_{\{s^{(ij)} < \beta\}}$. We use the smooth gate rather than the indicator because the indicator makes the model's right-hand side, and therefore the training loss, discontinuous in $\beta$ and in the state; with the indicator the derivative of the loss with respect to $\beta$ is zero almost everywhere (see Section \ref{sec:optimization_problems}). The width $w$ is held fixed and is not a trainable parameter. A positive value of $\alpha$ pushes colonies apart and a negative value pulls them together; we do not constrain its sign.

Adding the nonlocal transport mechanism to \eqref{eq:full_model} yields the model with transport
\begin{equation}\begin{split}\label{eq:full_model_transport}
&\frac{\partial}{\partial t} \bm{\gamma}^{(i)} (u, t) 
= \,\mathbf{n}^{(i)}(u,t) \left[ \mu \, \kappa_{\bm{\gamma}^{(i)}}(u,t)+c^{(i)} - \right. \\
&\int_{\mathbb{T}^1} f\left(\frac{1}{2} \left\lvert \bm{\gamma}^{(i)}(u,t) - \bm{\gamma}^{(i)}(v,t) \right\rvert^2\right) \left\lvert \partial_u \bm{\gamma}^{(i)}(v,t) \right\rvert \dd v  \\
&\left. - \sum_{j \neq i} \int_{\mathbb{T}^1} g\left(\frac{1}{2} \left\lvert \bm{\gamma}^{(i)}(u,t) - \bm{\gamma}^{(j)}(v,t) \right\rvert^2\right) \left\lvert \partial_u \bm{\gamma}^{(j)}(v,t)\right\rvert \dd v\right] \\
&+ \alpha\sum_{j\neq i} \frac{\bm{\mu}^{(i)}(t) - \bm{\mu}^{(j)}(t)}{|\bm{\mu}^{(i)}(t) - \bm{\mu}^{(j)}(t)|}\, \chi_w\!\left(s^{(ij)}(t) - \beta\right), \qquad i \in [m]; \\
    &s^{(ij)}(t) = \min\limits_{u, v\in \mathbb{T}^1}\left\lvert\bm{\gamma}^{(i)}(u, t) - \bm{\gamma}^{(j)}(v, t)\right\rvert, \qquad i,j \in [m].
\end{split}\end{equation}
Setting $\alpha = 0$ in \eqref{eq:full_model_transport} recovers the model \eqref{eq:full_model} of \cite{Scott2016Fronts}, which we refer to as the \emph{standard model}. Note that the standard model carries per-colony growth rates $c^{(i)}$ and allows $f \neq g$ as in \cite{Scott2016Fronts}. Section \ref{sec:results_prior} analyzes the cases where $c^{(i)} = c$ (shared growth rates) or $f = g$ (shared interaction kernels).

\textbf{Effect on the governing equations.} Since $\mathbf{h}^{(i)}(t)$ is a rigid translation, it changes neither the arc length nor the shape of a colony directly, and the evolution of the curve traced by $\bm{\gamma}^{(i)}(\cdot, t)$ still does not depend on its parametrization. In the tangent-angle representation, the equations for $\theta^{(i)}$ and $\sigma^{(i)}$ are therefore unchanged, and the transport mechanism enters only through the centroid equation (Supplementary Section~\ref{sec:deriv_transport}). Nondimensionalizing as in Section \ref{sec:nondim} multiplies $\mathbf{h}^{(i)}$ by $\tau$; renaming $\alpha' = \tau \alpha$ to $\alpha$ (the threshold $\beta$ and width $w$ are lengths and are unaffected), the governing equations with transport are
\begin{equation}\label{eq:tangent-angle-governing-equations-transport}
\begin{split}
\theta_t^{(i)} 
&= 
\frac{1}{{\sigma^{(i)}}^2} \left[ \theta^{(i)}_{uu}  + P_0[a^{(i)}] \theta^{(i)}_u 
- \mathcal{F}^{(i)}_u\right] \\
\partial_t \left[ {\sigma^{(i)}}^2 \right] &= -2\mu_{a^{(i)}} \\
\partial_t \bm{\mu}^{(i)} &= \frac{1}{2 \pi \sigma^{(i)}} \int_{\mathbb{T}^1} P_0[a^{(i)}] \bm{\tau}^{(i)} + \mathcal{F}^{(i)} \mathbf{n}^{(i)} \dd v + \mathbf{h}^{(i)} \\
\mathbf{h}^{(i)} &= \alpha\sum_{j\neq i} \frac{\bm{\mu}^{(i)} - \bm{\mu}^{(j)}}{|\bm{\mu}^{(i)} - \bm{\mu}^{(j)}|}\, \chi_w\!\left(s^{(ij)} - \beta\right) \\
s^{(ij)} &= \min_{u, v \in \mathbb{T}^1} \left\lvert \bm{\psi}^{(i)}(u, t) - \bm{\psi}^{(j)}(v, t) \right\rvert \\
a^{(i)} &= (\theta^{(i)}_u)^2 - \mathcal{F}^{(i)} \theta^{(i)}_u \\
\mathcal{F}^{(i)}(u, t) &= \sigma^{(i)} \left( c^{(i)} -\right. \\&\tau\int_{\mathbb{T}^1} f\left(\frac{1}{2} \left\lvert \bm{\psi}^{(i)}(u,t) - \bm{\psi}^{(i)}(v,t) \right\rvert^2\right) \sigma^{(i)} \dd v  \\
&\left.- \tau\sum_{j \neq i} \int_{\mathbb{T}^1} g\left(\frac{1}{2} \left\lvert \bm{\psi}^{(i)}(u,t) - \bm{\psi}^{(j)}(v,t) \right\rvert^2\right) \sigma^{(j)} \dd v \right) \\
\bm{\psi}^{(i)} &= \bm{\mu}^{(i)} + \sigma^{(i)} P_c[\bm{\tau}^{(i)}],
\end{split}
\end{equation}
with the same winding number and periodicity constraints as before. The transport mechanism translates each colony rigidly, but it still affects the shapes and growth of the colonies indirectly, because the cross-interaction term in $\mathcal{F}^{(i)}$ depends on the distances between colonies. As before, setting $\alpha = 0$ recovers \eqref{eq:tangent-angle-governing-equations}.

\section{Model Training}\label{sec:mt}

\subsection{Trainable Model Parameters}\label{sec:parameters}

Let $\mathcal{C} = \{ \bm{\gamma}^{(i)}\}_{i\in [m]}$ be a collection of $m$ parametrized smooth, simple, counterclockwise oriented contours $\bm{\gamma}^{(i)}: \mathbb{T}^1 \times \mathbb{R}^+ \to \mathbb{R}^2$ representing boundaries of colonies of \emph{Paenibacillus dendritiformis} bacteria. The dynamics of these interfaces over time are described by the nonlocal equations \eqref{eq:full_model_transport}. Since these reduce to the standard model \eqref{eq:full_model} when $\alpha = 0$, we work with the more general form throughout this section, so that one set of equations covers both cases. 
These governing equations contain $\mu, c^{(1)}, \ldots, c^{(m)} \in \mathbb{R}^+$ as model parameters, together with $\alpha \in \mathbb{R}$ and $\beta \in \mathbb{R}^+$ when the transport mechanism is included (the gate width $w$ is held fixed). Furthermore, the functions $f$ and $g$ are assumed to be functions from $\mathbb{R}^+$ to $\mathbb{R}^+$ which are members of a chosen parametrized class of interaction kernels. In this work, two classes of interaction kernels, each parametrized using two parameters, are considered:
\begin{align*}
\text{Gaussian} := \Big\{ f: \mathbb{R}^+ \to \mathbb{R} \quad \Big| &\quad f(s) = \frac{c}{\sigma \sqrt{2 \pi}} \exp \left( - \frac{s}{\sigma^2} \right),  \\
&\quad (c, \sigma) \in (\mathbb{R}^+)^2 \Big\}
\end{align*}
\begin{align*}
\text{Ramp} 
:= \Big\{ f: \mathbb{R}^+ \to \mathbb{R} \quad \Big| &\quad f(s) = \max(a -b\sqrt{s}, \,0), \\&\quad (a, b) \in (\mathbb{R}^+)^2 \Big\}
\end{align*}
Note that because the interaction kernels $f$ and $g$ use one-half of the squared distance as their argument, a function $f \in \text{Ramp}$ will have the shape of a downward piecewise linear ramp in terms of distance, hence the name.

Since both the Ramp and Gaussian classes of interaction kernels are closed under nonnegative scalar multiplication, the rescaled kernels $\tau f$ and $\tau g$ in the nondimensionalized equations (Section \ref{sec:nondim}) remain within the same class for any $\tau > 0$, so we train them directly. In the same way, $c^{(i)}$ and $\alpha$ denote the nondimensional growth rates and transport strength of Sections \ref{sec:nondim} and \ref{sec:transport_force}.

Altogether, the standard model ($\alpha = 0$) is parametrized by a set of $m+5$ parameters: the scaling parameters $\{c^{(i)}\}_{i \in [m]}$ and $\tau$, along with two parameters for each of the interaction kernels $f$ and $g$. In the case of the Gaussian kernels, these parameters may be written as
\[
\{c^{(1)}, \ldots, c^{(m)}, \tau, c_f, \sigma_f, c_g, \sigma_g\} \in (\mathbb{R}^+)^{m+5},
\]
and in the case of the Ramp kernels, these parameters may be written as
\[
\{c^{(1)}, \ldots, c^{(m)}, \tau, a_f, b_f, a_g, b_g\} \in (\mathbb{R}^+)^{m+5}.
\]

If a shared growth rate is used, then the number of parameters decreases to six in each case, and if shared interaction kernels are used, then the number of parameters decreases by two (to $m+3$ if using varying growth rates, and to four if using a constant growth rate). 

Including the nonlocal transport mechanism adds two parameters, the strength $\alpha \in \mathbb{R}$ and the threshold $\beta \in \mathbb{R}^+$, for a total of $m+7$ (the width $w$ of the gate \eqref{eq:smoothstep} is held fixed) if the rest of the model is standard. Because the gate is continuously differentiable, the loss depends continuously and differentiably on $\beta$ (see Section \ref{sec:optimization_problems}); this would not be the case if the gate were replaced by the indicator function $\mathbf{1}_{\{s^{(ij)} < \beta\}}$. 

\subsection{Target Data}

\begin{figure*}
        \centering
        \begin{subfigure}{0.22\textwidth}
          \centering
          \includegraphics[width=\linewidth]{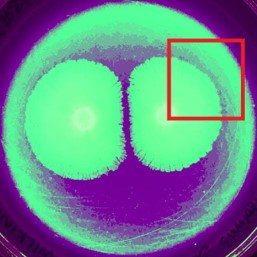}
          \caption{Similar colors in background} \label{difficultiesa}
        \end{subfigure} 
        \begin{subfigure}{0.22\textwidth}
          \includegraphics[width=\linewidth]{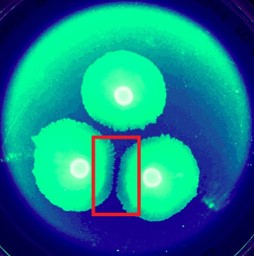}
          \centering
          \caption{False edges within the colony} \label{difficultiesb}
        \end{subfigure}
        \begin{subfigure}{0.22\textwidth}
          \includegraphics[width=\linewidth]{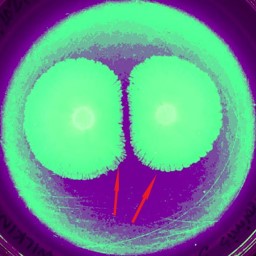}
          \centering
          \caption{Feathering effect on colony edges}  \label{difficultiesc}
        \end{subfigure} 
        \caption{Difficulties in Contour Extraction} \label{difficulties}
      \end{figure*}

This work uses data from \emph{Paenibacillus dendritiformis} bacterial colony growth experiments in \cite{dud22}. Details about the experiments can be found in Appendix \ref{sec:experiments_appendix} and in \cite{dud22}. The experiments yielded seven image sequences of Petri dishes containing growing collections of bacterial colonies.

\begin{figure}[H]
  \centering
  \begin{subfigure}[H]{0.22\textwidth}
    \centering
    \includegraphics[width=\linewidth]{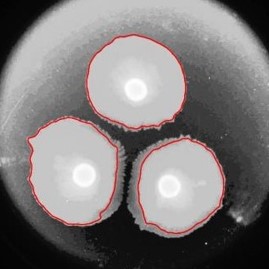}
    \caption{Thresholding} \label{threshold}
  \end{subfigure} 
  \hfill 
  \begin{subfigure}[H]{0.22\textwidth}
    \includegraphics[width=\linewidth]{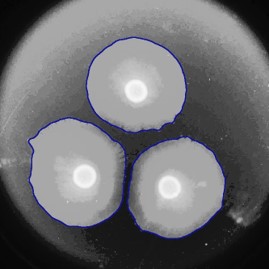}
    \centering
    \caption{Seeded region growing} \label{seeded-reg}
  \end{subfigure}
  \caption{This figure compares two methods of contour extraction (thresholding and seeded region growing) on the \nth{130} image in sequence 4. Here, thresholding is tricked by false edges within the colony and does not correctly capture the boundaries. } 
  \label{comparison}
\end{figure}

Next, an image segmentation procedure was used to identify the regions in each image containing the colonies and the background, after which the boundaries between regions were extracted as contours containing 100 points each. Since a simple thresholding algorithm was insufficient to accurately capture the boundaries of the colonies due to the presence of false edges, the image segmentation procedure was based on seeded region growing (see Figure \ref{comparison}). 

Another crucial goal of the image processing was to extract smooth contours. In the images of \emph{P. dendritiformis} colonies, fine structures sometimes called ``feathering'' appear on the edges of the colonies (see Figure \ref{difficultiesc}). This is undesirable since the nonlocal model equations do not capture this level of intricate detail, but instead treat the contours as smooth curves. To smooth out the segmented regions, a bilateral filter and median blur were applied to the images before segmentation.

After contour extraction, three of the highest-quality sequences of contours were chosen to form the target data to train the model parameters. Dataset 1 was taken from the seventh sequence, Dataset 2 was taken from the fifth sequence, and Dataset 3 was taken from the third sequence. Datasets 1 and 2 are two-colony datasets, the former having two colonies of similar size and the latter having two colonies of differing size, while Dataset 3 is a three-colony dataset. The target data for each contour consists of 100 points in $\mathbb{R}^2$, oriented counterclockwise, which represent locations of pixels corresponding to the associated contour. If we let $u_j = -\pi + \frac{2\pi}{n}(j-1)$ for $j \in [n]$ with $n = 100$, then to define the $i$th target contour at time $t_k$ we let 
\[
\tilde{\bm{\gamma}}^{(i)}_{j, k} := \tilde{\bm{\gamma}}^{(i)}(u_j, t_k) \in \mathbb{R}^2
\]
represent the location of the $j$th extracted point in the contour.

To form a continuous representation of the target data, $\tilde{\bm{\gamma}}^{(i)}(u, t_k)$ can be defined at any other position $u \in \mathbb{T}^1$ using a WENO (Weighted Essentially Non-Oscillatory) interpolation scheme based on quadratic Lagrange polynomials (see Algorithm \ref{alg:weno3}).

The extracted contour points are resampled so that the contour they represent has constant speed (see Algorithm \ref{alg:reparametrize}). Since the sampling points in $\mathbb{T}^1$ are equally spaced, this means that the distance between successive contour points, along the contour (in arc length), should be equal. This enables the speed of the contour and the distance between contour points to be expressed as constants, rather than varying functions on $\mathbb{T}^1$, simplifying the modeling and allowing the use of the tangent-angle representation.

The resampling is done by computing a change of variables function $\tilde{\xi}^{(i)}: \mathbb{T}^1 \times \mathbb{R} \to \mathbb{T}^1$ which satisfies the ODE initial value problem  
\[\partial_u \tilde{\xi}^{(i)}(u)  = \frac{\tilde{\sigma}^{(i)}}{|\partial_u \tilde{\bm{\gamma}}^{(i)}(u)|}, \qquad \tilde{\xi}^{(i)}(-\pi) = -\pi.
\]
The initial value problem is solved using the explicit fourth-order Runge--Kutta scheme, and then a new constant-speed representation $\tilde{\bm{\psi}}^{(i)}: \mathbb{T}^1 \times \mathbb{R} \to \mathbb{R}^2$ is created by interpolating at the new sample locations given through mapping by $\tilde{\xi}^{(i)}$ (see Algorithm \ref{alg:reparametrize} for more detail).

Lastly, the contour points are recalibrated so that the first point is oriented in the $(1, 0)$ direction from the centroid of the contour (see Algorithm \ref{alg:recalibrate}). Once again, we assume that the contours are oriented in a counterclockwise direction, and are star-shaped with respect to their centroids so that this operation is well-defined. The recalibration is done in order to facilitate easy calculation of training metrics by comparing distances between corresponding points on target and simulated contours. 

\subsection{Training Metrics}\label{sec:training_metrics}

\textbf{Continuous Training Metrics.} Let $\{\bm{\psi}^{(i)}\}_{i\in[m]}$ be a set of $m$ continuous constant-speed simulated contours, and let $\{\tilde{\bm{\psi}}^{(i)}\}_{i \in [m]}$ be the continuous constant-speed target contours, with average speed $\tilde{\sigma}^{(i)}(t)$ and centroids $\bm{\mu}^{(i)}(t)$. The absolute loss at each time step is defined as 
\[
L_{\text{abs}}(t) = \frac{1}{m} \sum_{i=1}^m \int_{\mathbb{T}^1} |\bm{\psi}^{(i)}(u, t) - \tilde{\bm{\psi}}^{(i)}(u, t) |^2 \,\tilde{\sigma}^{(i)}(t) \dd u
\]
and the relative loss is defined as 
\[
L_{\text{rel}}(t) = \frac{\sum_{i=1}^m \int_{\mathbb{T}^1} |\bm{\psi}^{(i)}(u, t) - \tilde{\bm{\psi}}^{(i)}(u, t) |^2 \,\tilde{\sigma}^{(i)}(t) \dd u  }{\sum_{i=1}^m \int_{\mathbb{T}^1} |\tilde{\bm{\psi}}^{(i)}(u, t) - \bm{\mu}^{(i)}(t) |^2 \,\tilde{\sigma}^{(i)}(t) \dd u }.
\]
Note that in this metric, the simulated contour at a point $u \in \mathbb{T}^1$, $\bm{\psi}^{(i)}(u, t)$, is compared to $\tilde{\bm{\psi}}^{(i)}(u, t)$, not $\tilde{\bm{\psi}}^{(i)}(v,t)$ for some $v \neq u$, so the metric requires that $\bm{\psi}^{(i)}(u, t)$ and $\tilde{\bm{\psi}}^{(i)}(u, t)$ correspond to the same part of their respective contours. Ideally, 
\[
u = \underset{v \in \mathbb{T}^1}{\text{argmin}}  \, |\bm{\psi}^{(i)}(v, t) - \tilde{\bm{\psi}}^{(i)}(u, t) |
\]
but this may be impossible to accomplish if the contours have different shapes. Instead, we ensure comparability of the parametrizations of the contours by calibrating each so that $\bm{\psi}^{(i)}(-\pi, t)$ and $\tilde{\bm{\psi}}^{(i)}(-\pi, t)$ correspond to the points directly to the right of the respective centroids (assuming the contours are star-shaped with respect to each centroid). See Algorithm \ref{alg:recalibrate} for further details.

The (absolute or relative) loss over an interval $[0, T]$ is calculated as the mean of the (absolute or relative) loss at each time step. For instance,
\[
L_{\text{abs}} = \frac{1}{T} \int_0^T \frac{1}{m} \sum_{i=1}^m \int_{\mathbb{T}^1} |\bm{\psi}^{(i)}(u, t) - \tilde{\bm{\psi}}^{(i)}(u, t) |^2 \,\tilde{\sigma}^{(i)}(t) \dd u \dd t
\]
\[
L_{\text{rel}} = \frac{1}{T} \int_0^T \frac{\sum_{i=1}^m \int_{\mathbb{T}^1} |\bm{\psi}^{(i)}(u, t) - \tilde{\bm{\psi}}^{(i)}(u, t) |^2 \,\tilde{\sigma}^{(i)}(t) \dd u  }{\sum_{i=1}^m \int_{\mathbb{T}^1} |\tilde{\bm{\psi}}^{(i)}(u, t) - \bm{\mu}^{(i)}(t) |^2 \,\tilde{\sigma}^{(i)}(t) \dd u } \dd t
\]

\textbf{Discrete Training Metrics.} As before, let $\{\bm{\psi}^{(i)}\}_{i\in[m]}$ be a set of $m$ continuous constant-speed simulated contours, and let $\{\tilde{\bm{\psi}}^{(i)}\}_{i \in [m]}$ be the continuous constant-speed target contours, with average speed $\tilde{\sigma}^{(i)}(t)$ and centroids $\bm{\mu}^{(i)}$. Now if we let 
\[
u_j = -\pi + \frac{2\pi}{n}(j-1), \qquad j \in [n],
\]
we may define
\[
\bm{\psi}^{(i)}_j = \bm{\psi}^{(i)}(u_j), \qquad \tilde{\bm{\psi}}^{(i)}_j = \tilde{\bm{\psi}}^{(i)}(u_j) \qquad i \in [m], \,\, j \in [n].
\]
When running a numerical simulation, these collections of $n$ points constitute the full representation of the contour, so it is worth explaining the implementation of the training metrics in this discrete case. The absolute loss is defined as 
\[
L_{\text{abs}} = \frac{1}{m n} \sum_{i=1}^m \sum_{j=1}^n |\bm{\psi}_{j}^{(i)}-\tilde{\bm{\psi}}_{j}^{(i)} |^2 \, \tilde{\sigma}^{(i)}
\]
and the relative loss is defined as 
\[
L_{\text{rel}} = \frac{\sum_{i=1}^m \sum_{j=1}^n |\bm{\psi}_{j}^{(i)}-\tilde{\bm{\psi}}_{j}^{(i)} |^2 \,\tilde{\sigma}^{(i)} }{\sum_{i=1}^m \sum_{j=1}^n | \tilde{\bm{\psi}}_{j}^{(i)} - \bm{\mu}^{(i)} |^2\,\tilde{\sigma}^{(i)} }
\]

Now suppose we have $K$ snapshots of a set of $m$ target and simulated constant-speed contours, consisting of $n$ points each, at times $\{t_1, \ldots, t_K\}$. We may write the associated simulated points as 
\[
\{\bm{\psi}_{j, k}^{(i)},  \quad i\in[m], \quad j \in [n], \quad k \in [K] \} \subseteq \mathbb{R}^2 
\]
and the associated target points as 
\[
\{\tilde{\bm{\psi}}_{j, k}^{(i)},  \quad i\in[m], \quad j \in [n], \quad k \in [K] \} \subseteq \mathbb{R}^2.
\]
Suppose that the average speeds and centroids of the target contours are given by 
\[
\{ \tilde{\sigma}^{(i)}_k, \quad i \in [m], \quad k \in [K]\}
\]
and
\[
\{ \bm{\mu}^{(i)}_k, \quad i \in [m], \quad k \in [K]\}.
\]
Then we can define the absolute and relative loss over the time sequence $\{t_1, \ldots, t_K\}$ by 
\[
L_{\text{abs}} = \frac{1}{Kmn} \sum_{k=1}^K \sum_{i=1}^m \sum_{j=1}^n |\bm{\psi}_{j, k}^{(i)} -\tilde{\bm{\psi}}_{j, k}^{(i)} |^2 \, \tilde{\sigma}^{(i)}_k
\]
\begin{equation}\label{eq:l_rel}
L_{\text{rel}} = \frac{1}{K} \sum_{k=1}^K \frac{\sum_{i=1}^m \sum_{j=1}^n |\bm{\psi}_{j, k}^{(i)}-\tilde{\bm{\psi}}_{j, k}^{(i)} |^2 \,\tilde{\sigma}^{(i)}_k }{\sum_{i=1}^m \sum_{j=1}^n | \tilde{\bm{\psi}}_{j, k}^{(i)} - \bm{\mu}^{(i)}_k |^2\,\tilde{\sigma}^{(i)}_k }.
\end{equation}

\textbf{Comparison with Prior Work.} The training approach of \cite{dud22} fit a version of the standard model with shared Gaussian self- and cross-interaction kernels and a shared growth rate, as described above. Results on that class of models are in Section \ref{sec:results_prior}. 

The work of \cite{dud22} also used a different loss: each boundary point on the target (observed) boundary contour was compared with the nearest point on the simulated contour, weighted by arc length and summed over the images. The loss used here instead compares points with the same parameter value $u$, after both contours have been given constant-speed parametrizations and calibrated to a common reference direction. This makes the loss at each image a smooth quadratic function of the simulated contour points, whereas a nearest-neighbor distance has kinks whenever the nearest neighbor of a point changes. The relative loss is also normalized by the spread of the target contours at each image, so that losses can be compared across datasets whose colonies differ in size. 

\subsection{Optimization Problems}\label{sec:optimization_problems}

Training consists of an iterative optimization method. Each evaluation of the loss requires simulating discrete contours over an interval $[0, T]$ and comparing them with target discrete contours extracted from images at a series of times $\{0, t_1, \ldots, t_K\} \subseteq [0, T]$. Model parameters are then updated based on an optimization scheme. The objective is the relative loss $L_{\text{rel}}$ (see Eq.~\eqref{eq:l_rel}) of the previous section.

This is a PDE-constrained optimization problem. Computing the gradient of the objective exactly would require either the adjoint of the nonlinear, nonlocal governing equations or automatic differentiation through the specialized numerical scheme used to solve them. Since the model has at most ten parameters, it is instead practical to approximate the gradient by central finite differences, at a cost of $2d+1$ forward simulations per gradient for $d$ parameters. These simulations are independent of each other and can be run in parallel. For this to work, the discrete objective must be differentiable in the parameters, at least on the scale of the finite-difference step; the end of this section explains why this holds after two small modifications to the numerical scheme. We use two optimization methods: the derivative-free Nelder--Mead method (Section \ref{sec:nelder_mead}) and the quasi-Newton method L-BFGS-B with finite-difference gradients (Section \ref{sec:lbfgs}).

Simply computing the simulated evolution of the contours on an interval $[0, T]$ with a prescribed initial condition is already expensive, so each evaluation of the objective is costly. The wall-clock time for an evaluation of the objective in the standard model can be as low as $15.5$ seconds if $\tau$ is large, but as long as $423$ seconds if $\tau$ is small and more time steps are required.

This section describes the constrained optimization problem at three levels of complexity, starting with the optimization problem in the continuous planar contour formulation, then moving to the continuous tangent-angle formulation, and then moving to the discrete tangent-angle formulation (which is the one actually implemented in code to train the model). Details on this formulation are further explained in Supplementary Sections~\ref{sec:optimization_algorithms} and~\ref{sec:numerical_algorithms}.

All of the contours are assumed to have constant speed, in the sense that the derivative of the parametrization they define has a constant magnitude. Hence they are denoted by $\bm{\psi}$ and $\tilde{\bm{\psi}}$ instead of $\bm{\gamma}$ and $\tilde{\bm{\gamma}}$. Furthermore, in each formulation, the class of kernels $\mathcal{F}$ may be chosen to be Ramp or Gaussian, which are each two-parameter families. The simulated contours are calibrated as described in the previous section before they are compared with the target contours. To shorten the notation, let
\[
D(t) := \sum_{i=1}^m \int_{\mathbb{T}^1} |\tilde{\bm{\psi}}^{(i)}(u, t) - \bm{\mu}^{(i)}(t) |^2 \,\tilde{\sigma}^{(i)}(t) \dd u
\]
denote the spread of the target contours at time $t$ (the denominator of the relative loss), which does not depend on the model parameters. The problems are stated for the model with transport; for the standard model, $\alpha = 0$ is fixed and the minimization is over the remaining $m + 5$ parameters.

\textbf{Optimization Problem for Continuous Planar Contours. } The optimization problem for continuous planar contours may be formulated as follows:
\begin{align*}
&\min_{\substack{f, g \in \mathcal{F};\; c^{(1)}, \ldots, c^{(m)}, \tau \in \mathbb{R}^+ \\ \alpha \in \mathbb{R};\; \beta \in \mathbb{R}^+ }} \frac{1}{T}\int_0^T  \frac{1}{D(t)} \\
&\qquad\qquad \times \sum_{i=1}^m \int_{\mathbb{T}^1} |\bm{\psi}^{(i)}(u, t) - \tilde{\bm{\psi}}^{(i)}(u, t) |^2 \,\tilde{\sigma}^{(i)}(t) \dd u \dd t\\
&\text{subject to } \quad \frac{\partial}{\partial t} \bm{\psi}^{(i)} (u, t)
= \mathbf{n}_{\bm{\psi}^{(i)}}(u, t) \left[ (\kappa_{\bm{\psi}^{(i)}}(u, t)+c^{(i)})\right. \\
&\left. -\int_{\mathbb{T}^1} \tau f\left(\frac{1}{2} \left\lvert \bm{\psi}^{(i)}(u, t) - \bm{\psi}^{(i)}(v,t) \right\rvert^2\right) \left\lvert \partial_u \bm{\psi}^{(i)}(v,t) \right\rvert \dd v \right.\\
& \left. -\sum_{j \neq i} \int_{\mathbb{T}^1} \tau g\left(\frac{1}{2} \left\lvert \bm{\psi}^{(i)}(u, t) - \bm{\psi}^{(j)}(v,t) \right\rvert^2\right) \left\lvert \partial_u \bm{\psi}^{(j)}(v,t)\right\rvert \dd v  \right]\\
&+ \alpha\sum_{j\neq i} \frac{\bm{\mu}^{(i)}(t) - \bm{\mu}^{(j)}(t)}{\|\bm{\mu}^{(i)}(t) - \bm{\mu}^{(j)}(t)\|_2}\, \chi_w\!\left(s^{(ij)}(t) - \beta\right),  \\
& \bm{\psi}^{(i)}(u, 0) = \tilde{\bm{\psi}}^{(i)}(u, 0)
\qquad i \in [m]
\end{align*}

\textbf{Optimization Problem for Continuous Tangent-Angle Representation of Contours. } The optimization problem using the equivalent governing equations expressed in the tangent-angle representation may be formulated as follows:

\begin{align*}
&\min_{\substack{f, g \in \mathcal{F};\; c^{(1)}, \ldots, c^{(m)}, \tau \in \mathbb{R}^+ \\ \alpha \in \mathbb{R};\; \beta \in \mathbb{R}^+ }} \frac{1}{T}\int_0^T  \frac{1}{D(t)} \\
&\qquad\qquad \times \sum_{i=1}^m \int_{\mathbb{T}^1} |\bm{\psi}^{(i)}(u, t) - \tilde{\bm{\psi}}^{(i)}(u, t) |^2 \,\tilde{\sigma}^{(i)}(t) \dd u \dd t\\
&\text{subject to } \partial_t \theta^{(i)}(u, t)
= \varepsilon^{(i)}(t) \left(\partial_{u}^2 \theta^{(i)} - \partial_u \mathcal{F}^{(i)} \right) \\
&+ \varepsilon^{(i)}(t)\left[ \int_{-\pi}^u a^{(i)}(v) \dd v - (u + \pi) \fint_{\mathbb{T}^1} a^{(i)}(v) \dd v \right] \partial_u \theta^{(i)}, \\
& \partial_t \varepsilon^{(i)} = 2 \varepsilon^{(i)}(t)^2 \fint_{\mathbb{T}^1} a^{(i)}(v) \dd v \\
&  \partial_t \bm{\mu}^{(i)}(t) = \varepsilon^{(i)}(t)^{1/2} \fint_{\mathbb{T}^1} \mathbf{v}^{(i)}(u) \dd u + \mathbf{h}^{(i)}(t) \\
&\mathcal{F}^{(i)}(u, t) =  \sigma^{(i)}(t) \left[ c^{(i)} - \tau\int_{\mathbb{T}^1} f\left(\frac{1}{2} \left\lvert \bm{\psi}^{(i)}(u, t) - \bm{\psi}^{(i)}(v,t) \right\rvert^2\right) \sigma^{(i)}(t) \dd v  \right. \\
&\left.  - \tau\sum_{j \neq i} \int_{\mathbb{T}^1} g\left(\frac{1}{2} \left\lvert \bm{\psi}^{(i)}(u, t) - \bm{\psi}^{(j)}(v,t) \right\rvert^2\right) \sigma^{(j)}(t)\dd v   \right] \\
&a^{(i)}(u, t) = (\partial_u \theta^{(i)}(u, t))^2 - \mathcal{F}^{(i)}(u, t) \partial_u \theta^{(i)}(u, t) \\
&\mathbf{v}^{(i)}(u, t) = \left[\int_{-\pi}^u a^{(i)}(v) \dd v - (u + \pi) \fint_{\mathbb{T}^1} a^{(i)}(v) \dd v \right]\bm{\tau}^{(i)} + \mathcal{F}^{(i)} \mathbf{n}^{(i)} \\
& \bm{\psi}^{(i)}(u ,t) = \bm{\mu}^{(i)}(t) + \sigma^{(i)}(t) P_c[\bm{\tau}^{(i)}(\cdot, t)](u) \\
& \bm{\psi}^{(i)}(u, 0) = \tilde{\bm{\psi}}^{(i)}(u, 0)
\qquad i \in [m]
\end{align*}
where $\varepsilon^{(i)} = 1/{\sigma^{(i)}}^2$ and $\mathbf{h}^{(i)}$ is the nonlocal transport mechanism \eqref{eq:transport_force}.

\textbf{Optimization Problem for Discrete Tangent-Angle Representation of Contours. } Let $\{0, t_1, \ldots, t_K \} \subseteq [0, T]$ be the sequence of times at which the images are taken, and let
\[
D_k := \sum_{i=1}^m \sum_{j=1}^n | \tilde{\bm{\psi}}_{j, k}^{(i)} - \bm{\mu}^{(i)}_k |^2\,\tilde{\sigma}^{(i)}_k.
\]
Then the governing equations may be simulated numerically (see Algorithms \ref{alg:timestep} and \ref{alg:timestepping} for the implementation used in this work), and the optimization problem using the discrete formulation is as follows:

\begin{align*}
&\min_{\substack{f, g \in \mathcal{F};\; c^{(1)}, \ldots, c^{(m)}, \tau \in \mathbb{R}^+ \\ \alpha \in \mathbb{R};\; \beta \in \mathbb{R}^+ }}  \sum_{k=1}^K \frac{1}{D_k} \sum_{i=1}^m \sum_{j=1}^n |\bm{\psi}^{(i)}_{j,k} - \tilde{\bm{\psi}}^{(i)}_{j,k} |^2 \,\tilde{\sigma}^{(i)}_k\\
&\text{subject to } \{\bm{\psi}^{(i)}_{j,k}, i \in [m], j \in [n], k \in [K] \} \text{ generated by } \\ &\text{Algorithm \ref{alg:timestepping} with } N_{\text{sub}} \text{ substeps per image and } \bm{\psi}^{(i)}_{j,0} = \tilde{\bm{\psi}}^{(i)}_{j,0}, \\
&\qquad i \in [m], j \in [n]
\end{align*}

\textbf{Time Discretization.} Algorithm \ref{alg:timestepping} (Supplementary Section~\ref{sec:optimization_algorithms}) advances the solution from one image to the next in a fixed number $N_{\text{sub}}$ of equal substeps. Since consecutive images are $1/\tau$ units of nondimensional time apart (Section \ref{sec:nondim}), each substep has length
\[
\Delta t = \frac{1}{\tau N_{\text{sub}}}.
\]
The number $N_{\text{sub}}$ is chosen once, before an optimization run begins, and is held fixed during the run, so that a change in $\tau$ changes the length of each substep but not the number of substeps. We choose $N_{\text{sub}}$ so that $\Delta t / \Delta x \leq 0.75$ at the initial value of $\tau$, where $\Delta x = 2\pi/n$ is the grid spacing on $\mathbb{T}^1$. As $\tau$ changes during an optimization run, $\Delta t / \Delta x$ may drift; in runs where it drifted too high, we restarted the run from its best point with a larger value of $N_{\text{sub}}$. 

Optimization runs with small $\tau$ values, which need many substeps, were sometimes run with coarser steps (up to $\Delta t/\Delta x \approx 3$) to reduce their cost; but every loss and parameter set reported in Section \ref{sec:r} was re-evaluated with $\Delta t / \Delta x \leq 0.75$. 

The time-stepping scheme is second-order accurate in $\Delta t$ (see Supplementary Section~\ref{sec:numerical_algorithms}), so the discretization error falls by a factor of about four each time $N_{\text{sub}}$ is doubled. For example, at the optimized parameters of the standard model for Dataset 1 with Gaussian kernels, the relative difference between the objective and its value computed with 400 substeps per image was $3.6 \times 10^{-6}$, $1.4 \times 10^{-5}$, $5.8 \times 10^{-5}$, and $2.7 \times 10^{-4}$ at $\Delta t / \Delta x = 0.18$, $0.36$, $0.71$, and $1.48$ respectively. The error is therefore negligible when $\Delta t / \Delta x \leq 0.75$, and the fitted parameters are insensitive to the precise choice of $N_{\text{sub}}$ as long as it is not too small.

The time-stepping scheme also contains a time step restriction which keeps the average speeds $\sigma^{(i)}$ positive (Algorithm \ref{alg:timestep}). With a fixed number of substeps, this restriction is only a safeguard, and it never became active in any of our runs: at the optimized parameters of the configurations we checked, the maximum allowed step exceeded $\Delta t$ by a factor of at least $43$ (Supplementary Section~\ref{sec:numerical_algorithms}).

\textbf{Differentiability of the Objective.} Gradient-based optimization requires the objective to be differentiable in the model parameters, at least on the scale of the finite-difference step used to estimate its gradient (Section \ref{sec:lbfgs}). Two features of earlier versions of our numerical scheme made the discrete objective discontinuous, and both have been removed:
\begin{enumerate}
    \item \emph{The transport gate.} Using the smoothstep $\chi_w$ in place of the indicator function $\mathbf{1}_{\{s^{(ij)} < \beta\}}$ makes the objective continuously differentiable in both $s^{(ij)}$ and $\beta$. Using the indicator function makes the objective piecewise constant in $\beta$: its derivative with respect to $\beta$ is zero wherever it exists, and jumps occur whenever a change in the parameters moves the time at which a pair of colonies crosses the threshold $\beta$ past a time step. 
    \item \emph{The number of time steps.} Earlier versions used a fixed time step $\Delta t$ and took a shorter final step onto each image time. The number of steps between images, $\lceil 1/(\tau \Delta t) \rceil$, then changes by one whenever $1/\tau$ crosses a multiple of $\Delta t$, and the objective jumps. Near the optimum for Dataset 1 with Gaussian kernels, the jump across one such boundary was $1.6 \times 10^{-4}$ of the objective, and the forward difference quotient with respect to $1/\tau$ across the jump was about $7.8$, compared with values of order $0.1$ on either side. With a fixed number of substeps per image, the time grid scales with $1/\tau$ and the objective is smooth.
\end{enumerate}

The remaining operations in the numerical scheme are continuous and piecewise smooth. The FFT-based differentiation and integration (Algorithms \ref{alg:derivative} and \ref{alg:primitive}) are linear, and the exponential time-differencing update (Algorithm \ref{alg:propagate2}) and the Newton iteration used for recentering (Algorithm \ref{alg:normalize}) are smooth. The WENO interpolation (Algorithm \ref{alg:weno3}) branches on the ratio of its two smoothness indicators for numerical stability, but both branches evaluate the same rational function of the indicators, which are bounded away from zero; the interpolation weights are therefore smooth functions of the data. Resampling (Algorithm \ref{alg:reparametrize}) is applied only to the target data, since the tangent-angle representation yields a constant-speed contour at every step. Recalibration (Algorithm \ref{alg:recalibrate}) is applied to the simulated contours only when they are compared with the target contours to calculate the loss, and does not feed back into the dynamics; it depends on the parameters through the location of a zero crossing found by linear interpolation, which moves continuously, although its derivative jumps when the crossing moves from one grid interval to the next. Finally, the Ramp kernels $\max(a - b\sqrt{s}, 0)$ and the discrete minimum distance $s^{(ij)}$ in the transport gate are Lipschitz but not differentiable everywhere: the former has a kink when the distance between a pair of grid points crosses the end of the ramp, and the latter when the closest pair of grid points changes. 

The discrete objective is therefore continuous and piecewise smooth, and each of its kinks comes from a single pair of grid points (or a single grid interval) and changes the slope of the objective only slightly. In practice the objective behaves as a smooth function on the scale of our finite-difference step. We checked this directly by computing central-difference estimates of the gradient with steps $h$ ranging from $10^{-6}$ to $10^{-2}$ (relative steps, in the log parametrization of Section \ref{sec:lbfgs}). In a check done with the shared growth rate model with Gaussian kernels at its initial point for Dataset 1, the estimates of all six partial derivatives agreed to within $0.01\%$ for $h \in [10^{-5}, 10^{-3}]$. With the transport mechanism and the smoothstep gate, the estimates of the partial derivatives with respect to $\alpha$ and $\beta$ agreed to within $0.1\%$ and $0.3\%$ respectively. We therefore use $h = 10^{-4}$. Differentiability does not rule out very steep slopes, however: near the small-$\tau$ shared growth rate optima of Section \ref{sec:results_prior}, the objective is steep enough in some directions that refining them with L-BFGS-B was ineffective. 

The Ramp kernels are only Lipschitz, and for them the nonlocal forces are continuously differentiable only under additional conditions on the set of pairs of points at distance $\sqrt{2}\,a/b$; we do not pursue this here. We do not attempt a proof for the discrete model, whose objective is only piecewise smooth for the reasons given above; the numerical checks above show that it behaves as a smooth function at the scale that matters for our optimization.

\subsection{Nelder--Mead Method}\label{sec:nelder_mead}

We use the Nelder--Mead method (sometimes called the amoeba method), a direct search method based on comparing function values at points on a simplex, for optimizing the objective function \cite{gao_implementing_2012}. This method works with a simplex (a polytope of $d+1$ vertices in $\mathbb{R}^d$) by iteratively updating the point with the worst value in the simplex via reflection, expansion, contraction, or shrinkage of the simplex. Algorithm \ref{alg:nelder-mead} (Supplementary Section~\ref{sec:optimization_algorithms}) describes \texttt{scipy}'s implementation of the Nelder--Mead method \cite{virtanen2020scipy}. While the Nelder--Mead method has descent guarantees in the case of a uniformly convex function, our problem is not uniformly convex and these guarantees do not apply.

We apply the Nelder--Mead method to the model parameters directly (not to the log parametrization of Section \ref{sec:lbfgs}), with \texttt{scipy}'s default convergence tolerances ($10^{-4}$ on the size of the simplex and on the spread of the function values across its vertices). When the simplex converges, we restart the method from the best point found with a fresh simplex, and we stop once a restart improves the best loss by less than a relative amount of $10^{-4}$, or when the time budget for the run is exhausted.

Because the Nelder--Mead method moves a simplex of parameters, the loss is evaluated many times. The corresponding loss curve appears somewhat oscillatory as the algorithm tests points and finds that some do not have as low an objective function value as those previously tested.

\subsection{L-BFGS-B with Finite-Difference Gradients}\label{sec:lbfgs}

Since the objective is differentiable on the scale of our finite-difference step (Section \ref{sec:optimization_problems}), it can also be minimized with a gradient-based quasi-Newton method. We use L-BFGS-B \cite{byrd1995limited, zhu1997algorithm}, a limited-memory BFGS method which handles bound constraints, as implemented in \texttt{scipy} \cite{virtanen2020scipy}, with the gradient approximated by central differences \cite{nocedal2006numerical}. The only bounds are the wide box constraints described below, which serve as a safeguard and are not active at any of the solutions we report. The limited memory usage is not strictly necessary for an optimization problem with few parameters; nevertheless, L-BFGS-B was chosen for its modern implementation in \texttt{scipy}. Algorithm \ref{alg:lbfgs} (Supplementary Section~\ref{sec:optimization_algorithms}) summarizes the method.

\textbf{Log Parametrization.} The model parameters differ in magnitude by several orders of magnitude, and all of them except $\alpha$ must be positive. We therefore optimize over variables $\mathbf{u}$ with $x_i = e^{u_i}$ for every parameter $x_i$ except the transport parameters $\alpha$ and $\beta$, which are optimized directly ($\alpha$ because it may change sign; $\beta$ is bounded below by $10^{-3}$). In these variables positivity is automatic, a step in $u_i$ is a relative change in $x_i$, and a single finite-difference step size is appropriate for all of the log-parametrized parameters. We also bound each log-parametrized variable to within $5$ of its initial value, i.e.\ to within a factor of $e^5 \approx 150$ of the initial parameter.

\textbf{Central Differences.} The partial derivative with respect to $u_i$ is approximated by
\[
\hat{g}_i(\mathbf{u}) = \frac{L(\mathbf{u} + h_i \mathbf{e}_i) - L(\mathbf{u} - h_i \mathbf{e}_i)}{2 h_i},
\]
with $h_i = h$ for the log-parametrized variables and $h_i = h\, |x_{0,i}|$ for $\alpha$ and $\beta$, where $\mathbf{x}_0$ is the initial point. One gradient therefore costs $2d$ evaluations of the objective in addition to the evaluation at $\mathbf{u}$ itself (13 in the six-parameter case), all of which are independent and are run in parallel. We use $h = 10^{-4}$, which lies well inside the range of steps over which the estimates agree (Section \ref{sec:optimization_problems}): it is large enough that rounding errors in the difference of two nearly equal losses are negligible, and small enough that the $O(h^2)$ truncation error is not visible.

\textbf{Scaling of the First Step.} On its first iteration, L-BFGS-B has no curvature information and uses the identity as its approximation of the Hessian, so its first trial step is the negative gradient itself. For our objective, the gradient at the initial point has a norm of order $10^2$--$10^3$ in the log variables, so an unscaled first step would change the parameters by enormous factors and drive every variable to its bound. We therefore multiply the objective by the constant $\kappa = \rho_0 / \|\hat{\mathbf{g}}(\mathbf{u}_0)\|$, which does not change its minimizers, so that the first trial step has length $\rho_0$ in the variables $\mathbf{u}$. We use $\rho_0 = 0.05$, i.e.\ a first step which changes the parameters by about $5\%$. Later iterations use the curvature information accumulated by L-BFGS-B.

\textbf{Failed Simulations.} Occasionally a trial point causes the forward simulation to fail, for example when an interface degenerates. At such points we replace the objective by the best value found so far plus the penalty $10^3 (1 + \|\mathbf{u} - \mathbf{u}_{\text{best}}\|^2)$. Because the penalty grows with the distance from the best point, the finite-difference gradient points back toward feasible parameters and the line search backtracks, rather than seeing a flat region that it would mistake for a minimum.

\begin{figure}[H]
    \centering
    \includegraphics[width=\linewidth]{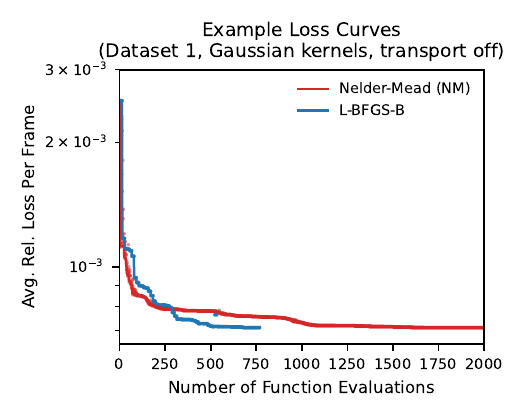}
    \caption{This figure shows the loss at each function evaluation in the Nelder--Mead (NM) algorithm and the L-BFGS-B algorithm with finite differences during training on Dataset 1 with Gaussian kernels, a single growth rate $c$ shared by both colonies, and transport turned off. This run illustrates the general behavior of the two methods; the fits reported in Section \ref{sec:r} use per-colony growth rates. The $x$-axis represents the number of function evaluations; small dots show the actual function evaluations and the solid lines show the cumulatively smallest function evaluation. For this dataset, both NM and L-BFGS-B converge to the same minimum, but L-BFGS-B reaches it more quickly.}
    \label{fig:nelder-mead-loss-curve}
\end{figure}

\textbf{Combining the Two Methods.} Nelder--Mead and L-BFGS-B have complementary strengths. Because the Nelder--Mead simplex takes large exploratory steps, it is more flexible and less likely to get stuck. L-BFGS-B descends to the bottom of a basin with far fewer evaluations of the objective, even though each gradient costs $2d + 1$ evaluations when there are $d$ parameters. Additionally, these evaluations can be run in parallel, so L-BFGS-B can also be much faster in wall-clock time. We therefore mainly rely upon Nelder--Mead to find a basin, and use L-BFGS-B to converge quickly to its minimum. For example, in the last round of optimization for Dataset 1 with Gaussian kernels in a shared growth rate model, Nelder--Mead took about 2298 function evaluations to converge, with 39 evaluations from L-BFGS-B used to descend to a true local minimum. Starting from the same initial point as Nelder--Mead, L-BFGS-B took 767 evaluations to descend to the same minimum. These loss curves are shown in Figure \ref{fig:nelder-mead-loss-curve}. 

\section{Results}\label{sec:r}

In this section we give quantitative results for fitting models to target contours extracted from experimental data. We fit models to three datasets: Dataset 1 (two colonies, similar size), Dataset 2 (two colonies, different size), and Dataset 3 (three colonies), using both the Ramp and the Gaussian classes of interaction kernels. Dataset 1, where the two colonies are approximately the same size, is the least challenging dataset, while Dataset 3, with three colonies of different sizes, is the most challenging. In every case the objective is the relative loss (Algorithm \ref{alg:loss}), computed with a fixed number of substeps per image chosen as in Section \ref{sec:optimization_problems}. Section \ref{sec:results_standard} reports the fits of the standard model, Section \ref{sec:results_prior} reports results for models with shared growth rate and/or shared interaction kernels, and Section \ref{sec:results_transport} reports results for models including the nonlocal transport mechanism. 

Throughout this section, colonies are numbered as follows: in Datasets 1 and 2, colony 1 is the left colony and colony 2 the right colony; in Dataset 3, colony 1 is the left colony, colony 2 the upper colony, and colony 3 the lower-right colony, which is also the largest. 

\subsection{Standard Model}\label{sec:results_standard}

For each configuration we fit the $m+5$ parameters of the standard model \eqref{eq:full_model}, following the strategy of Section \ref{sec:lbfgs} using both Nelder--Mead and L-BFGS-B. The best results for each configuration along with the associated parameters are displayed in Table \ref{tab:standard}. Notably, the best fits have a large range in $\tau$, and the values of $\tau$ in different datasets do not overlap: $\tau = 0.88$--$0.89$ (Dataset 1), $0.19$--$0.40$ (Dataset 2), $0.027$--$0.029$ (Dataset 3).

\begin{table*}
\centering
\footnotesize
\resizebox{\textwidth}{!}{
\begin{tabular}{|c|c|c|c|c|c|c|}
\hline
 & \multicolumn{2}{c|}{Dataset 1} & \multicolumn{2}{c|}{Dataset 2} & \multicolumn{2}{c|}{Dataset 3} \\
 & Gaussian & Ramp & Gaussian & Ramp & Gaussian & Ramp \\
\hline
$c$ & \begin{tabular}{@{}c@{}}$c_1 = 1.93 \times 10^{0}$ \\ $c_2 = 1.87 \times 10^{0}$\end{tabular} & \begin{tabular}{@{}c@{}}$c_1 = 1.84 \times 10^{0}$ \\ $c_2 = 1.79 \times 10^{0}$\end{tabular} & \begin{tabular}{@{}c@{}}$c_1 = 2.80 \times 10^{-1}$ \\ $c_2 = 3.02 \times 10^{-1}$\end{tabular} & \begin{tabular}{@{}c@{}}$c_1 = 5.97 \times 10^{-1}$ \\ $c_2 = 6.44 \times 10^{-1}$\end{tabular} & \begin{tabular}{@{}c@{}}$c_1 = 6.62 \times 10^{-2}$ \\ $c_2 = 6.43 \times 10^{-2}$ \\ $c_3 = 6.96 \times 10^{-2}$\end{tabular} & \begin{tabular}{@{}c@{}}$c_1 = 6.22 \times 10^{-2}$ \\ $c_2 = 6.05 \times 10^{-2}$ \\ $c_3 = 6.53 \times 10^{-2}$\end{tabular} \\ \hline
$\tau$ & $8.94 \times 10^{-1}$ & $8.77 \times 10^{-1}$ & $1.87 \times 10^{-1}$ & $4.03 \times 10^{-1}$ & $2.86 \times 10^{-2}$ & $2.66 \times 10^{-2}$ \\ \hline
\multirow{2}{*}{Self-interaction} & $c_f = 6.01 \times 10^{-1}$ & $a_f = 1.21 \times 10^{-1}$ & $c_f = 7.61 \times 10^{-3}$ & $a_f = 5.84 \times 10^{-2}$ & $c_f = 4.90 \times 10^{-1}$ & $a_f = 1.72 \times 10^{-1}$ \\
 & $\sigma_f = 1.87 \times 10^{0}$ & $b_f = 4.03 \times 10^{-2}$ & $\sigma_f = 5.42 \times 10^{-2}$ & $b_f = 2.44 \times 10^{-2}$ & $\sigma_f = 1.13 \times 10^{0}$ & $b_f = 1.04 \times 10^{-1}$ \\ \hline
\multirow{2}{*}{Cross-interaction} & $c_g = 1.16 \times 10^{0}$ & $a_g = 1.37 \times 10^{-2}$ & $c_g = 9.78 \times 10^{-1}$ & $a_g = 1.04 \times 10^{-2}$ & $c_g = 8.01 \times 10^{-1}$ & $a_g = 6.72 \times 10^{-3}$ \\
 & $\sigma_g = 3.31 \times 10^{1}$ & $b_g = 2.39 \times 10^{-4}$ & $\sigma_g = 3.44 \times 10^{1}$ & $b_g = 1.67 \times 10^{-4}$ & $\sigma_g = 4.35 \times 10^{1}$ & $b_g = 8.69 \times 10^{-5}$ \\ \hline
Avg.\ rel.\ loss & $6.07 \times 10^{-4}$ & $\mathbf{6.05 \times 10^{-4}}$ & $8.19 \times 10^{-4}$ & $\mathbf{7.91 \times 10^{-4}}$ & $2.23 \times 10^{-3}$ & $\mathbf{2.20 \times 10^{-3}}$ \\ \hline
\end{tabular}
}
\caption{Optimized parameters and average relative losses of the standard model (per-colony growth rates, $f \neq g$, no transport) for each dataset and kernel class, irrespective of basin. Losses are the average over images of the relative loss defined in Section \ref{sec:training_metrics}. The lowest loss for each dataset is indicated in bold.}
\label{tab:standard}
\end{table*}

\begin{figure*}
        \centering
        \begin{subfigure}{0.25\textwidth}
            \centering
            \includegraphics[width=0.85\linewidth]{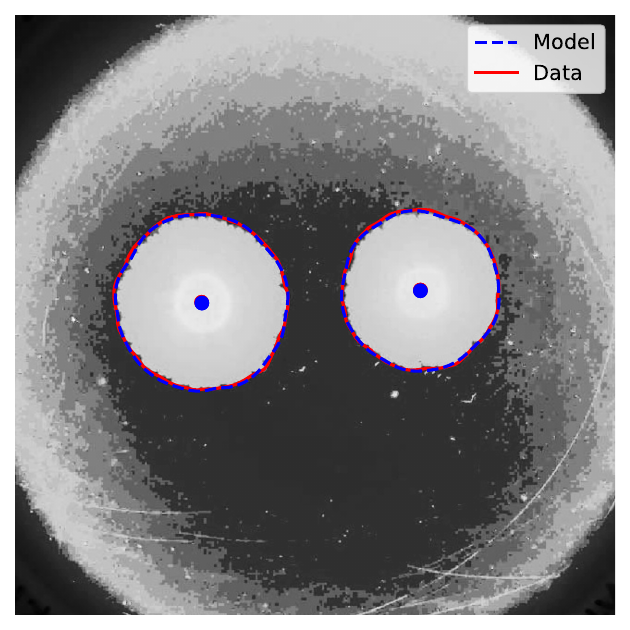}
            \caption{Ramp kernel, step 30}
            \label{DS1.ramp.30}
        \end{subfigure}
        \begin{subfigure}{0.25\textwidth}
            \centering
            \includegraphics[width=0.85\linewidth]{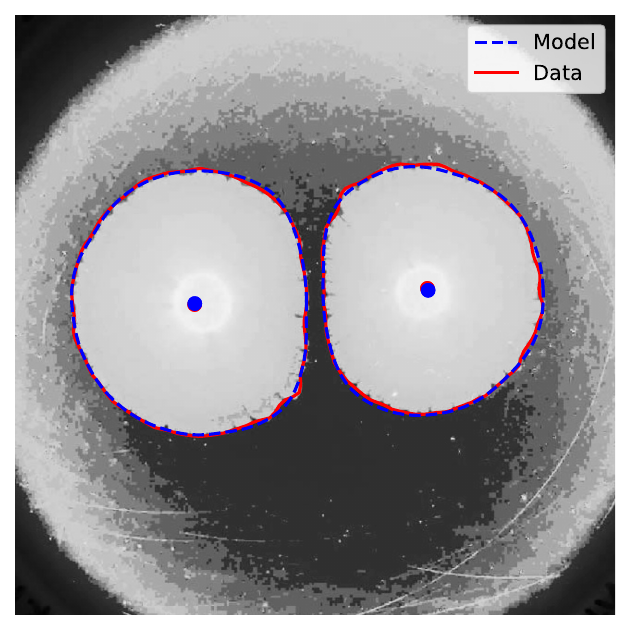}
            \caption{Ramp kernel, step 70}
            \label{DS1.ramp.70}
        \end{subfigure}
        \begin{subfigure}{0.25\textwidth}
            \centering
            \includegraphics[width=0.85\linewidth]{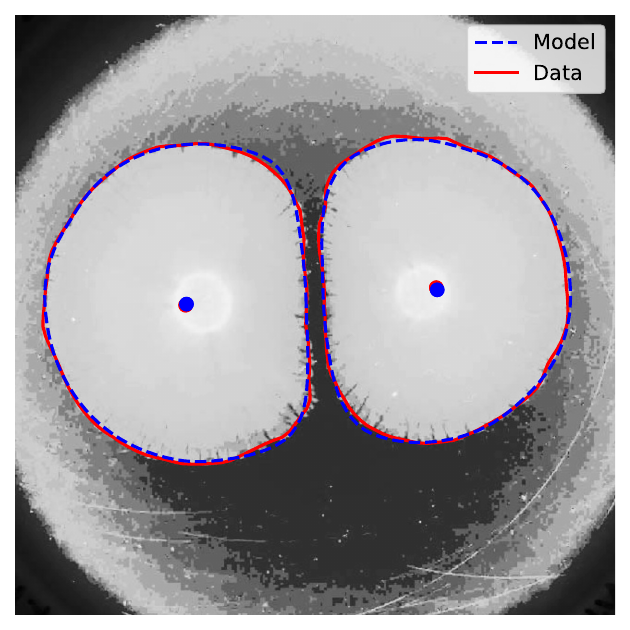}
            \caption{Ramp kernel, step 101}
            \label{DS1.ramp.101}
        \end{subfigure}
        \begin{subfigure}{0.25\textwidth}
            \centering
            \includegraphics[width=0.85\linewidth]{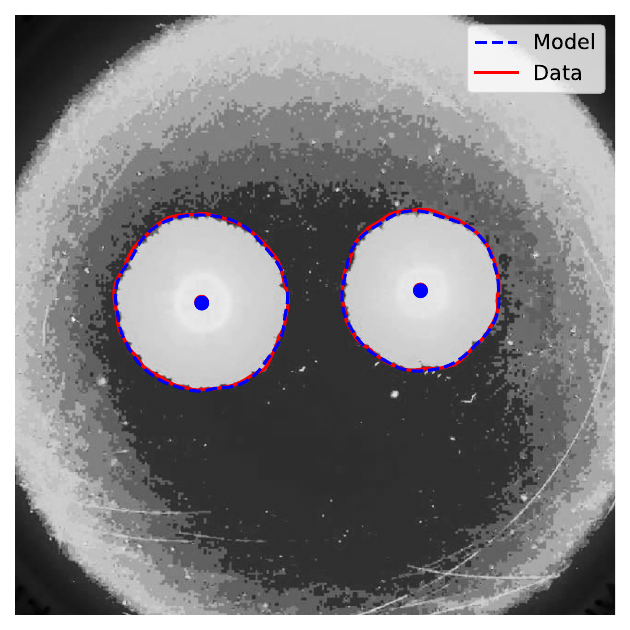}
            \caption{Gaussian kernel, step 30}
            \label{DS1.gaussian.30}
        \end{subfigure}
        \begin{subfigure}{0.25\textwidth}
            \centering
            \includegraphics[width=0.85\linewidth]{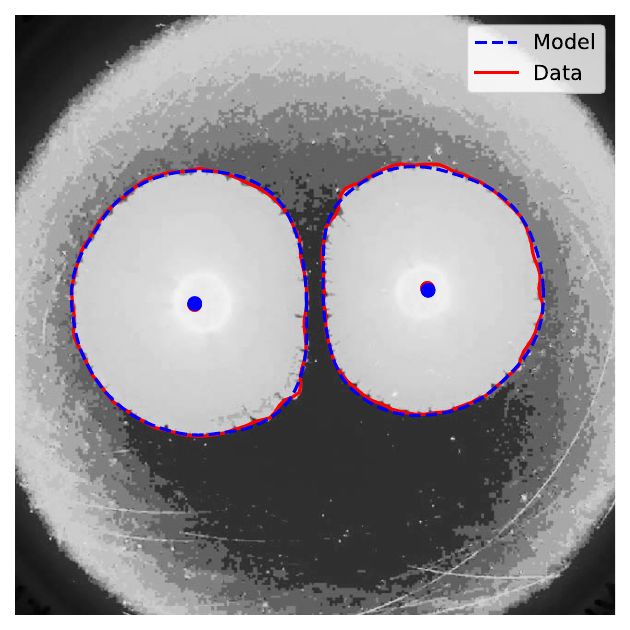}
            \caption{Gaussian kernel, step 70}
            \label{DS1.gaussian.70}
        \end{subfigure}
        \begin{subfigure}{0.25\textwidth}
            \centering
            \includegraphics[width=0.85\linewidth]{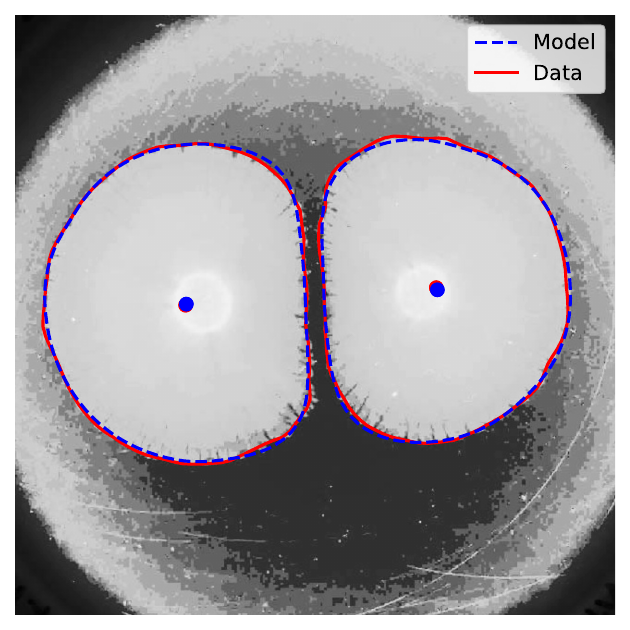}
            \caption{Gaussian kernel, step 101}
            \label{DS1.gaussian.101}
        \end{subfigure}
        \caption{Dataset 1 (two colonies, similar size), standard model. Simulated contours of the best fits of the standard model compared with the target contours. Colony 1 is the left colony and colony 2 the right colony. Here Gaussian and Ramp kernels both perform very well, capturing the shape and positions of the colonies accurately.}
        \label{DS1}
\end{figure*}

\begin{figure*}
        \centering
        \begin{subfigure}{0.25\textwidth}
            \centering
            \includegraphics[width=0.85\linewidth]{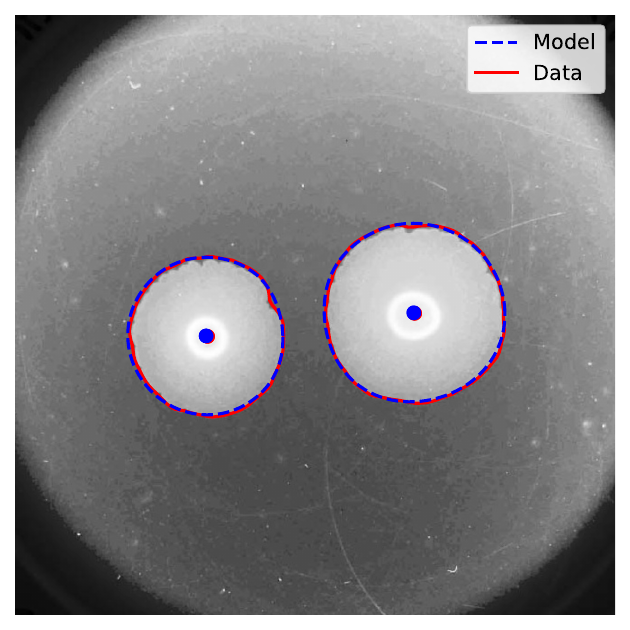}
            \caption{Ramp kernel, step 50}
            \label{DS2.ramp.50}
        \end{subfigure}
        \begin{subfigure}{0.25\textwidth}
            \centering
            \includegraphics[width=0.85\linewidth]{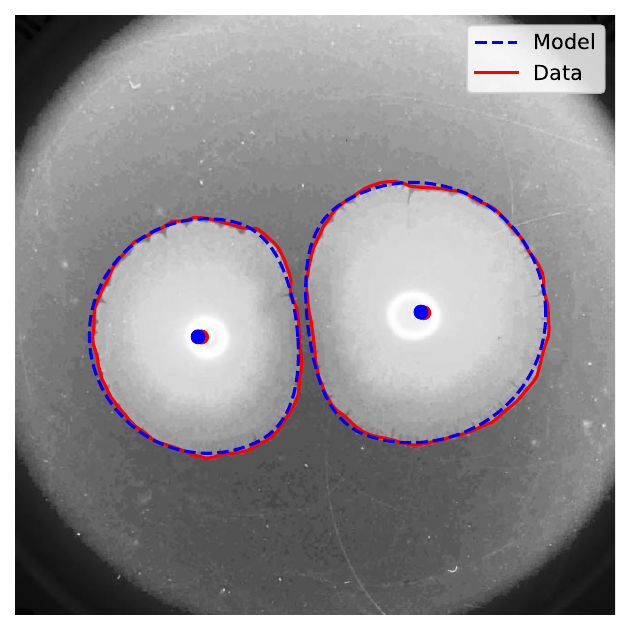}
            \caption{Ramp kernel, step 90}
            \label{DS2.ramp.90}
        \end{subfigure}
        \begin{subfigure}{0.25\textwidth}
            \centering
            \includegraphics[width=0.85\linewidth]{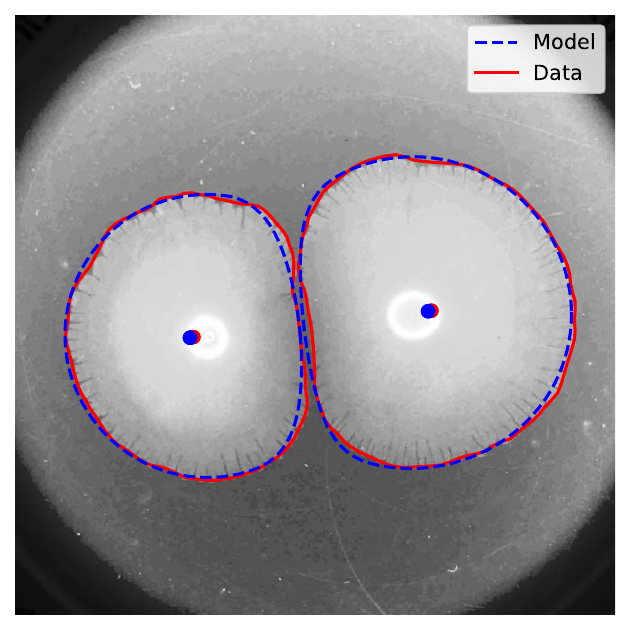}
            \caption{Ramp kernel, step 118}
            \label{DS2.ramp.118}
        \end{subfigure}
        \begin{subfigure}{0.25\textwidth}
            \centering
            \includegraphics[width=0.85\linewidth]{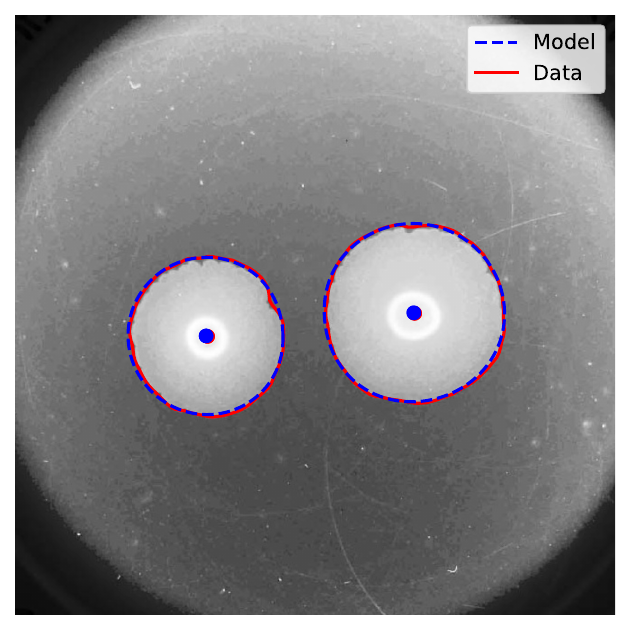}
            \caption{Gaussian kernel, step 50}
            \label{DS2.gaussian.50}
        \end{subfigure}
        \begin{subfigure}{0.25\textwidth}
            \centering
            \includegraphics[width=0.85\linewidth]{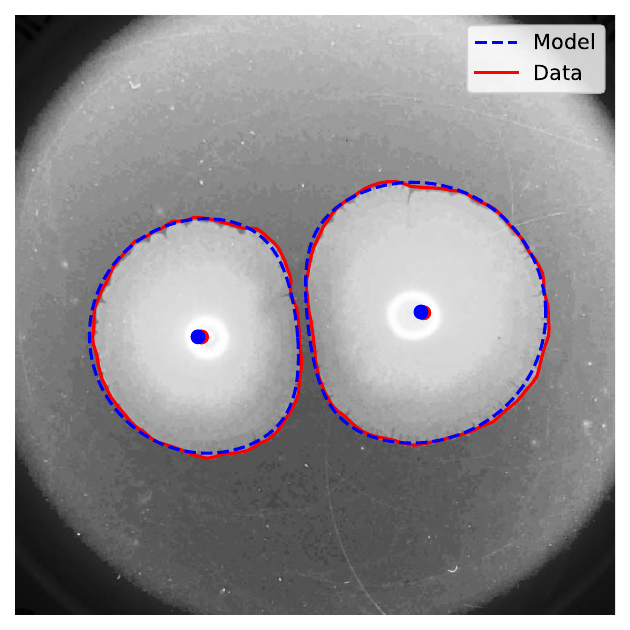}
            \caption{Gaussian kernel, step 90}
            \label{DS2.gaussian.90}
        \end{subfigure}
        \begin{subfigure}{0.25\textwidth}
            \centering
            \includegraphics[width=0.85\linewidth]{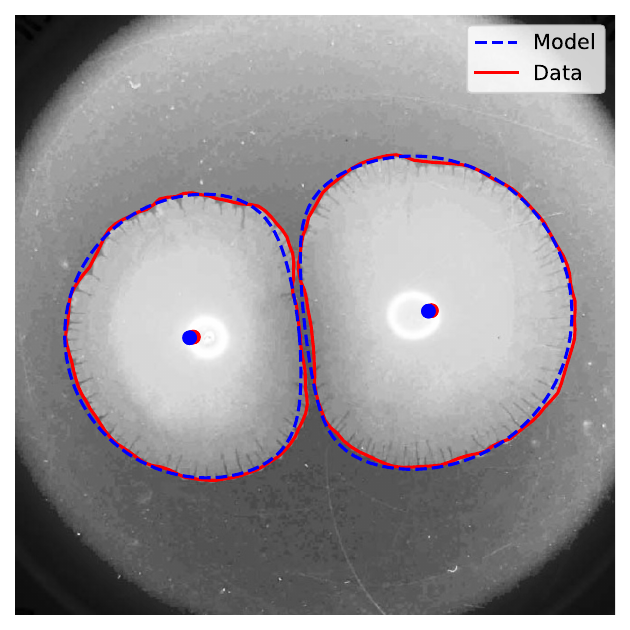}
            \caption{Gaussian kernel, step 118}
            \label{DS2.gaussian.118}
        \end{subfigure}
        \caption{Dataset 2 (two colonies, different size), standard model. Simulated contours of the best fits of the standard model compared with the target contours. Colony 1 is the left colony and colony 2 the right colony. Here Gaussian and Ramp kernels both perform very well, capturing the shape and positions of the colonies accurately.}
        \label{DS2}
\end{figure*}

\begin{figure*}
        \centering
        \begin{subfigure}{0.25\textwidth}
            \centering
            \includegraphics[width=0.85\linewidth]{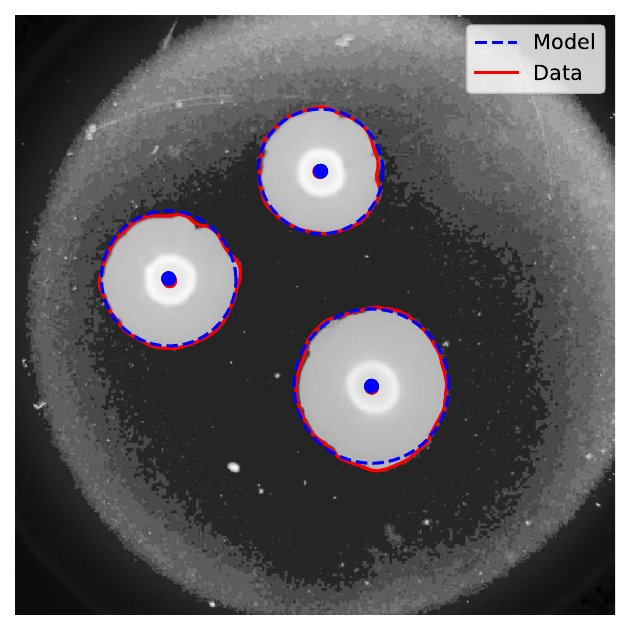}
            \caption{Ramp kernel, step 50}
            \label{DS3.ramp.50}
        \end{subfigure}
        \begin{subfigure}{0.25\textwidth}
            \centering
            \includegraphics[width=0.85\linewidth]{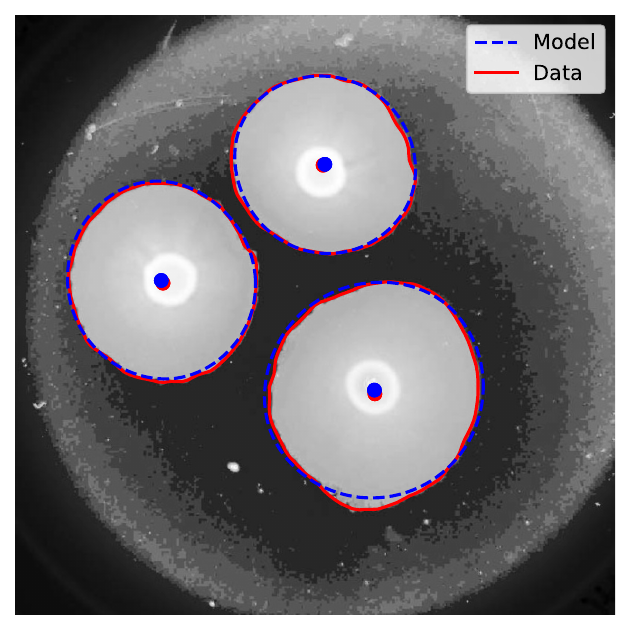}
            \caption{Ramp kernel, step 90}
            \label{DS3.ramp.90}
        \end{subfigure}
        \begin{subfigure}{0.25\textwidth}
            \centering
            \includegraphics[width=0.85\linewidth]{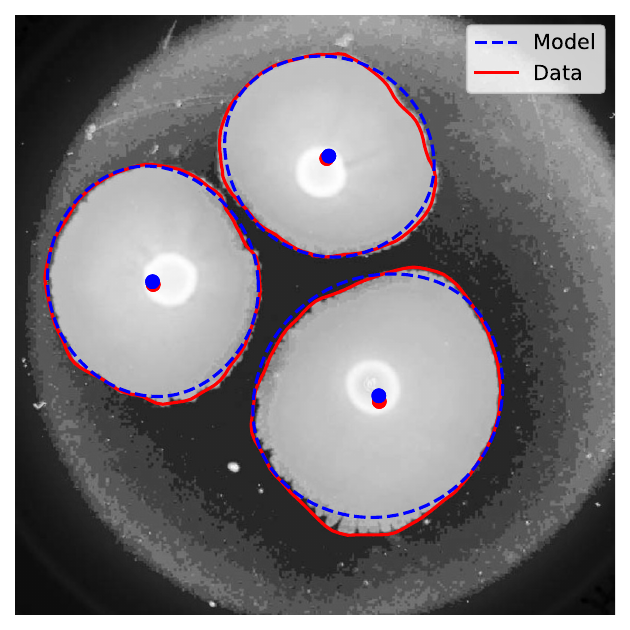}
            \caption{Ramp kernel, step 118}
            \label{DS3.ramp.118}
        \end{subfigure}
        \begin{subfigure}{0.25\textwidth}
            \centering
            \includegraphics[width=0.85\linewidth]{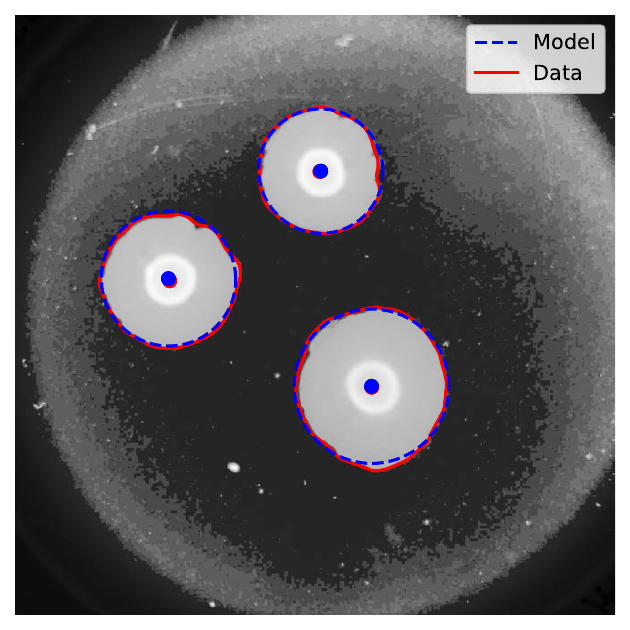}
            \caption{Gaussian kernel, step 50}
            \label{DS3.gaussian.50}
        \end{subfigure}
        \begin{subfigure}{0.25\textwidth}
            \centering
            \includegraphics[width=0.85\linewidth]{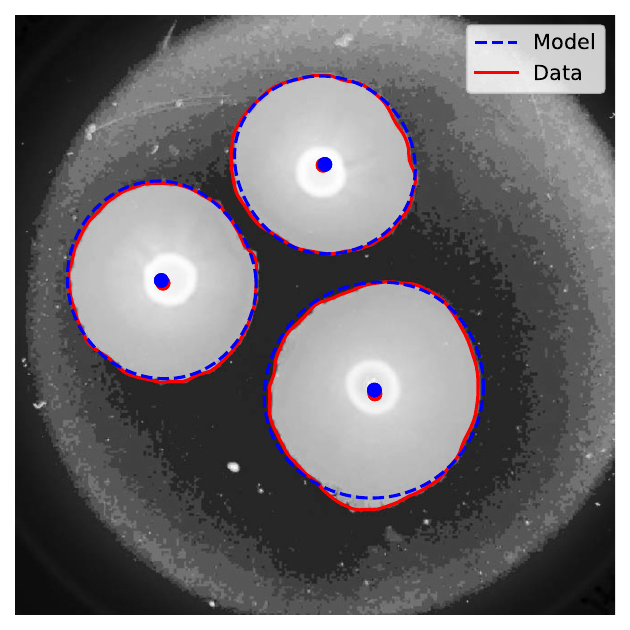}
            \caption{Gaussian kernel, step 90}
            \label{DS3.gaussian.90}
        \end{subfigure}
        \begin{subfigure}{0.25\textwidth}
            \centering
            \includegraphics[width=0.85\linewidth]{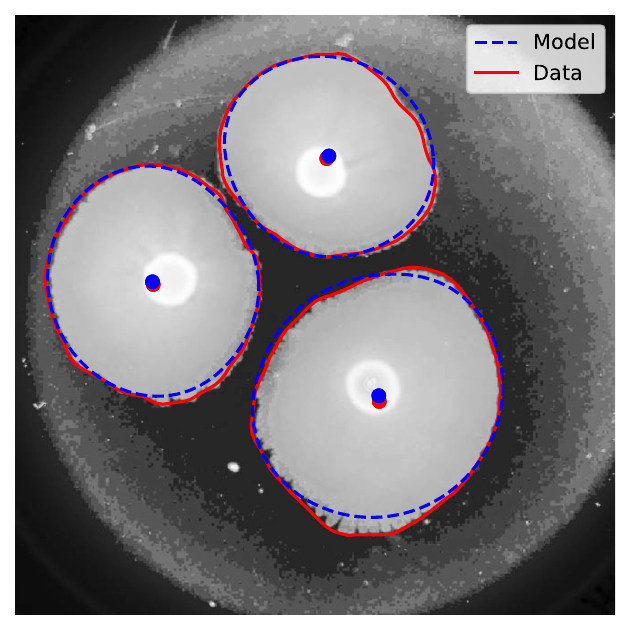}
            \caption{Gaussian kernel, step 118}
            \label{DS3.gaussian.118}
        \end{subfigure}
        \caption{Dataset 3 (three colonies), standard model. Simulated contours of the best fits of the standard model compared with the target contours. Colony 1 is the left colony, colony 2 the upper colony, and colony 3 the lower-right (largest) colony. Here, the model captures the stationary fronts and predicts the sizes of the colonies well. It struggles more in capturing the shapes of the colonies, especially colony 3.}
        \label{DS3}
\end{figure*}

Next we visually compare predictions made by numerically simulating contours with the target contours extracted from experimental data. The simulation begins from the initial extracted contours, and parameters are chosen based on the best fits given by Table \ref{tab:standard}. For each dataset and kernel class, a sequence of three images shows the evolution of the model predictions and the target data over time. Each image shows the raw grayscale image of the bacterial colonies growing in a Petri dish, overlaid with a solid red line and a dotted blue line. The solid red line depicts the contour extracted from the image using the segmentation procedure described in Section \ref{sec:image_segmentation}. The dotted blue line represents the simulated contour at the relevant time step. The centroids of the predicted and target contours are shown as blue and red dots, respectively. Figure \ref{DS1} corresponds to Dataset 1 (two colonies, similar size), Figure \ref{DS2} to Dataset 2 (two colonies, different size), and Figure \ref{DS3} to Dataset 3 (three colonies). Overall, the Gaussian and Ramp kernels perform well and capture the dynamics of colony growth accurately, but they perform best on Dataset 1, where the two colonies are approximately the same size. Dataset 3, with three colonies of different sizes, is the most challenging dataset and has the worst performance.

\begin{figure*}
    \centering
    \includegraphics[width=\linewidth]{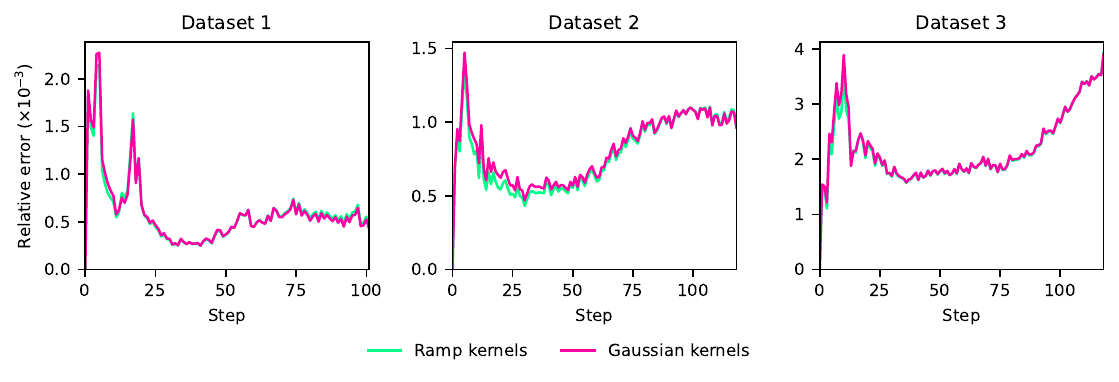}
    \caption{Relative error at each image of the best fits of the standard model for Dataset 1 (two colonies, similar size), Dataset 2 (two colonies, different size), and Dataset 3 (three colonies), for each kernel class.}
    \label{fig:errors}
\end{figure*}

Next we illustrate the relative error at each image of the best fits of the standard model for Datasets 1, 2, and 3 (Figure \ref{fig:errors}). Because the extracted contours are relatively noisy, each dataset contains frames where all best-fit models struggle. The figure shows that the performance of the Gaussian and Ramp kernels is very similar across datasets, with the Ramp kernels performing slightly better in all three datasets.

\begin{figure*}
    \centering
    \includegraphics[width=\linewidth]{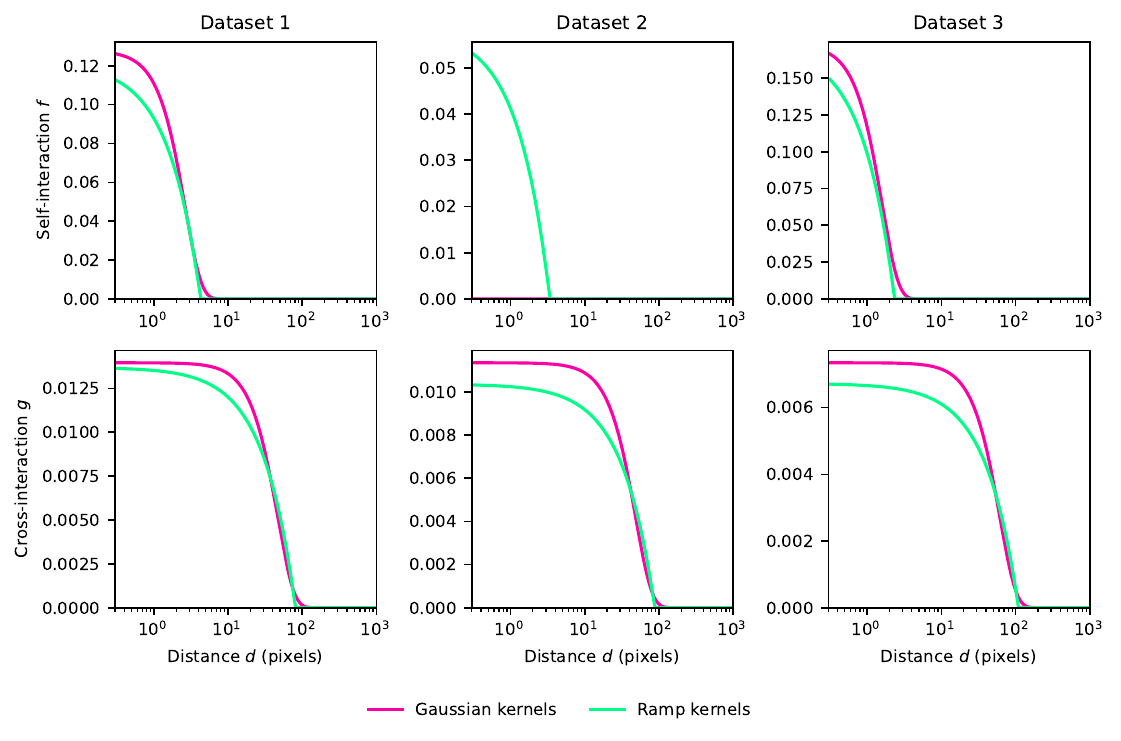}
    \caption{Fitted interaction kernels of the standard model for each dataset (columns): self-interaction kernels $f$ (top row) and cross-interaction kernels $g$ (bottom row), for the best fit of each kernel class. Note that the logarithmic horizontal axis distorts the shapes of Gaussian and Ramp kernels. In every fit, the self-interaction kernel has a higher peak but a much shorter range than the cross-interaction kernel. For Dataset 2 the Gaussian self-interaction kernel ($\sigma_f \approx 0.05$ pixels) is too narrow to be visible on this axis.}
    \label{fig:kernels}
\end{figure*}

Lastly, we display the fitted interaction kernels in the best models for each configuration in Figure \ref{fig:kernels}. The kernels are shown in physical units (in contrast to Table \ref{tab:standard}) as functions of the distance $d$ in pixels between two boundary points; the logarithmic horizontal axis makes both short- and long-range kernels visible but distorts their shapes. Across datasets, the self-interaction kernels have a higher peak but have a much shorter range, reducing their effective strength. Cross-interaction kernels reach over a much larger range, meaning that inhibition between colonies is active from very early times, while self-interaction is much more local and weak. Notably, the best-fit Gaussian self-interaction kernel for Dataset 2 is barely distinguishable from zero. 

\subsection{Comparison to Models from Prior Work}\label{sec:results_prior}

This subsection compares the standard model with per-colony growth rates and distinct self- and cross-interaction kernels with the model class of \cite{dud22} (shared constant growth rate and shared interaction kernels), as well as with the two intermediate models (per-colony growth rates with shared interaction kernels, and shared growth rates with distinct self- and cross-interaction kernels). Optimization is done following the strategy of Section \ref{sec:lbfgs} using both Nelder--Mead and L-BFGS-B. The best parameters and losses are shown in Table \ref{tab:prior_work}. Forcing interaction kernels to be shared ($f = g$) raises the loss more than sharing growth rates, but both raise it substantially. 

\begin{table*}
\centering
\footnotesize
\resizebox{\textwidth}{!}{
\begin{tabular}{|c|c|c|c|c|c|c|c|c|c|}
\hline
 & & \multicolumn{2}{c|}{per-colony $c$, $f\neq g$} & \multicolumn{2}{c|}{shared $c$, $f\neq g$} & \multicolumn{2}{c|}{per-colony $c$, $f=g$} & \multicolumn{2}{c|}{shared $c$, $f=g$} \\
 & & Avg.\ rel.\ loss & $\tau$ & Avg.\ rel.\ loss & $\tau$ & Avg.\ rel.\ loss & $\tau$ & Avg.\ rel.\ loss & $\tau$ \\
\hline
\multirow{2}{*}{Dataset 1} & Gaussian & $\mathbf{6.07 \times 10^{-4}}$ & $8.94 \times 10^{-1}$ & \begin{tabular}{@{}c@{}}$7.14 \times 10^{-4}$ \\ ($+18\%$)\end{tabular} & $7.12 \times 10^{-1}$ & \begin{tabular}{@{}c@{}}$2.44 \times 10^{-3}$ \\ ($+302\%$)\end{tabular} & $1.35 \times 10^{0}$ & \begin{tabular}{@{}c@{}}$2.57 \times 10^{-3}$ \\ ($+323\%$)\end{tabular} & $6.39 \times 10^{-1}$ \\ \cline{2-10}
 & Ramp & $\mathbf{6.05 \times 10^{-4}}$ & $8.77 \times 10^{-1}$ & \begin{tabular}{@{}c@{}}$7.00 \times 10^{-4}$ \\ ($+16\%$)\end{tabular} & $7.05 \times 10^{-1}$ & \begin{tabular}{@{}c@{}}$2.40 \times 10^{-3}$ \\ ($+296\%$)\end{tabular} & $1.33 \times 10^{0}$ & \begin{tabular}{@{}c@{}}$2.52 \times 10^{-3}$ \\ ($+317\%$)\end{tabular} & $6.33 \times 10^{-1}$ \\ \hline
\multirow{2}{*}{Dataset 2} & Gaussian & $\mathbf{8.19 \times 10^{-4}}$ & $1.87 \times 10^{-1}$ & \begin{tabular}{@{}c@{}}$1.50 \times 10^{-3}$ \\ ($+83\%$)\end{tabular} & $3.83 \times 10^{-2}$ & \begin{tabular}{@{}c@{}}$1.71 \times 10^{-3}$ \\ ($+109\%$)\end{tabular} & $7.60 \times 10^{-1}$ & \begin{tabular}{@{}c@{}}$2.36 \times 10^{-3}$ \\ ($+188\%$)\end{tabular} & $7.76 \times 10^{-1}$ \\ \cline{2-10}
 & Ramp & $\mathbf{7.91 \times 10^{-4}}$ & $4.03 \times 10^{-1}$ & \begin{tabular}{@{}c@{}}$1.46 \times 10^{-3}$ \\ ($+84\%$)\end{tabular} & $3.66 \times 10^{-2}$ & \begin{tabular}{@{}c@{}}$1.80 \times 10^{-3}$ \\ ($+128\%$)\end{tabular} & $9.67 \times 10^{-1}$ & \begin{tabular}{@{}c@{}}$2.30 \times 10^{-3}$ \\ ($+191\%$)\end{tabular} & $7.95 \times 10^{-1}$ \\ \hline
\multirow{2}{*}{Dataset 3} & Gaussian & $\mathbf{2.23 \times 10^{-3}}$ & $2.86 \times 10^{-2}$ & \begin{tabular}{@{}c@{}}$3.07 \times 10^{-3}$ \\ ($+37\%$)\end{tabular} & $1.01 \times 10^{-2}$ & \begin{tabular}{@{}c@{}}$3.17 \times 10^{-3}$ \\ ($+42\%$)\end{tabular} & $5.45 \times 10^{0}$ & \begin{tabular}{@{}c@{}}$3.20 \times 10^{-3}$ \\ ($+43\%$)\end{tabular} & $1.76 \times 10^{-2}$ \\ \cline{2-10}
 & Ramp & $\mathbf{2.20 \times 10^{-3}}$ & $2.66 \times 10^{-2}$ & \begin{tabular}{@{}c@{}}$3.07 \times 10^{-3}$ \\ ($+39\%$)\end{tabular} & $5.81 \times 10^{-3}$ & \begin{tabular}{@{}c@{}}$3.13 \times 10^{-3}$ \\ ($+42\%$)\end{tabular} & $2.90 \times 10^{0}$ & \begin{tabular}{@{}c@{}}$5.28 \times 10^{-3}$ \\ ($+140\%$)\end{tabular} & $8.11 \times 10^{-1}$ \\ \hline
\end{tabular}
}
\caption{Average relative losses and values of $\tau$ of the best fits of four model classes, without transport, for each dataset and kernel class: per-colony or shared growth rates, with separate ($f \neq g$) or identical ($f = g$) self- and cross-interaction kernels. The best loss in each configuration (row) is shown in bold, and percentages are changes relative to the first column. The last column, with Gaussian kernels, is the model class of \cite{dud22}.}
\label{tab:prior_work}
\end{table*}

\begin{figure*}
        \centering
        \begin{subfigure}{0.4\textwidth}
            \centering
            \includegraphics[width=0.85\linewidth]{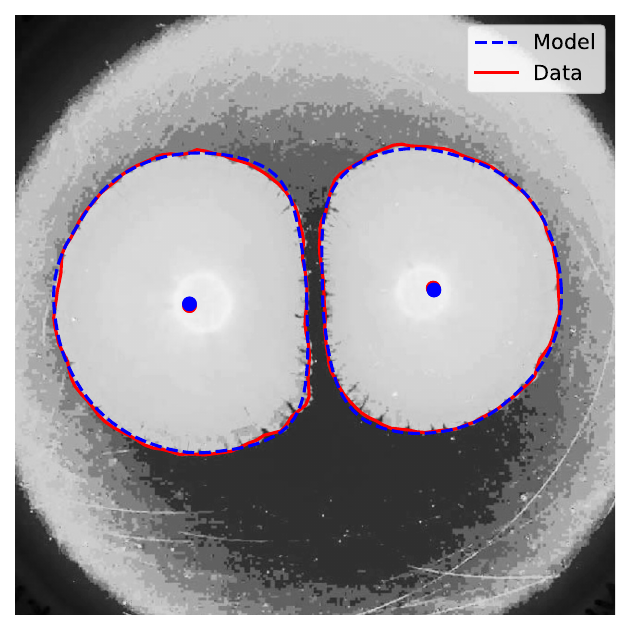}
            \caption{Per-colony $c$, $f\neq g$ (Ramp kernels)}
            \label{DS1_prior.a}
        \end{subfigure}
        \begin{subfigure}{0.4\textwidth}
            \centering
            \includegraphics[width=0.85\linewidth]{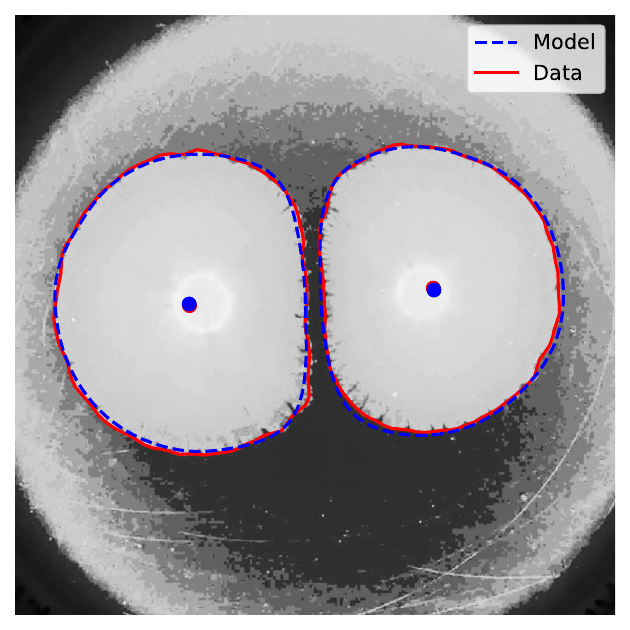}
            \caption{Shared $c$, $f\neq g$ (Ramp kernels)}
            \label{DS1_prior.b}
        \end{subfigure}
        \\
        \begin{subfigure}{0.4\textwidth}
            \centering
            \includegraphics[width=0.85\linewidth]{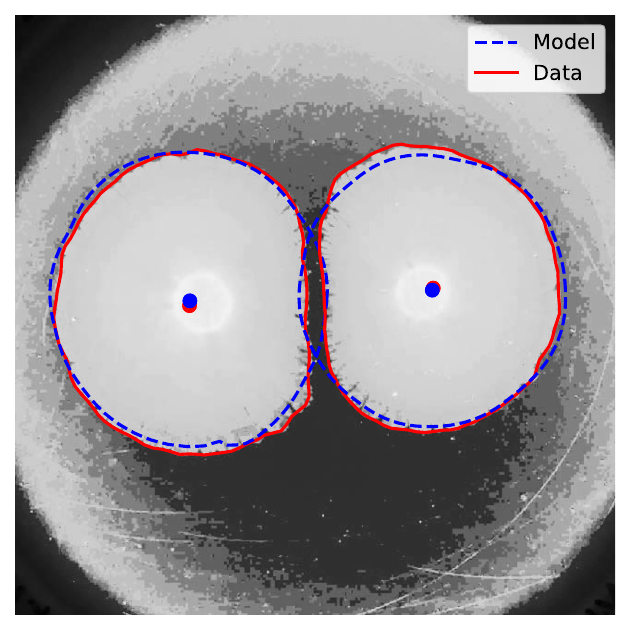}
            \caption{Per-colony $c$, $f=g$ (Ramp kernels)}
            \label{DS1_prior.c}
        \end{subfigure}
        \begin{subfigure}{0.4\textwidth}
            \centering
            \includegraphics[width=0.85\linewidth]{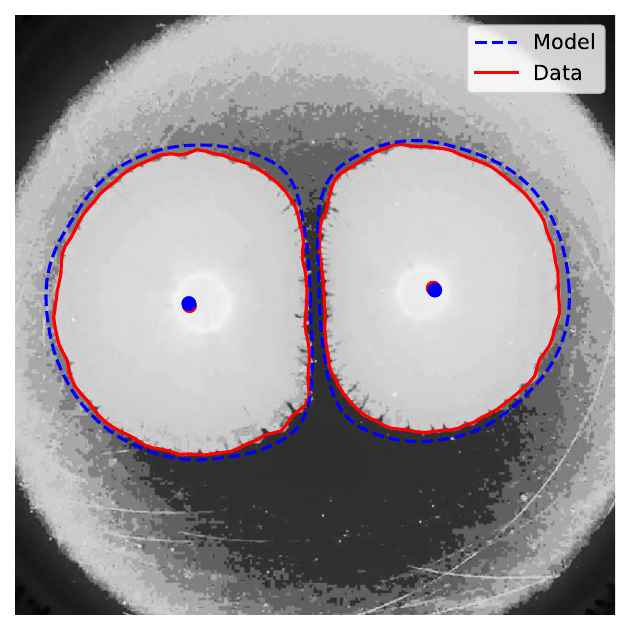}
            \caption{Shared $c$, $f=g$ (Ramp kernels)}
            \label{DS1_prior.d}
        \end{subfigure}
        \caption{Dataset 1 (two colonies, similar size), step 90: simulated contours of the best fit of each model class without transport, compared with the target contours. In each case the better of the Gaussian and Ramp kernels is shown. In the model with $f=g$ and shared $c$, the growth rate of the colonies is overpredicted and the separation between colonies is underpredicted. When $f=g$ but $c$ is allowed to vary, the  best fit has contours that cross, so it does not reproduce the stationary fronts phenomenon. }
        \label{DS1_prior}
\end{figure*}

\begin{figure*}
        \centering
        \begin{subfigure}{0.4\textwidth}
            \centering
            \includegraphics[width=0.85\linewidth]{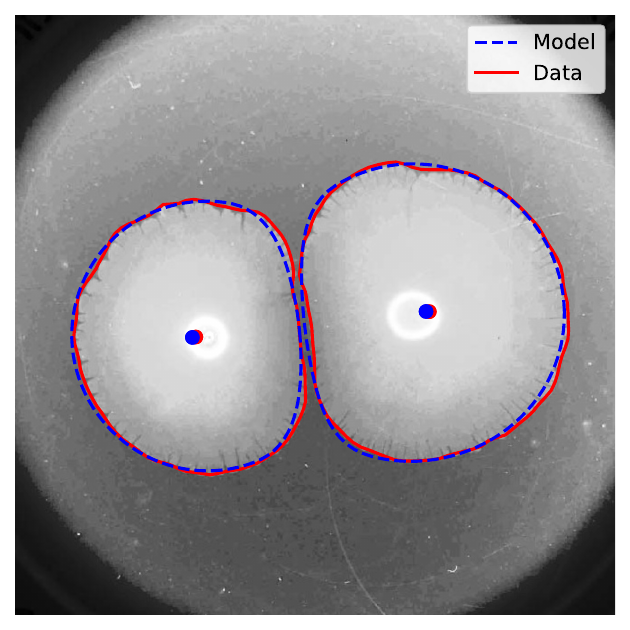}
            \caption{Per-colony $c$, $f\neq g$ (Ramp kernels)}
            \label{DS2_prior.a}
        \end{subfigure}
        \begin{subfigure}{0.4\textwidth}
            \centering
            \includegraphics[width=0.85\linewidth]{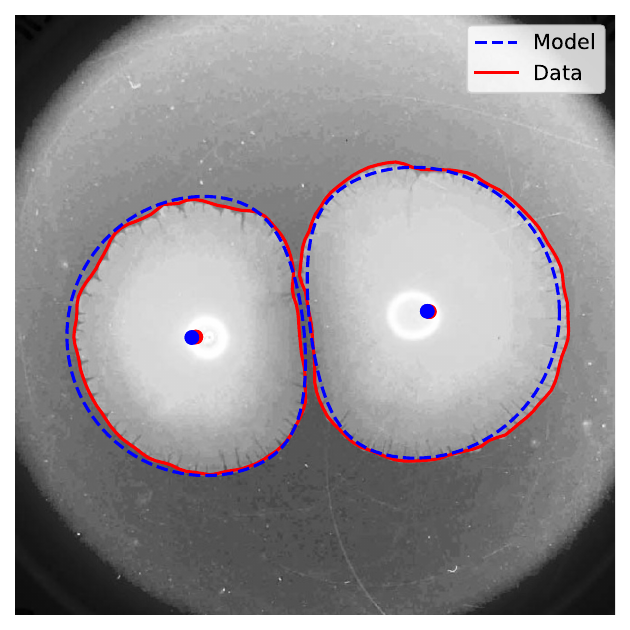}
            \caption{Shared $c$, $f\neq g$ (Ramp kernels)}
            \label{DS2_prior.b}
        \end{subfigure}
        \\
        \begin{subfigure}{0.4\textwidth}
            \centering
            \includegraphics[width=0.85\linewidth]{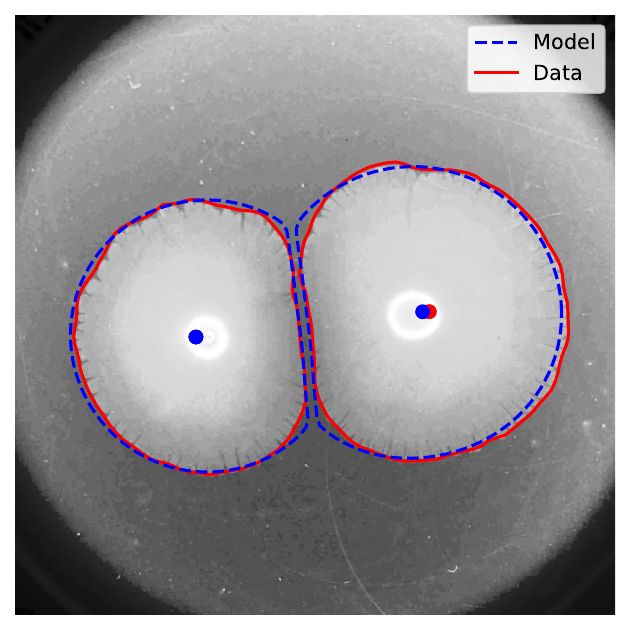}
            \caption{Per-colony $c$, $f=g$ (Gaussian kernels)}
            \label{DS2_prior.c}
        \end{subfigure}
        \begin{subfigure}{0.4\textwidth}
            \centering
            \includegraphics[width=0.85\linewidth]{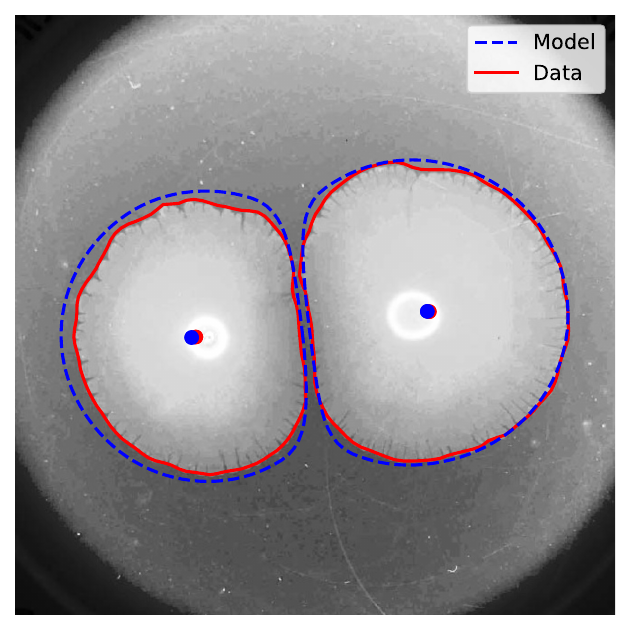}
            \caption{Shared $c$, $f=g$ (Ramp kernels)}
            \label{DS2_prior.d}
        \end{subfigure}
        \caption{Dataset 2 (two colonies, different size), step 110: simulated contours of the best fit of each model class without transport, compared with the target contours. In each case the better of the Gaussian and Ramp kernels is shown. Sharing a growth rate (right column) causes the size of Colony 1 (left) to be overpredicted and the size of Colony 2 (right) to be underpredicted, and forcing $f=g$ worsens the fit. Notably, the lower left configuration ($f=g$ with varying $c$) shows corners in the colonies near the edge of the stationary front.}
        \label{DS2_prior}
\end{figure*}

\begin{figure*}
        \centering
        \begin{subfigure}{0.4\textwidth}
            \centering
            \includegraphics[width=0.85\linewidth]{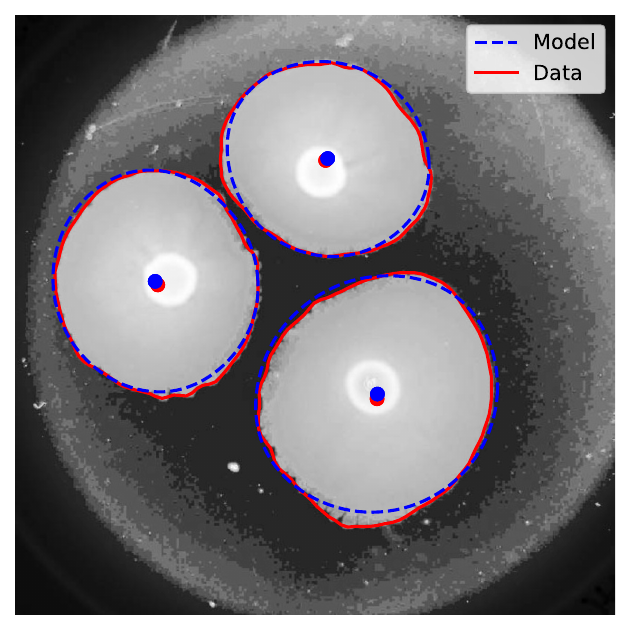}
            \caption{Per-colony $c$, $f\neq g$ (Ramp kernels)}
            \label{DS3_prior.a}
        \end{subfigure}
        \begin{subfigure}{0.4\textwidth}
            \centering
            \includegraphics[width=0.85\linewidth]{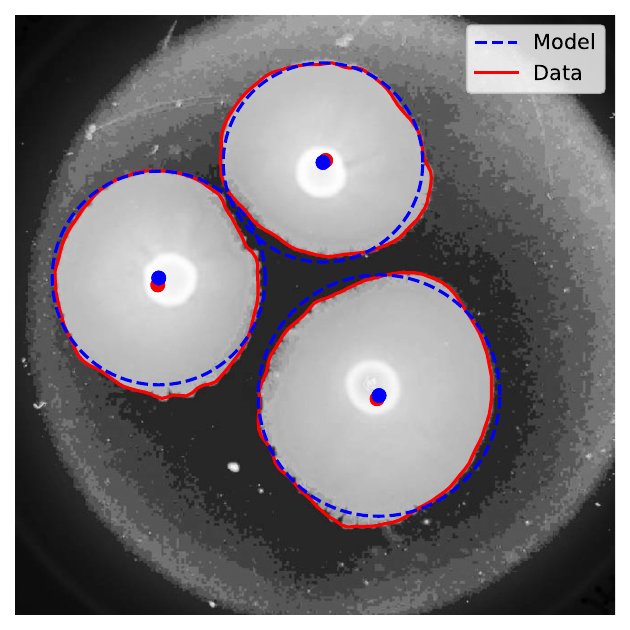}
            \caption{Shared $c$, $f\neq g$ (Gaussian kernels)}
            \label{DS3_prior.b}
        \end{subfigure}
        \\
        \begin{subfigure}{0.4\textwidth}
            \centering
            \includegraphics[width=0.85\linewidth]{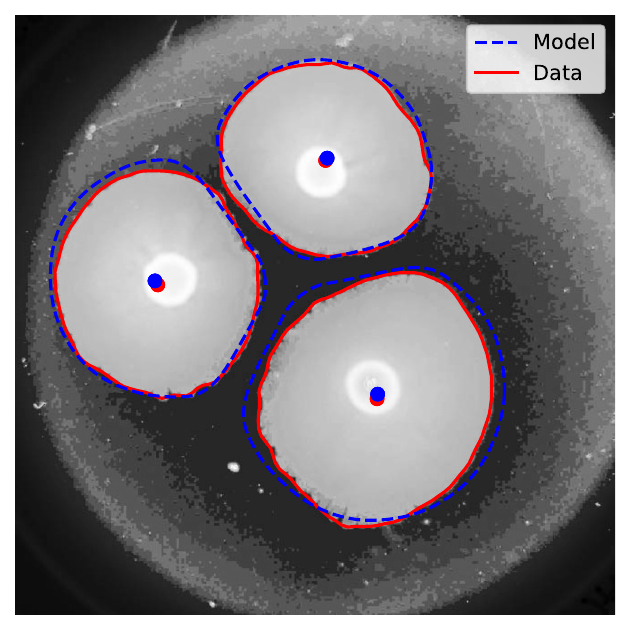}
            \caption{Per-colony $c$, $f=g$ (Ramp kernels)}
            \label{DS3_prior.c}
        \end{subfigure}
        \begin{subfigure}{0.4\textwidth}
            \centering
            \includegraphics[width=0.85\linewidth]{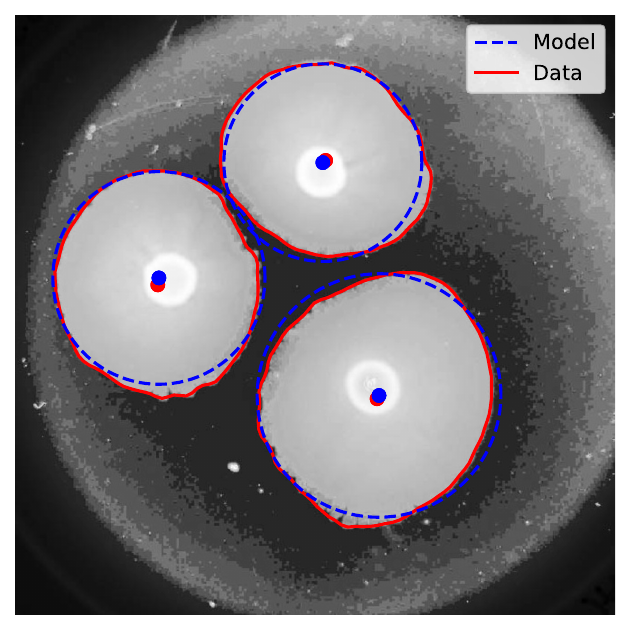}
            \caption{Shared $c$, $f=g$ (Gaussian kernels)}
            \label{DS3_prior.d}
        \end{subfigure}
        \caption{Dataset 3 (three colonies), step 110: simulated contours of the best fit of each model class without transport, compared with the target contours. In each case the better of the Gaussian and Ramp kernels is shown. Sharing a growth rate (right column) causes the sizes of the colonies to be off and causes the separation between colonies to be underpredicted, and forcing $f=g$ further worsens the fit. Notably, the lower left configuration ($f=g$ with varying $c$) shows the model overpredicting the separation between Colonies 1 and 2 (left side) and producing a straighter front than the data.} 
        \label{DS3_prior}
\end{figure*}

Next we visually compare predictions made by numerically simulating contours with the target contours extracted from experimental data, as in the previous subsection. Parameters are chosen based on the best fits given by Table \ref{tab:prior_work}. For each dataset, we compare the four model configurations in Table \ref{tab:prior_work} (taking the best of either Ramp or Gaussian), displaying a single still image for each configuration. Figure \ref{DS1_prior} corresponds to Dataset 1 (two colonies, similar size), Figure \ref{DS2_prior} to Dataset 2 (two colonies, different size), and Figure \ref{DS3_prior} to Dataset 3 (three colonies). For each dataset, an image near the end of the sequence is chosen. 

For Dataset 1, sharing the growth rate does not significantly impact the model fit since the two colonies are very close to the same size. However, sharing interaction kernels causes the best-fitting model to either overpredict the sizes of the colonies or introduce crossing contours, increasing the loss by $296$--$323\%$. For Dataset 2, sharing the growth rate leads to inaccurate predictions of colony size, increasing the loss by over $80\%$, and sharing interaction kernels distorts shapes even further, resulting in losses $109$--$191\%$ larger than the standard model. For Dataset 3, sharing the growth rate leads to inaccurate predictions of colony size, and sharing interaction kernels worsens the fit, but the effect is smaller than in Datasets 1 and 2, only increasing the loss by around $40\%$. 

\subsection{Nonlocal Transport Mechanism}\label{sec:results_transport}

This subsection quantitatively analyzes the performance of model configurations with and without the nonlocal transport mechanism introduced in Section \ref{sec:transport_force}. Optimization is done following the strategy of Section \ref{sec:lbfgs} using both Nelder--Mead and L-BFGS-B. In the Nelder--Mead runs, the initial simplex vertex for $\alpha$ was placed at $-\alpha_0$ rather than $1.05\,\alpha_0$, so that both repulsive and attractive transport were explored from the start. All runs with transport used the smoothstep gate with width $w = 10\sqrt{2}$ in pixel units. 

The best parameters and losses are shown in Table \ref{tab:transport}. The addition of the nonlocal transport mechanism does not significantly improve the standard model, indicating that the colony-scale transport induced by the nonlocal interaction force is already sufficiently captured by the standard model. However, in models with shared growth rates, the addition of the nonlocal transport mechanism can compensate somewhat for the missing degrees of freedom. (All models shown in this section use distinct self- and cross-interaction kernels, i.e. $f \neq g$). Notably, in the shared growth rate configurations where the nonlocal transport mechanism helps most, the $\beta$ parameter is large enough that the transport gate is open for most of the frames, so it acts more like a constant attractive force than one that switches on when colonies get close enough. 

Almost all of the best transport fits for each dataset, in both shared growth rate and varying growth rate model configurations, have a negative $\alpha$ value, indicating that the nonlocal transport mechanism causes attraction rather than repulsion (with the only exception being the shared growth rate, transport-on, Dataset 3 Gaussian case). However, the magnitude of this effect is small for the standard (varying growth rate) model. 

\begin{table*}
\centering
\footnotesize
\resizebox{\textwidth}{!}{%
\begin{tabular}{|c|c|c|c|c|c|c|c|c|c|}
\hline
 & & \multicolumn{2}{c|}{per-colony $c$, transport off} & \multicolumn{2}{c|}{per-colony $c$, transport on} & \multicolumn{2}{c|}{shared $c$, transport off} & \multicolumn{2}{c|}{shared $c$, transport on} \\
 & & Avg.\ rel.\ loss & $\tau$ & Avg.\ rel.\ loss & $\tau$ & Avg.\ rel.\ loss & $\tau$ & Avg.\ rel.\ loss & $\tau$ \\
\hline
\multirow{2}{*}{Dataset 1} & Gaussian & $6.07 \times 10^{-4}$ & $8.94 \times 10^{-1}$ & \begin{tabular}{@{}c@{}}$6.03 \times 10^{-4}$ \\ ($-0.6\%$)\end{tabular} & $8.70 \times 10^{-1}$ & \begin{tabular}{@{}c@{}}$7.14 \times 10^{-4}$ \\ ($+18\%$)\end{tabular} & $7.12 \times 10^{-1}$ & \begin{tabular}{@{}c@{}}$7.04 \times 10^{-4}$ \\ ($+16\%$)\end{tabular} & $6.94 \times 10^{-1}$ \\ \cline{2-10}
 & Ramp & $6.05 \times 10^{-4}$ & $8.77 \times 10^{-1}$ & \begin{tabular}{@{}c@{}}$6.05 \times 10^{-4}$ \\ ($-0.1\%$)\end{tabular} & $8.69 \times 10^{-1}$ & \begin{tabular}{@{}c@{}}$7.00 \times 10^{-4}$ \\ ($+16\%$)\end{tabular} & $7.05 \times 10^{-1}$ & \begin{tabular}{@{}c@{}}$6.96 \times 10^{-4}$ \\ ($+15\%$)\end{tabular} & $6.92 \times 10^{-1}$ \\ \hline
\multirow{2}{*}{Dataset 2} & Gaussian & $8.19 \times 10^{-4}$ & $1.87 \times 10^{-1}$ & \begin{tabular}{@{}c@{}}$7.96 \times 10^{-4}$ \\ ($-2.8\%$)\end{tabular} & $2.27 \times 10^{-1}$ & \begin{tabular}{@{}c@{}}$1.50 \times 10^{-3}$ \\ ($+83\%$)\end{tabular} & $3.83 \times 10^{-2}$ & \begin{tabular}{@{}c@{}}$1.03 \times 10^{-3}$ \\ ($+25\%$)\end{tabular} & $7.35 \times 10^{-1}$ \\ \cline{2-10}
 & Ramp & $7.91 \times 10^{-4}$ & $4.03 \times 10^{-1}$ & \begin{tabular}{@{}c@{}}$7.74 \times 10^{-4}$ \\ ($-2.1\%$)\end{tabular} & $3.79 \times 10^{-1}$ & \begin{tabular}{@{}c@{}}$1.46 \times 10^{-3}$ \\ ($+84\%$)\end{tabular} & $3.66 \times 10^{-2}$ & \begin{tabular}{@{}c@{}}$8.97 \times 10^{-4}$ \\ ($+13\%$)\end{tabular} & $7.40 \times 10^{-1}$ \\ \hline
\multirow{2}{*}{Dataset 3} & Gaussian & $2.23 \times 10^{-3}$ & $2.86 \times 10^{-2}$ & \begin{tabular}{@{}c@{}}$2.21 \times 10^{-3}$ \\ ($-1.0\%$)\end{tabular} & $2.71 \times 10^{-2}$ & \begin{tabular}{@{}c@{}}$3.07 \times 10^{-3}$ \\ ($+37\%$)\end{tabular} & $1.01 \times 10^{-2}$ & \begin{tabular}{@{}c@{}}$2.66 \times 10^{-3}$ \\ ($+19\%$)\end{tabular} & $1.01 \times 10^{-2}$ \\ \cline{2-10}
 & Ramp & $2.20 \times 10^{-3}$ & $2.66 \times 10^{-2}$ & \begin{tabular}{@{}c@{}}$2.18 \times 10^{-3}$ \\ ($-0.9\%$)\end{tabular} & $2.59 \times 10^{-2}$ & \begin{tabular}{@{}c@{}}$3.07 \times 10^{-3}$ \\ ($+39\%$)\end{tabular} & $5.81 \times 10^{-3}$ & \begin{tabular}{@{}c@{}}$2.36 \times 10^{-3}$ \\ ($+7.0\%$)\end{tabular} & $9.81 \times 10^{-3}$ \\ \hline
\end{tabular}
}
\caption{Average relative losses and values of $\tau$ of the best fits with and without the nonlocal transport mechanism, for per-colony and shared growth rates, for each dataset and kernel class (note that $f \neq g$ throughout). Percentages are changes relative to the first column (standard model). }
\label{tab:transport}
\end{table*}

\begin{figure*}
        \centering
        \begin{subfigure}{0.4\textwidth}
            \centering
            \includegraphics[width=0.85\linewidth]{OctoberFigures/frames/Dataset2_ramp_largeK_notransport_percolony_step110.pdf}
            \caption{Per-colony $c$, transport off (Ramp kernels)}
            \label{DS2_transport.a}
        \end{subfigure}
        \begin{subfigure}{0.4\textwidth}
            \centering
            \includegraphics[width=0.85\linewidth]{OctoberFigures/frames/Dataset2_ramp_largeK_notransport_shared_step110.pdf}
            \caption{Shared $c$, transport off (Ramp kernels)}
            \label{DS2_transport.b}
        \end{subfigure}
        \\
        \begin{subfigure}{0.4\textwidth}
            \centering
            \includegraphics[width=0.85\linewidth]{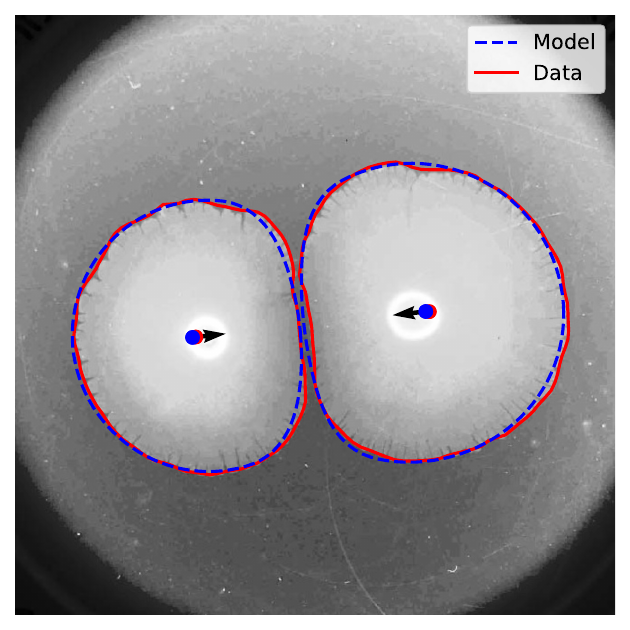}
            \caption{Per-colony $c$, transport on (Ramp kernels)}
            \label{DS2_transport.c}
        \end{subfigure}
        \begin{subfigure}{0.4\textwidth}
            \centering
            \includegraphics[width=0.85\linewidth]{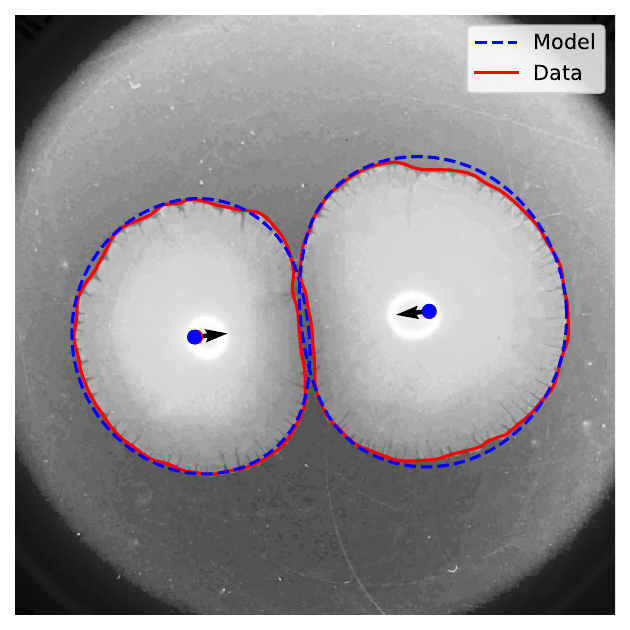}
            \caption{Shared $c$, transport on (Ramp kernels)}
            \label{DS2_transport.d}
        \end{subfigure}
        \caption{Dataset 2 (two colonies, different size), step 110: simulated contours of the best fits with and without the nonlocal transport mechanism, for per-colony and shared growth rates, compared with the target contours. In each case the better of the Gaussian and Ramp kernels is shown. In the shared $c$ configuration, turning on transport somewhat compensates for the lack of per-colony growth rates, improving the loss at intermediate times. However, it pushes the two fronts past each other in the final frames (thus not reproducing the ``stationary front'' phenomenon).}
        \label{DS2_transport}
\end{figure*}

Next we visually compare predictions made by numerically simulating contours with the target contours extracted from experimental data, as in the previous subsection. Parameters are chosen based on the best fits given by Table \ref{tab:transport}. We compare the four model configurations in Table \ref{tab:transport} (taking the best of either Ramp or Gaussian), displaying a single still image for each configuration. Because the fits are broadly similar and to save space, we just display Dataset 2 (Figure \ref{DS2_transport}), which is emblematic of the other two datasets. In the shared growth rate configuration, transport lowers the error at intermediate times as well as the total loss, but the attraction pushes the two fronts past each other in the final images, so the stationary front is not reproduced. In the varying growth rate configuration, adding transport only improves the loss by $2\%$. 

\begin{figure*}
    \centering
    \includegraphics[width=0.75\linewidth]{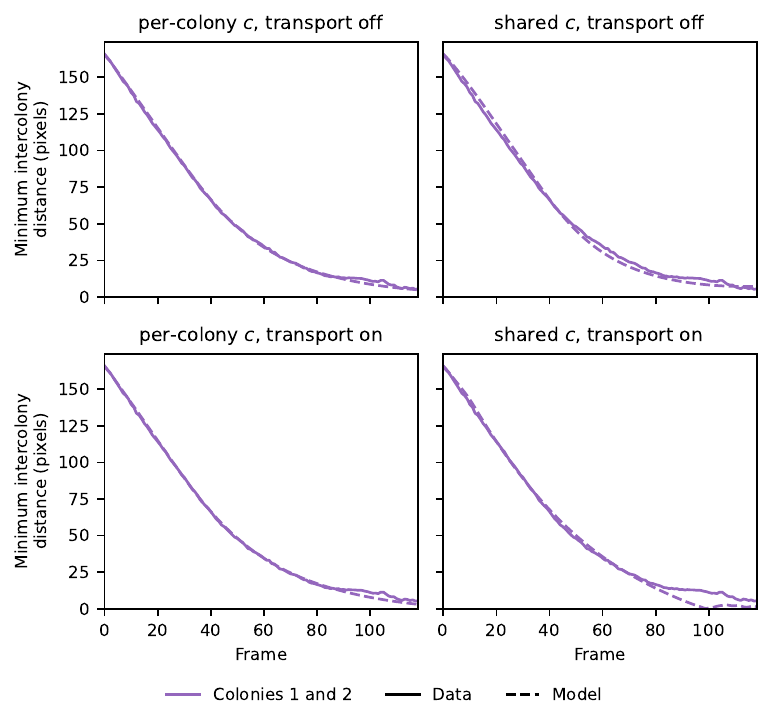}
    \caption{Minimum intercolony distance in pixels for Dataset 2 (two colonies, different size), for the best fits with and without the nonlocal transport mechanism, for per-colony and shared growth rates, compared to data. In each case, the single model with the best overall fit is chosen, whether Gaussian or Ramp. While adding transport improves the loss in the shared-$c$ configuration by 38\% (see Table \ref{tab:transport}), it results in the intercolony distance collapsing to zero (see lower right subfigure). 
    }
    \label{fig:min_distance_transport_ds2}
\end{figure*}

\begin{figure*}
    \centering
    \includegraphics[width=0.75\linewidth]{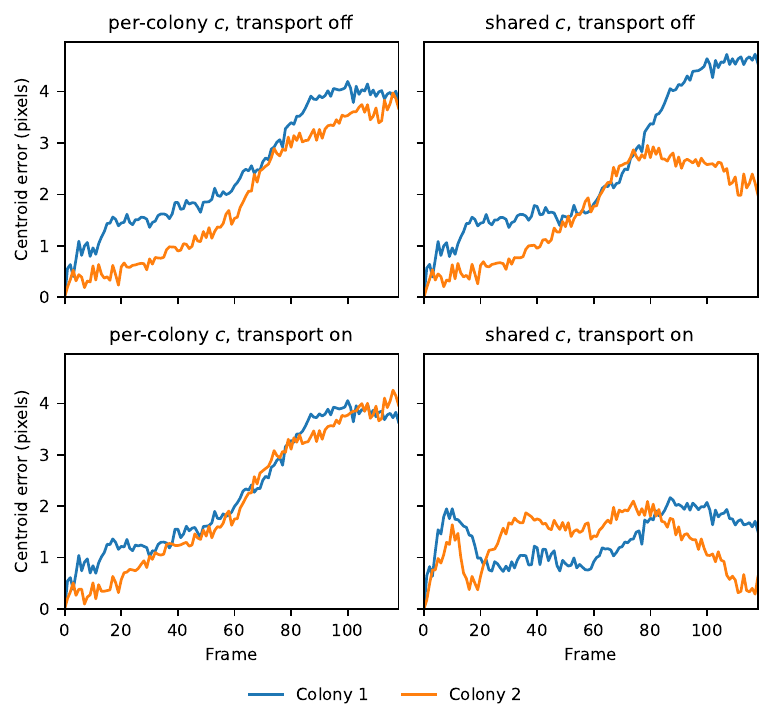}
    \caption{Centroid error (compared to data) in pixels for Dataset 2 (two colonies, different size), for the best fits with and without the nonlocal transport mechanism, for per-colony and shared growth rates. In each case, the single model with the best overall fit is chosen, whether Gaussian or Ramp. In the shared-$c$ configuration, adding transport allows the centroid error to be kept under 2.2 pixels, a large improvement. However, adding transport to the varying-$c$ configuration does not result in a large improvement. }
    \label{fig:centroid_transport_ds2}
\end{figure*}

Next, we show the minimum intercolony distance and centroid errors (as before, for Dataset 2 only). The minimum intercolony distance is shown in Figure \ref{fig:min_distance_transport_ds2} and the centroid errors are shown in Figure \ref{fig:centroid_transport_ds2}. On Dataset 2 in the shared growth rate configuration, using the nonlocal transport mechanism allows the loss to decrease, the colony sizes to more closely match, and the centroid error to decrease, but does not as faithfully predict the minimum intercolony distance, which touches zero as the fronts cross.

\section{Discussion and Future Work}\label{sec:c}

This work uses a PDE-constrained optimization procedure to learn interaction kernels for a nonlocal bacterial growth model from real experimental image sequences. Taken as a whole, this demonstrates an effective scientific machine learning framework for systems with a priori unknown governing dynamics and limited data. We consider two different forms of parametrized interaction kernels (Gaussian kernels, used in earlier work \cite{Scott2016Fronts,dud22}, and Ramp kernels, introduced here), and demonstrate a small but notable difference in performance between these two classes, with Ramp kernels performing better (by $0.25$--$3.4\%$ on two-colony image sequences and $1.4\%$ on a three-colony image sequence). 

The fitted interaction kernels differ considerably between datasets and between types (self- vs cross-interaction). In every fit, the cross-interaction kernel has a much longer range than the self-interaction kernel, and attempting to fit shared interaction kernels results in $42$--$323\%$ higher losses, supporting the modeling choice to parametrize the kernels separately. 

Another contribution of this work is the demonstration of the robustness of the model of \cite{Scott2016Fronts} with respect to colony-scale drift. To test this, we compare the standard model with an extension including a nonlocal transport mechanism. Even though it includes two new parameters, the transport mechanism only reduces the loss of the best fits by $0.1$--$2.8\%$. On the other hand, in a model where growth rates are shared, transport reduces the loss of the best fits by $13$--$38\%$ on Datasets 2 and 3, suggesting that the nonlocal transport mechanism is compensating for a missing degree of freedom in the model in this case, providing support for using distinct per-colony growth rates. 

On the methodological side, we find that the objective becomes differentiable in the model parameters, at least on the scale relevant to optimization, after two small changes to the numerical scheme: a fixed number of substeps per image and a smooth transport gate. This allows L-BFGS-B with finite-difference gradients to complement Nelder--Mead. The main bottlenecks of the approach are the time required to evaluate the objective with the PDE solver (up to several minutes per evaluation when $\tau$ is small) and the difficult loss landscape.

There are several directions for future work. For one, all of the results reported here are fits to the training data. Further analyses could test the predictive ability of the model by held-out prediction (training on early images and predicting later ones) or by generalization between datasets. Because the fitted kernels differ significantly across datasets, fitting all image sequences jointly would test how well a single set of kernels describes all of the experiments. Learning the interaction kernels nonparametrically, rather than within the Ramp and Gaussian classes, is another natural extension, although the sparsity of the data makes this challenging.

\section*{Funding}
We acknowledge support from the following grants: NSF 1813654, 2112085, 2331100 and ARO W911NF-19-1-0288.

\section*{Acknowledgements}
We would like to acknowledge participants in the 2023 UCLA CAM REU (Yuxuan Liu, Kyle Alkire, Yuntong Ju, Kevin Phan, Ryan Schoenburg, Yue Yang, and Liangjiawa Yuan) for their help in generating and testing some of the early ideas and software in this direction, including the image processing software.

\printbibliography

\newpage
\section{Appendix}\label{sec:appendix}

\subsection{Experiments}\label{sec:experiments_appendix}

This work uses data from \emph{Paenibacillus dendritiformis} bacterial colony growth experiments in \cite{dud22}. Experiments were conducted using sets of six Petri dishes of bacterial colonies placed into a single roundtable in order to ensure a consistent temperature of 29.5 \textdegree C and 30\% humidity for each set. Rotation of the roundtable allowed a camera to take a picture of a different Petri dish every five minutes; thus for a single colony, images are taken at 30-minute intervals. To further improve the pictures taken, lighting conditions were adjusted so that only the plate was illuminated. Because the bacterial colonies often did not grow during an initial period of $24$ hours, further processing of the images was done to discard images that did not show bacterial growth.

The experiments yielded seven image sequences which were amenable to segmentation, which were numbered $1$--$7$. Image sequences $1$--$6$ have $120$ images each, while image sequence $7$ has a sequence of $103$ images. The size of each image in pixels is the same within each sequence but differs across different sequences, ranging from $1080$ pixels to $1110$ pixels in both height and width. The time between images in each sequence is long enough that the size of the bacterial colonies can be considered non-decreasing.

\subsection{Image Processing}\label{sec:image_segmentation}

In order to model the behavior of the bacterial colonies, boundaries of the colonies were extracted using image segmentation and used as the target data in an optimization routine. The aim of the image segmentation was to capture accurate and smooth contours so that the data can be used to train the unknown parameters and kernels of the nonlocal PDE model. Smooth contours are preferred because the nonlocal PDE model does not account for fine-scale structures and intricate details on the colony boundaries.

However, difficulties arose from the similarity between the color of the colonies and the color of the background, making it challenging for contour extraction algorithms to distinguish between the two. In Figure \ref{difficultiesa}, the color of the colonies is too close to the color of the background, which can result in over-segmentation (where the extracted contours include regions which are not within the bacterial colony).  In contrast, Figure \ref{difficultiesb} shows false edges where the color of the edges of the colonies is more similar to the color of the background than the color of the interior of the colonies. This can lead to under-segmentation (where the extracted contours do not include these edge regions which are part of the colony). A third challenge is the ``feathering'', or intricate fractal-like structures present along the contours of bacterial colonies (see Figure \ref{difficultiesc}). This can lead to a lack of smoothness in the extracted contours.

\subsubsection{Pre-processing}

Previous work used a simple thresholding method to perform the contour extraction. However, due to the difficulties stated above, the extracted contours displayed both over-segmentation and under-segmentation \cite{dud22}. The approach taken in this work first uses various preprocessing strategies to handle these difficulties: cropping, bilateral and median blurring, and frame differencing. The preprocessing makes it easier for the image segmentation algorithm to correctly identify the contours of the bacterial colonies.

Cropping was done to eliminate unnecessary information and to speed up the rest of the preprocessing by reducing the amount of work that needs to be done. In particular, 40\% of each side of each image is cropped to retain the center of the original image.

Next, bilateral and median blurring are applied in order to obtain smooth contours. Filtering was done by using \texttt{OpenCV}'s \texttt{bilateralFilter} and \texttt{medianBlur} in Python \cite{opencv_library}. For bilateral blurring, a value of $15$ is used for the pixel neighborhood, filter sigma in the color space, and filter sigma in the coordinate space. For median blurring, a value of $15$ is used for the size of the kernel.

\begin{figure}[H]
        \centering
        \begin{subfigure}[H]{0.24\textwidth}
          \centering
          \includegraphics[width=\linewidth]{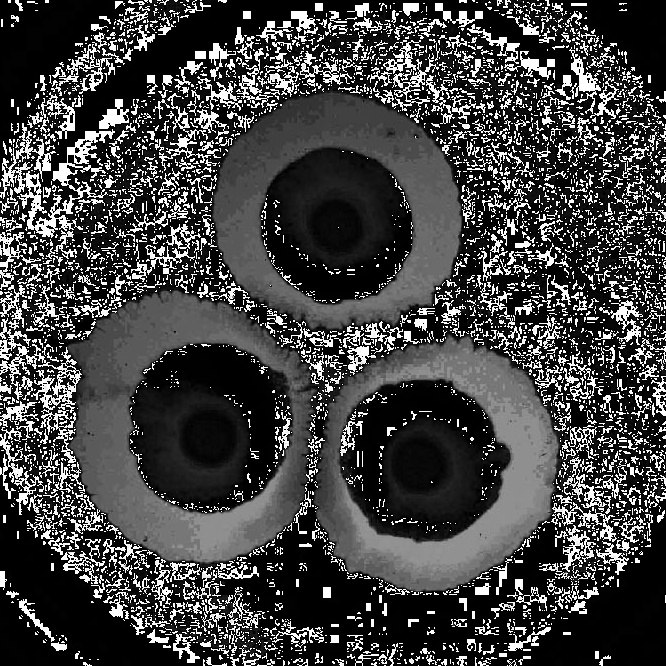}
          \caption{Before thresholding} \label{frame-before}
        \end{subfigure} 
        \hfill 
        \begin{subfigure}[H]{0.24\textwidth}
          \includegraphics[width=\linewidth]{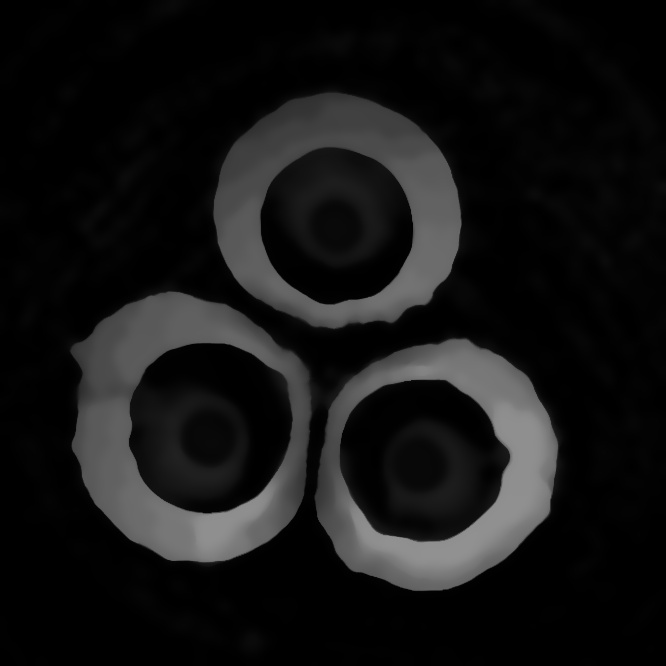}
          \centering
          \caption{After thresholding} \label{frame-after}
        \end{subfigure}
        \caption{Frame differences, before and after thresholding} \label{frame-diff}
\end{figure}

Lastly, frame differencing was done to isolate only the growth of the bacterial colonies. Frame differencing is a method used to detect the motion of objects \cite{Zhang2012AVD}. Two images are converted into grayscale and the difference between the two images is taken pixel by pixel. Afterward, thresholding is applied to remove any unwanted artifacts. In Figure \ref{frame-before}, the image is produced by taking the difference between the \nth{120} image and the \nth{60} image. Thresholding is applied to get rid of any white pixels as seen in Figure \ref{frame-after}. In grayscale, any pixel with a value of $200$ or less is converted into a black pixel. Any non-black pixels in Figure \ref{frame-after} represent the growth of the colony from the \nth{60} image in the sequence to the \nth{120} image in the sequence.

For each image sequence, the initial $10$--$15$ images are preprocessed using only cropping and a bilateral filter. Since issues with under-segmentation, over-segmentation, and rough edges were much more prevalent with the later images than the earlier images, the rest of the images were preprocessed by a combination of cropping, bilateral and median blurring, and frame differencing.

\subsubsection{Segmentation} 

Seeded region growing was chosen for segmenting the preprocessed image since it is fast and parameter-free \cite{seededregion, MEHNERT19971065}. To segment an image using seeded region growing, seeds, or labels attached to specific pixels, must first be initialized. Based on a similarity criterion, these seeds iteratively expand into regions which form the segments of the image. In particular, for this application, the mean color of the region is used as the similarity criterion.

The number of desired segments in each image is controlled by the number of seed initializations. For instance, in Figure \ref{frame-after}, a minimum of four seeds would be placed: one seed for each of the bacterial colonies and one seed for the background. This ensures that the number of regions in the segmented image will be four. Furthermore, since the size of the bacterial colonies is non-decreasing, for every image except the first in a dataset, new seeds can be initialized from the previous segmented image in the sequence. This is done by using the pixels that are labeled as being part of the bacterial colonies as new seeds. This also ensures that the size of the bacterial colonies in the segmented image is non-decreasing.

However, if an image is not properly segmented, this often means that the segmentation will not be successful for the rest of the images in the sequence. In most cases, if under-segmentation occurs, the segmented images will be sufficient to be used for training data. If over-segmentation occurs, the error in segmentation will propagate as most of the background will be considered as part of the bacterial colonies. One approach to handle this is to place more seeds for labeling the background to ensure that the seeded region growing algorithm does not label them as part of the bacterial colonies.

The algorithm to extract the contours from the images is written in Python and Julia. The preprocessing and contour extraction use Python's \verb|OpenCV| package \cite{opencv_library} and the seeded region growing is in Julia's \verb|ImageSegmentation.jl| package \cite{imageseg.jl}. The final step of extracting the contours from a segmented image uses \texttt{findContours} from OpenCV, although any contour extraction algorithm should be sufficient for this last step.

\begin{algorithm}[H]
  \caption{Extracting Contours from Images of Bacterial Colonies}\label{alg:contours}
  \begin{algorithmic}
  \REQUIRE Sequence of $K$ images, each containing $m$ colonies 
  \ENSURE Sequence of $K$ sets of $m$ contours 

  \FOR{$k = 1$ to $K$}
  
    \STATE Preprocess image $k$ by cropping, blurring, and frame differencing as needed

    \ENDFOR
    
    \STATE Initialize $m$ seeds for image $1$ in the sequence

    \FOR{$k = 1$ to $K$}
    
      \STATE Apply seeded region growing to image $k$ to get segmented image $k$ \\
      \IF{$k < K$}
        \STATE Initialize $m$ new seeds for image $k+1$ from segmented image $k$
     \ENDIF

     \ENDFOR

     \FOR{$k = 1$ to $K$}
    
      \STATE Extract $m$ contours from segmented image $k$\;

    \ENDFOR
    \end{algorithmic}
\end{algorithm}

\subsubsection{Results}
Figure \ref{threshold} features contour extraction using a threshold-based method \cite{dud22}. Figure \ref{seeded-reg}, in contrast, shows contour extraction using seeded region growing. The combination of frame differencing and seeded region growing successfully captures the false edges of the bottom two colonies. The extracted contours in Figure \ref{seeded-reg} generally have smoother contours than the contours in Figure \ref{threshold} due to the bilateral and median filters used. For a single image sequence, the algorithm takes about 37 seconds on an AMD Ryzen{\textsuperscript{TM}} 7 5825U with 24 gigabytes of RAM. 

\clearpage
\section*{Supplementary Material}
\addcontentsline{toc}{section}{Supplementary Material}
\setcounter{section}{0}
\setcounter{algorithm}{0}
\setcounter{figure}{0}
\setcounter{table}{0}
\renewcommand{\thesection}{S\arabic{section}}
\renewcommand{\theHsection}{S\arabic{section}}
\renewcommand{\thealgorithm}{S\arabic{algorithm}}
\renewcommand{\theHalgorithm}{S\arabic{algorithm}}
\renewcommand{\thefigure}{S\arabic{figure}}
\renewcommand{\theHfigure}{S\arabic{figure}}
\renewcommand{\thetable}{S\arabic{table}}
\renewcommand{\theHtable}{S\arabic{table}}

Section, equation, and algorithm numbers with the prefix S refer to this supplementary material; all other numbers refer to the main text.

\section{Derivation of the Governing Equations}\label{sec:derivation}

This section derives the tangent-angle form \eqref{eq:tangent-angle-governing-equations} of the governing equations, stated in Section \ref{sec:dynamics_main}, from the nondimensionalized model \eqref{eq:full_model_nondim}, using the shape-scaling representation as an intermediate step, and then describes how the nonlocal transport mechanism modifies the result. Much of the derivation is similar to that in \cite{vonbrecht2023frontsarrestiianalytical}.

\subsection{Constant-Speed and Shape-Scaling Representations}\label{sec:deriv_representations}

This section gives the details of the constant-speed and shape-scaling representations summarized in Section \ref{sec:representations}.

Given a planar contour $\bm{\gamma}: \mathbb{T}^1\to \mathbb{R}^2$, define its average speed $\sigma$ as 
\[
\sigma = \frac{1}{2\pi} \int_{\mathbb{T}^1} |\bm{\gamma}'(u)| \dd u,
\]
the arc length of $\bm{\gamma}$ divided by $2\pi$. Then we may construct a change of variables $\xi: \mathbb{T}^1 \to \mathbb{T}^1$ to convert $\bm{\gamma}$ to a constant-speed representation. Let $\eta: \mathbb{T}^1 \to \mathbb{T}^1$ be defined by
\[
\eta(u) := -\pi + \frac{1}{\sigma} \int_{-\pi}^u|\bm{\gamma}'(v)| \dd v.
\]
Note that $\eta$ is well defined as a function from $\mathbb{T}^1$ to $\mathbb{T}^1$ since $\eta(-\pi) = -\pi$ and $\eta(\pi) = \pi$. As long as the contour $\bm{\gamma}$ is regular, i.e. $|\bm{\gamma}'(u)| \neq 0$ for any $u \in \mathbb{T}^1$, we have $\eta'(u) > 0$. Using this fact along with the definition of $\sigma$ we can  see that $\eta$ is injective and surjective, so it is an invertible function on $\mathbb{T}^1$ and we may let $\xi = \eta^{-1}$. Then $\bm{\psi} := \bm{\gamma} \circ \xi$ is the constant speed representation of the contour $\bm{\gamma}$, since 
\[
|\bm{\psi}'(u)| 
= |\bm{\gamma}'(\xi(u))| |\xi'(u)| 
= |\bm{\gamma}'(\xi(u))|\frac{1}{|\eta'(\xi(u))|} = \sigma. 
\]
Alternatively, $\xi$ may be computed as the solution to the initial value problem 
\begin{equation*}
\xi'(u) = \frac{\sigma}{|\bm{\gamma}'(\xi(u))|}, \qquad \xi(-\pi) = -\pi
\end{equation*}
(see also Eq. \eqref{eq:reparametrize_ivp}). This procedure is more suitable for numerical algorithms and is the implementation actually used to convert discrete contours to a constant-speed representation (see Algorithm \ref{alg:reparametrize} for details on the numerical implementation of this algorithm).

Dividing by $\sigma$ yields the unit-speed representation $\bm{\phi}$, otherwise called the ``shape'' of the interface:
\begin{equation}\label{eq:shape}
\bm{\phi}(u) := \frac{\bm{\psi}(u)}{\sigma} = \frac{\bm{\gamma}(\xi(u))}{\sigma}.
\end{equation}
Thus, if one has access to the shape $\bm{\phi}$ and the average speed $\sigma$ (the ``shape-scaling representation''), one may recover $\bm{\gamma}$ using 
\[
\bm{\gamma}(u) = \sigma \bm{\phi}(\eta(u)).
\]
Note that in \cite{Scott2016Fronts} the scale is defined as $\varepsilon := 1/\sigma^2$, which gives another equivalent representation.

Recall that the unit tangent is $\bm{\tau} = \bm{\phi}' = (-\sin\theta, \cos\theta)^T$ and the outward normal is $\mathbf{n} = \bm{\tau}^\perp = (\cos\theta, \sin\theta)^T$ (Section \ref{sec:representations}). Differentiation relates the shape function $\bm{\phi}$ to the tangent-angle function $\theta$ as follows:
\begin{equation}\label{eq:shape-tangent-identities}
\begin{split}
\frac{\partial^2 \bm{\phi}}{\partial u^2}(u,t) 
&= - \frac{\partial \theta}{\partial u}(u,t) \mathbf{n}(u, t) \\
\frac{\partial^3 \bm{\phi}}{\partial u^3}(u,t) 
&= - \frac{\partial^2 \theta}{\partial u^2}(u,t) \mathbf{n}(u, t) - \left( \frac{\partial \theta}{\partial u}(u,t) \right)^2 \bm{\tau}(u, t) \\
\frac{\partial^2 \bm{\phi}}{\partial u \partial t}(u,t) 
&= - \frac{\partial \theta}{\partial t}(u,t) \mathbf{n}(u, t) \\
\end{split}
\end{equation}

\subsection{Identities for Evolving Planar Contours}

Consider an evolving regular planar contour $\bm{\gamma}(u, t) \in C^2(\mathbb{T}^1 \times [0, T], \mathbb{R}^2)$ which is twice differentiable in $u$ and $t$. Let $\sigma(t)$ be the average speed of $\bm{\gamma}(\cdot, t)$, given by 
\begin{equation}\label{eq:sigma}
\sigma(t) := \frac{1}{2\pi} \int_{\mathbb{T}^1} \Big| {\frac{\partial \bm{\gamma}}{\partial u}}(v, t) \Big| \dd v.
\end{equation}
Then, as in the previous section, define $\eta:\mathbb{T}^1\times[0,T] \to \mathbb{T}^1$ and $\xi:\mathbb{T}^1 \times [0,T] \to \mathbb{T}^1$ by
\begin{equation}\label{eq:eta}
\eta(u, t) = -\pi + \frac{1}{\sigma(t)} \int_{-\pi}^u \Big| \frac{\partial \bm{\gamma}}{\partial u}(v, t) \Big| \dd v
\end{equation}
\begin{equation}\label{eq:xi}
\eta(\xi(u, t), t) = \xi(\eta(u,t),t) = u
\end{equation}
for all $(u, t) \in \mathbb{T}^1 \times [0,T]$. Thus $\eta$ satisfies
\begin{equation}\label{eq:eta_ivp}
\eta(-\pi, t) = -\pi, \qquad \frac{\partial \eta}{\partial u}(u, t) = \frac{1}{\sigma(t)} \Big| \frac{\partial \bm{\gamma}}{\partial u}(u, t) \Big|.
\end{equation}
Thus 
\[
1 = \frac{\partial}{\partial u} \left[ \eta(\xi(u, t), t) \right] = \frac{\partial \eta}{\partial u}(\xi(u,t), t) \frac{\partial \xi}{\partial u}(u, t)
\]
so $\xi$ satisfies 
\begin{equation}\label{eq:xi_ivp}
\xi(-\pi, t) = -\pi, \qquad \frac{\partial \xi}{\partial u}(u, t) 
= \frac{\sigma(t)}{\Big|\frac{\partial \bm{\gamma}}{\partial u}(\xi(u, t), t)\Big|}.
\end{equation}
As before, $\bm{\psi}(u, t) := \bm{\gamma}(\xi(u,t),t)$ is the constant-speed representation of this evolving contour. Using the chain rule we have 
\begin{equation}\label{eq:psi_u}
\frac{\partial \bm{\psi}}{\partial u}(u, t)
= \frac{\partial \bm{\gamma}}{\partial u}(\xi(u, t), t) \frac{\partial \xi}{\partial u}(u, t)
\end{equation}
\begin{equation}\label{eq:psi_t}
\frac{\partial \bm{\psi}}{\partial t}(u, t) =\frac{\partial \bm{\gamma}}{\partial u}(\xi(u, t), t) \frac{\partial \xi}{\partial t}(u, t) + \frac{\partial \bm{\gamma}}{\partial t}(\xi(u, t), t).
\end{equation}
Define $\mathbf{v}(u, t) = \frac{\partial \bm{\gamma}}{\partial t}(\xi(u, t), t)$ to be the rate of change of $\bm{\gamma}$ evaluated in the constant-speed coordinate system. Then we have 
\begin{equation}\label{eq:v_u}
\frac{\partial \mathbf{v}}{\partial u}(u, t) = \frac{\partial^2 \bm{\gamma}}{\partial u \partial t}(\xi(u, t), t) \frac{\partial \xi}{\partial u}(u,t).
\end{equation}
Furthermore, by differentiating the average speed $\sigma(t)$ and applying \eqref{eq:xi_ivp} we obtain
\begin{equation}\label{eq:sigma_prime}
\begin{split}
\sigma'(t) %
&= \frac{1}{2 \pi} \int_{\mathbb{T}^1} \frac{ \frac{\partial \bm{\gamma}}{\partial u}(v, t)}{\Big| \frac{\partial \bm{\gamma}}{\partial u}(v, t) \Big|} \cdot \frac{\partial^2 \bm{\gamma}}{\partial u \partial t}(v, t) \dd v \\
&= \frac{1}{2\pi \sigma(t)} \int_{\mathbb{T}^1} \frac{\partial \bm{\gamma}}{\partial u}(\xi(v,t), t) \cdot \frac{\partial^2 \bm{\gamma}}{\partial u \partial t}(\xi(v,t), t) \left( \frac{\partial \xi}{\partial u}(v, t)\right)^2 \dd v \\
&= \frac{1}{2\pi \sigma(t)} \int_{\mathbb{T}^1} \frac{\partial \bm{\psi}}{\partial u}(v,t) \cdot \frac{\partial \mathbf{v}}{\partial u}(v,t) \dd v
\end{split}
\end{equation}
By differentiating $\xi(\eta(u, t), t) = u$ in time, we have 
\[
\frac{\partial \xi}{\partial t}(\eta(u,t) ,t) = - \frac{\partial \xi}{\partial u}(\eta(u,t),t) \frac{\partial \eta}{\partial t} (u,t),
\]
or, after changing variables, 
\begin{equation}
\frac{\partial \xi}{\partial t}(u ,t) = - \frac{\partial \xi}{\partial u}(u,t) \frac{\partial \eta}{\partial t} (\xi(u, t),t).
\end{equation}
Thus, using \eqref{eq:psi_t} and \eqref{eq:psi_u}, we have
\begin{equation}\label{eq:easy_terms}
\begin{split}
\frac{\partial \bm{\psi}}{\partial t}(u, t) - \mathbf{v}(u, t) 
&=\frac{\partial \bm{\gamma}}{\partial u}(\xi(u, t), t) \frac{\partial \xi}{\partial t}(u, t) \\
&= - \frac{\partial \bm{\gamma}}{\partial u}(\xi(u, t), t) \frac{\partial \xi}{\partial u}(u,t) \frac{\partial \eta}{\partial t} (\xi(u, t),t) \\
&= - \frac{\partial \bm{\psi}}{\partial u}(u, t) \frac{\partial \eta}{\partial t} (\xi(u, t),t).
\end{split}
\end{equation}
Using \eqref{eq:eta} we may compute (by way of \eqref{eq:xi_ivp}, \eqref{eq:sigma_prime}, \eqref{eq:psi_u}, \eqref{eq:v_u})
\begin{equation}\label{eq:eta_t1}
\begin{split}
&\frac{\partial \eta}{\partial t} (\xi(u, t),t) \\
&= \frac{1}{\sigma(t)} \int_{-\pi}^{\xi(u,t)} \frac{\frac{\partial \bm{\gamma}}{\partial u}(v, t)}{\Big| \frac{\partial \bm{\gamma}}{\partial u}(v, t) \Big|} \cdot \frac{\partial^2 \bm{\gamma}}{\partial u \partial t}(v, t) \dd v \\
&\qquad- \frac{\sigma'(t)}{\sigma(t)^2} \int_{-\pi}^{\xi(u,t)} \Big| \frac{\partial \bm{\gamma}}{\partial u}(v, t) \Big| \dd v \\
&=\frac{1}{\sigma(t)} \int_{-\pi}^{u} \frac{\frac{\partial \bm{\gamma}}{\partial u}(\xi(v, t), t)}{\Big| \frac{\partial \bm{\gamma}}{\partial u}(\xi(v,t), t) \Big|} \cdot \frac{\partial^2 \bm{\gamma}}{\partial u \partial t}(\xi(v, t), t) \frac{\partial \xi}{\partial u}(v,t) \dd v \\
&\qquad- \frac{\sigma(t)}{\sigma(t)^2} \left( \eta(\xi(u,t),t) + \pi \right) \sigma'(t)\\
&=\frac{1}{\sigma(t)^2} \int_{-\pi}^{u} \frac{\partial \bm{\psi}}{\partial u}(v,t) \cdot \frac{\partial \mathbf{v}}{\partial u}(v,t) \dd v \\
&\qquad- \frac{1}{\sigma(t)^2} \frac{u + \pi}{2 \pi} \int_{\mathbb{T}^1} \frac{\partial \bm{\psi}}{\partial u}(v,t) \cdot \frac{\partial \mathbf{v}}{\partial u}(v,t) \dd v  \\
\end{split}
\end{equation}
Now, recall the operator $P_0 : C(\mathbb{T}^1, \mathbb{R}^d) \to C(\mathbb{T}^1, \mathbb{R}^d)$ of Section \ref{sec:representations}, which maps $f$ to the zero-Dirichlet primitive of $f - \mu_f$:
\[
P_0[f](u) := \int_{-\pi}^u f(v)\dd v - (u + \pi) \mu_f.
\]
Then we have 
\begin{equation}\label{eq:eta_t2}
\begin{split}
\frac{\partial \eta}{\partial t} (\xi(u, t),t) 
&= \frac{1}{\sigma(t)^2} P_0\left[ \frac{\partial \bm{\psi}}{\partial u}(\cdot, t)\cdot \frac{\partial \mathbf{v}}{\partial u}(\cdot, t) \right](u),
\end{split}
\end{equation}
so that the transport relation
\begin{equation}\label{eq:transport}
\begin{split}
\frac{\partial \bm{\psi}}{\partial t}(u, t) + \frac{P_0\left[ \frac{\partial \bm{\psi}}{\partial u}(\cdot, t)\cdot \frac{\partial \mathbf{v}}{\partial u}(\cdot, t) \right](u)}{\sigma(t)^2}  \frac{\partial \bm{\psi}}{\partial u}(u, t) = \mathbf{v}(u, t)
\end{split}
\end{equation}
holds.

Now, let $\bm{\phi}(u, t) = \bm{\psi}(u,t)/\sigma(t)$ be the shape of the interface as in \eqref{eq:shape}, where 
\begin{equation}\label{eq:unit_speed}
\Big| \frac{\partial \bm{\phi}}{\partial u}(u,t) \Big| = \frac{\sigma(t)}{\sigma(t)} = 1.
\end{equation}
Using
\eqref{eq:unit_tangent} and \eqref{eq:unit_normal} we can derive that
\begin{equation}
 \frac{\partial \bm{\phi}}{\partial u}(u, t) 
= \bm{\tau}(\xi(u, t), t)
\end{equation}
\begin{equation}
\left(\frac{\partial \bm{\phi}}{\partial u}\right)^{\perp} = \mathbf{n}(\xi (u, t), t)
\end{equation}
Recalling the identity \eqref{eq:curvature}, we have
\begin{equation}\label{eq:shape_curvature}
\frac{\partial^2 \bm{\phi}}{\partial u^2}(u, t) 
= \sigma(t)\, \kappa_{\bm{\gamma}}(\xi(u, t), t)\, \mathbf{n}(\xi(u, t), t).
\end{equation}

Conversely, at each fixed $t$ the recovery formula of the previous section gives
\begin{equation}\label{eq:evolving_shape_recovery}
\bm{\gamma}(u, t) = \sigma(t)\, \bm{\phi}(\eta(u, t), t).
\end{equation}
Since $\eta$ depends on the parametrization of $\bm{\gamma}$ through \eqref{eq:eta}, the evolving pair $(\bm{\phi}, \sigma)$ determines the evolving contour only up to a time-dependent reparametrization, namely $\bm{\psi}(u, t) = \sigma(t) \bm{\phi}(u, t)$. This is sufficient for our purposes: the governing equations \eqref{eq:full_model} prescribe only the normal velocity of each contour, so the evolution of the curve traced by $\bm{\gamma}^{(i)}(\cdot, t)$ does not depend on its parametrization. Note also that, unlike the tangent-angle representation, the shape-scaling representation loses no centroid information, since $\bm{\psi}$ is a position rather than a derivative.

\subsection{Governing Equations in Shape-Scaling Representation}\label{sec:dynamics1}

The goal of this section is to use the shape-scaling representation to derive an equivalent form of the governing equations. To avoid carrying the curvature coefficient $\mu$ through the derivation only to eliminate it later (see Section \ref{sec:nondim}), we derive directly from the nondimensionalized governing equations \eqref{eq:full_model_nondim}, in which the curvature coefficient has already been set to $1$ and the interaction kernels $f, g$ carry an explicit factor of $\tau$. Much of the derivation is similar to that found in \cite{vonbrecht2023frontsarrestiianalytical}. The nonlocal transport mechanism is treated separately in Section \ref{sec:deriv_transport}.

Consider a family of evolving contours $\{\bm{\gamma}^{(i)}(u, t), i \in [m]\} \subseteq C^2(\mathbb{T}^1 \times [0, T], \mathbb{R}^2)$ which are twice differentiable in $u$ and $t$. Let $\sigma^{(i)}(t)$ be the average speed of the $i$th contour, given by \eqref{eq:sigma}, and define 
\begin{equation*}
\begin{split}
&\eta^{(i)}(u, t), \qquad
\xi^{(i)}(u,t) , \qquad 
\bm{\psi}^{(i)}(u, t), \qquad
\bm{\phi}^{(i)}(u, t), \\
& \qquad \qquad \qquad \qquad \qquad \qquad i \in [m], u \in \mathbb{T}^1, t \in [0, T]
\end{split}
\end{equation*}
by \eqref{eq:eta}, \eqref{eq:xi}, $\bm{\psi}^{(i)}(u,t) = \bm{\gamma}^{(i)}(\xi^{(i)}(u,t), t)$, and \eqref{eq:shape}, respectively.

Now consider the constant growth/decay and normal-direction nonlocal force terms.  These may be encapsulated in
\[
F^{(i)}(u, t) = c^{(i)} - \tau v_{\text{NL}}^{(i)}(u, t).
\]
Once again considering the identity \eqref{eq:curvature}, the governing equations can be written as
\begin{equation}\label{eq:model_step1}
\begin{split}
\partial_t \bm{\gamma}^{(i)}(u, t) &= \frac{\partial_u\bm{\tau}^{(i)}(u, t)}{|\partial_u \bm{\gamma}^{(i)}(u, t)|} + F^{(i)}(u, t) \mathbf{n}^{(i)}(u, t)
\end{split}
\end{equation}
Let $\mathcal{F}^{(i)}$ be a scaled and shifted version of  $F^{(i)}$:
\begin{equation*}
\begin{split}
&\frac{\mathcal{F}^{(i)}(u, t) }{\sigma^{(i)}(t)}
= c^{(i)} - \tau v_{\text{NL}}^{(i)}(\xi^{(i)}(u, t), t) \\
&= c^{(i)} -\tau\int_{\mathbb{T}^1} f\left(\frac{1}{2} \left\lvert \bm{\psi}^{(i)}(u,t) - \bm{\psi}^{(i)}(v,t) \right\rvert^2\right) \sigma^{(i)}(t) \dd v \\
&- \tau\sum_{j \neq i} \int_{\mathbb{T}^1} g\left(\frac{1}{2} \left\lvert \bm{\psi}^{(i)}(u,t) - \bm{\psi}^{(j)}(v,t) \right\rvert^2\right) \sigma^{(j)}(t) \dd v.
\end{split}
\end{equation*}
Thus, applying the change of variables $\xi^{(i)}$ and the identity \eqref{eq:shape_curvature} to the equation \eqref{eq:model_step1} yields
\begin{equation*} 
\begin{split}
\partial_t \bm{\gamma}^{(i)}(\xi^{(i)}(u, t), t)
&= \frac{1}{\sigma^{(i)}(t)}\left[ \partial^2_u \bm{\phi}^{(i)}(u, t) \right. \\&+ \left. \mathcal{F}^{(i)}(u, t) \left(\partial_u \bm{\phi}^{(i)}(u, t)\right)^{\perp} \right].
\end{split}
\end{equation*}
Following the previous section, define the left-hand side above as
\[
\mathbf{v}^{(i)}(u, t) =\partial_t \bm{\gamma}^{(i)}(\xi^{(i)}(u, t), t).
\]
Recalling \eqref{eq:unit_speed}, we know that $\partial^2_u \bm{\phi}^{(i)}(u, t) \perp \partial_u \bm{\phi}^{(i)}(u, t)$. Hence $\mathbf{v}^{(i)} \perp \partial_u \bm{\phi}^{(i)}(u, t)$. Suppressing function arguments, we can write 
\begin{equation*}
\begin{split}
\mathbf{v}^{(i)}
&= \langle \mathbf{v}^{(i)}, {\bm{\phi}^{(i)}_u}^\perp \rangle {\bm{\phi}^{(i)}_u}^\perp \\
\mathbf{v}^{(i)}_u 
&= \left( \langle \mathbf{v}^{(i)}_u, {\bm{\phi}^{(i)}_u}^\perp \rangle + \langle \mathbf{v}^{(i)},  {\bm{\phi}_{uu}^{(i)}}^\perp \rangle \right) {\bm{\phi}^{(i)}_u}^\perp \\&+ \langle \mathbf{v}^{(i)}, {\bm{\phi}^{(i)}_u}^\perp \rangle  {\bm{\phi}_{uu}^{(i)}}^\perp \\
\langle \mathbf{v}^{(i)}_u, \bm{\phi}^{(i)}_u \rangle &=  \langle \mathbf{v}^{(i)}, {\bm{\phi}^{(i)}_u}^\perp \rangle \langle  {\bm{\phi}_{uu}^{(i)}}^\perp, \bm{\phi}^{(i)}_u \rangle \\
&= -\langle \mathbf{v}^{(i)}, {\bm{\phi}^{(i)}_u}^\perp \rangle \langle  {\bm{\phi}^{(i)}_{uu}},  {\bm{\phi}^{(i)}_u}^\perp \rangle
\end{split}
\end{equation*}
Thus 
\begin{equation*}
\begin{split}
\langle \mathbf{v}^{(i)}_u, \bm{\phi}^{(i)}_u \rangle
&= -\langle \mathbf{v}^{(i)}, {\bm{\phi}^{(i)}_u}^\perp \rangle \langle {\bm{\phi}^{(i)}_{uu}}, {\bm{\phi}^{(i)}_u}^\perp \rangle \\
&= -\frac{1}{\sigma^{(i)}}  \langle \bm{\phi}^{(i)}_{uu} + \mathcal{F}^{(i)} {\bm{\phi}^{(i)}_u}^\perp, {\bm{\phi}^{(i)}_u}^\perp \rangle \langle {\bm{\phi}^{(i)}_{uu}}, {\bm{\phi}^{(i)}_u}^\perp \rangle \\
&= -\frac{1}{\sigma^{(i)}} \left( \langle \bm{\phi}^{(i)}_{uu}, {\bm{\phi}^{(i)}_u}^\perp  \rangle^2 + \mathcal{F}^{(i)}\langle {\bm{\phi}^{(i)}_{uu}}, {\bm{\phi}^{(i)}_u}^\perp \rangle \right) \\
&= -\frac{1}{\sigma^{(i)}} \left( | \bm{\phi}^{(i)}_{uu}|^2 + \mathcal{F}^{(i)}\langle {\bm{\phi}^{(i)}_{uu}}, {\bm{\phi}^{(i)}_u}^\perp \rangle \right)
\end{split}
\end{equation*}
Let $a^{(i)}(u, t)$ be the term in parentheses on the right. Then 
\[
a^{(i)} = - \sigma^{(i)} \langle \mathbf{v}^{(i)}_u, \bm{\phi}^{(i)}_u \rangle = -\langle \mathbf{v}^{(i)}_u, \bm{\psi}^{(i)}_u \rangle
\]
so the transport relation \eqref{eq:transport} becomes
\begin{equation*}
\bm{\psi}^{(i)}_t = \mathbf{v}^{(i)} + \frac{P_0\left[ a^{(i)}\right]}{{\sigma^{(i)}}^2}  \bm{\psi}^{(i)}_u(u, t). 
\end{equation*}
Furthermore, recalling \eqref{eq:sigma_prime}, we have 
\[
{\sigma^{(i)}}'(t) = -\frac{1}{2 \pi \sigma^{(i)}(t)} \int_{\mathbb{T}^1} a^{(i)}(v, t) \dd v = - \frac{1}{\sigma^{(i)}(t)} \mu_{a^{(i)}}
\]
Using the product rule, we have 
\begin{equation}
\begin{split}
\partial_t \bm{\phi}^{(i)}(u, t) &= \frac{\partial_t \bm{\psi}^{(i)}(u, t)}{\sigma^{(i)}(t)} - \bm{\psi}^{(i)}(u, t) \frac{{\sigma^{(i)}}'(t)}{\sigma^{(i)}(t)^2} \\
&=  \frac{\mathbf{v}^{(i)}(u, t)}{\sigma^{(i)}(t)} + \frac{P_0\left[ a^{(i)}(\cdot, t)\right](u)}{{\sigma^{(i)}}(t)^2}  \bm{\phi}^{(i)}_u(u, t) \\&\qquad+ \frac{\bm{\phi}^{(i)}(u, t)}{\sigma^{(i)}(t)^2} \mu_{a^{(i)}}
\end{split}
\end{equation}
Thus, again suppressing function arguments, 
\begin{equation*} 
\begin{split}
 \bm{\phi}^{(i)}_t
&= \frac{1}{{\sigma^{(i)}}^2} \left[ \bm{\phi}^{(i)}_{uu} + P_0[a^{(i)}] \bm{\phi}^{(i)}_u  \right.\\&+\left. \mathcal{F}^{(i)}{\bm{\phi}^{(i)}_u}^{\perp} + \mu_{a^{(i)}} \bm{\phi}^{(i)} \right]
\end{split}
\end{equation*}
Thus the full dynamics can be described as follows (with the equation for $\sigma^{(i)}$ written in terms of ${\sigma^{(i)}}^2$ for convenience): 
\begin{equation}\label{eq:shape-scaling-governing-equations}
\begin{split}
 \bm{\phi}^{(i)}_t
&= \frac{1}{{\sigma^{(i)}}^2} \left[ \bm{\phi}^{(i)}_{uu} + P_0[a^{(i)}] \bm{\phi}^{(i)}_u  \right.\\&+\left. \mathcal{F}^{(i)}{\bm{\phi}^{(i)}_u}^{\perp} + \mu_{a^{(i)}} \bm{\phi}^{(i)} \right] \\
\partial_t \left[ {\sigma^{(i)}}^2 \right] &= -2\mu_{a^{(i)}} \\
a^{(i)} &= |\bm{\phi}^{(i)}_{uu}|^2 + \langle \bm{\phi}^{(i)}_{uu}, {\bm{\phi}^{(i)}_{u}}^\perp \rangle \mathcal{F}^{(i)}, \\
\mathcal{F}^{(i)}(u, t) &= \sigma^{(i)} \left( c^{(i)} -\right. \\&\tau\int_{\mathbb{T}^1} f\left(\frac{1}{2} \left\lvert \bm{\psi}^{(i)}(u,t) - \bm{\psi}^{(i)}(v,t) \right\rvert^2\right) \sigma^{(i)} \dd v  \\
&\left.- \tau\sum_{j \neq i} \int_{\mathbb{T}^1} g\left(\frac{1}{2} \left\lvert \bm{\psi}^{(i)}(u,t) - \bm{\psi}^{(j)}(v,t) \right\rvert^2\right) \sigma^{(j)} \dd v \right) \\
\bm{\psi}^{(i)} &= \sigma^{(i)} \bm{\phi}^{(i)}
\end{split}
\end{equation}

This form of the equations has the advantage of having a constant in front of the diffusion term in the evolution for $\bm{\phi}^{(i)}$, but also introduces a nonlocal tangential (advection) term $P_0[a^{(i)}] \bm{\phi}^{(i)}_u$ which was not present before. This term only reparametrizes the contour, keeping its speed constant; it should not be confused with the nonlocal transport mechanism of Section \ref{sec:transport_force}. 

\subsection{Governing Equations in Tangent-Angle Representation}\label{sec:dynamics2}

The tangent-angle representation may also be used to obtain an equivalent form of the governing equations \eqref{eq:full_model_nondim}.

As in a previous section, let 
\begin{equation*}
\bm{\tau}^{(i)}(u, t) = \frac{\partial \bm{\phi}^{(i)}}{\partial u} (u, t)
\end{equation*}
be the unit tangent to the $i$th contour, and let $\mathbf{n}^{(i)}$ be the corresponding unit normal. Using the properties described by \eqref{eq:shape-tangent-identities} and differentiating the evolution equation \eqref{eq:shape-scaling-governing-equations}, we obtain
\begin{equation*} 
\begin{split}
\partial_t \bm{\phi}_u^{(i)}
&= \frac{1}{{\sigma^{(i)}}^2} \left[ \bm{\phi}^{(i)}_{uuu} + P_0[a^{(i)}] \bm{\phi}^{(i)}_{uu} + (a^{(i)}-\mu_{a^{(i)}}) \bm{\phi}_u^{(i)} 
 \right.\\&+\left. \mu_{a^{(i)}} \bm{\phi}^{(i)}_u +\mathcal{F}^{(i)}_u{\bm{\phi}^{(i)}_u}^{\perp} + \mathcal{F}^{(i)}{\bm{\phi}^{(i)}_{uu}}^{\perp}\right] \\
 &= \frac{1}{{\sigma^{(i)}}^2} \left[ -\theta^{(i)}_{uu} \mathbf{n}^{(i)} - \left(\theta^{(i)}_u\right)^2 \bm{\tau}^{(i)} - P_0[a^{(i)}] \theta^{(i)}_u \mathbf{n}^{(i)}
 \right.\\&+\left. a^{(i)} \bm{\tau}^{(i)}  +\mathcal{F}^{(i)}_u\mathbf{n}^{(i)} + \mathcal{F}^{(i)}\theta^{(i)}_u \bm{\tau}^{(i)} \right] \\
 &= \frac{1}{{\sigma^{(i)}}^2} \left[ -\theta^{(i)}_{uu}  - P_0[a^{(i)}] \theta^{(i)}_u 
+ \mathcal{F}^{(i)}_u\right] \mathbf{n}^{(i)}\\
\end{split}
\end{equation*}
Here we have used the fact that 
\[
a^{(i)} = |\bm{\phi}^{(i)}_{uu}|^2 + \langle \bm{\phi}^{(i)}_{uu}, {\bm{\phi}^{(i)}_{u}}^\perp \rangle \mathcal{F}^{(i)}
= {\theta_u^{(i)}}^2 - \mathcal{F}^{(i)} \theta_u^{(i)}.
\]
Taking a dot product with $- \mathbf{n}^{(i)}$ yields $\theta_t^{(i)}$: 
\begin{equation*}
\begin{split}
\theta_t^{(i)} = 
\frac{1}{{\sigma^{(i)}}^2} \left[ \theta^{(i)}_{uu}  + P_0[a^{(i)}] \theta^{(i)}_u 
- \mathcal{F}^{(i)}_u\right]
\end{split}
\end{equation*}
Furthermore, since the centroids are given by $\bm{\mu}^{(i)}= \mu_{\bm{\psi}^{(i)}}$, by integrating \eqref{eq:shape-scaling-governing-equations} multiplied by $\sigma^{(i)}$ we obtain 
\[
\partial_t \bm{\mu}^{(i)} = \frac{1}{2 \pi \sigma^{(i)}} \int_{\mathbb{T}^1} P_0[a^{(i)}] \bm{\tau}^{(i)} + \mathcal{F}^{(i)} \mathbf{n}^{(i)} \dd v.
\]
The evolution equation for $\sigma^{(i)}$ remains the same. Together, these give the system \eqref{eq:tangent-angle-governing-equations} stated in Section \ref{sec:dynamics_main}, along with the winding number and periodicity constraints for the tangent-angle function.

\subsection{Effect of the Nonlocal Transport Mechanism}\label{sec:deriv_transport}

Since $\mathbf{h}^{(i)}(t)$ is a rigid translation, the evolution of the curve traced by $\bm{\gamma}^{(i)}(\cdot, t)$ still does not depend on its parametrization, and the derivations of Sections \ref{sec:dynamics1} and \ref{sec:dynamics2} carry over with one change. The velocity $\mathbf{v}^{(i)}$ gains the $u$-independent term $\mathbf{h}^{(i)}$, so it is $\mathbf{v}^{(i)} - \mathbf{h}^{(i)}$, rather than $\mathbf{v}^{(i)}$, that is orthogonal to $\bm{\phi}^{(i)}_u$. Because $\partial_u \mathbf{h}^{(i)} = 0$, however, $\mathbf{v}^{(i)}_u$ is unchanged, and hence so are $a^{(i)}$, $P_0[a^{(i)}]$, and the evolution of $\sigma^{(i)}$. In the shape-scaling representation \eqref{eq:shape-scaling-governing-equations}, the equation for $\bm{\phi}^{(i)}$ gains the term $\mathbf{h}^{(i)}/\sigma^{(i)}$. In the tangent-angle representation, this term is annihilated by differentiation in $u$, so the equations for $\theta^{(i)}$ and $\sigma^{(i)}$ are unchanged. Integrating the equation for $\bm{\phi}^{(i)}$ multiplied by $\sigma^{(i)}$, as in Section \ref{sec:dynamics2}, shows that the centroid velocity gains the term $\mathbf{h}^{(i)}$. In the nondimensionalized variables, this gives the system \eqref{eq:tangent-angle-governing-equations-transport}. 

\section{Optimization Algorithms}\label{sec:optimization_algorithms}

This section collects the algorithms used to evaluate the training objective and to minimize it (Sections \ref{sec:optimization_problems}--\ref{sec:lbfgs}). Algorithm \ref{alg:timestepping} simulates the contours over the sequence of images with a fixed number of substeps per image, and Algorithm \ref{alg:loss} uses it to evaluate the relative loss. Algorithms \ref{alg:nelder-mead} and \ref{alg:lbfgs} are the two optimization methods: \texttt{scipy}'s implementation of the Nelder--Mead method, and L-BFGS-B with central-difference gradients in the log parametrization of Section \ref{sec:lbfgs}.

\begin{algorithm}[H]
\caption{Simulate Contours Numerically}
\label{alg:timestepping}
\begin{algorithmic}
\REQUIRE Starting contour points $\{ \bm{\psi}^{(i)}_{j,0}, i \in [m], j \in [n]\}$ \\
Number of images $K$ (images are one unit of physical time apart), \qquad Number of substeps per image $N_{\text{sub}}$ \\
Interaction kernels $f, g:\mathbb{R}^+ \to \mathbb{R}^+$ \\
Time scaling factor $\tau \in \mathbb{R}^+$,\qquad
Model parameters $c^{(1)}, \ldots, c^{(m)}, \beta \in \mathbb{R}^+$, $\alpha \in \mathbb{R}$ ($\alpha = 0$ for the standard model)

\ENSURE Calibrated contour points $\{\bm{\psi}^{(i)}_{j,k}, i \in [m], j \in [n], k \in [K] \}$ at the image times

\STATE Convert starting contour points to tangent-angle representation using Algorithm \ref{alg:gamtotht}, and recenter using Algorithm \ref{alg:normalize}

\STATE Store tangent-angle representation $\{ \theta_j^{(i)}, \bm{\mu}^{(i)}, \sigma^{(i)}, \; i \in [m], j \in [n] \}$ and set the substep length
\[
\Delta t \gets \frac{1}{\tau N_{\text{sub}}}
\]

\FOR{ $k = 1, \ldots, K$ }

\FOR{ $\ell = 1, \ldots, N_{\text{sub}}$ }

 \STATE Advance $\theta_{j}^{(i)}, \bm{\mu}^{(i)}, \sigma^{(i)}$ by $\Delta t$ using Algorithm \ref{alg:timestep}, which also recenters the tangent-angle function and reconstructs the contour points $\bm{\psi}_j^{(i)}$

 \STATE If the time step restriction in Algorithm \ref{alg:timestep} shortened the step, stop with an error (see Section \ref{sec:optimization_problems})

\ENDFOR

\STATE Recalibrate the contour points using Algorithm \ref{alg:recalibrate} and store them:
\[
\bm{\psi}_{j, k}^{(i)} \gets \bm{\psi}_{j}^{(i)}, \qquad i \in [m], j \in [n]
\]
(Recalibration only changes the parametrization used for comparison with the target contours; it does not affect the subsequent evolution.)

\ENDFOR

\end{algorithmic}
\end{algorithm}

\begin{algorithm}[H]
\caption{Evaluating the Relative Loss over Time Sequence}
\label{alg:loss}
\begin{algorithmic}
\REQUIRE Starting contour points $\{ \bm{\tilde{\psi}}^{(i)}_{j,0}, i \in [m], j \in [n]\}$ \\
Number of images $K$, \qquad Number of substeps per image $N_{\text{sub}}$ \\
Target contour points $\{\bm{\tilde{\psi}}^{(i)}_{j,k}, i \in [m], j \in [n], k \in [K] \}$, target average speeds $\{\tilde{\sigma}^{(i)}_{k}, i \in [m],k \in [K] \}$, and target centroids $\{\bm{\mu}^{(i)}_{k}, i \in [m],k \in [K] \}$  \\
Interaction kernels $f, g:\mathbb{R}^+ \to \mathbb{R}^+$, \\
Time scaling factor $\tau \in \mathbb{R}^+$, \qquad
Model parameters $c^{(1)}, \ldots, c^{(m)}, \beta \in \mathbb{R}^+$, $\alpha \in \mathbb{R}$

\ENSURE Loss $L_{\text{rel}}$

\STATE Simulate contours numerically using Algorithm \ref{alg:timestepping} to obtain calibrated simulated contour points
\[
\{\bm{\psi}^{(i)}_{j,k}, i \in [m], j \in [n], k \in [K]\}.
\]

\STATE Compute relative loss (assuming target and simulated contour points are calibrated appropriately using Algorithm \ref{alg:recalibrate}):
\[L_{\text{rel}} \gets \frac{1}{K} \sum_{k=1}^K \frac{\sum_{i=1}^m \sum_{j=1}^n |\bm{\psi}_{j, k}^{(i)}-\tilde{\bm{\psi}}_{j, k}^{(i)} |^2 \,\tilde{\sigma}^{(i)}_k }{\sum_{i=1}^m \sum_{j=1}^n | \tilde{\bm{\psi}}_{j, k}^{(i)} - \bm{\mu}^{(i)}_k |^2\,\tilde{\sigma}^{(i)}_k }
\]

\end{algorithmic}
\end{algorithm}

\begin{algorithm}[H]
\caption{Nelder--Mead method for Derivative-Free Optimization} \label{alg:nelder-mead}
\begin{algorithmic}
\REQUIRE Objective function to minimize, $f: \mathbb{R}^d \to \mathbb{R}$ \\
Number of iterations $N$ \\
Reflection coefficient $\rho > 0$ (default $1$), expansion multiplier $\chi > 1$ (default $2$), contraction coefficient $\psi \in (0, 0.5]$ (default $0.5$), shrink factor $\sigma \in (0, 1)$ (default $0.5$) \\
Initial guess $\mathbf{x}_0 \in \mathbb{R}^d$ \\
(Optional): Initial (additional) simplex vertices $\mathbf{x}_1, \ldots, \mathbf{x}_{d} \in \mathbb{R}^d$
\ENSURE Current point $\mathbf{x}^\ast \in \mathbb{R}^d$ with minimal objective function value

\IF{ simplex vertices $\mathbf{x}_1, \ldots, \mathbf{x}_{d} \in \mathbb{R}^d$ not provided} \STATE generate them by perturbing one coordinate of $\mathbf{x}_0$ at a time:
\[
\mathbf{x}_{i} \gets \begin{cases}
    \mathbf{x}_{0} + 0.05 \, (\mathbf{x}_{0})_i \, \mathbf{e}_i & \text{if } (\mathbf{x}_{0})_i \neq 0 \\
    \mathbf{x}_{0} + 0.00025 \, \mathbf{e}_i &\text{otherwise, }
\end{cases}
\qquad i \in [d].
\]
\ENDIF

\STATE Compute function values at vertices:
\[
f_0, \ldots, f_d \gets f(\mathbf{x}_0), \ldots, f(\mathbf{x}_d)
\]

\FOR{$j = 1$ to $N$}

\STATE Reindex vertices $\mathbf{x}_0, \ldots, \mathbf{x}_d$ and associated function values $f_0, \ldots, f_d$ to maintain ordering
\[
f_0 \leq f_1 \leq \cdots \leq f_d
\]

\STATE Compute centroid of the $d$ best vertices and reflected point
\[
\mathbf{x}_c \gets \frac{1}{d}\sum_{i=0}^{d-1}\mathbf{x}_i, \quad \mathbf{x}_r \gets \mathbf{x}_c + \rho  \, (\mathbf{x}_c  - \mathbf{x}_d), \quad f_r \gets f(\mathbf{x}_r)
\]

\textbf{if } $f_r < f_0$, compute expansion point:
\[
\mathbf{x}_e \gets \mathbf{x}_c + \rho \, \chi \, (\mathbf{x}_c - \mathbf{x}_d), \qquad f_e \gets f(\mathbf{x}_e)
\]
\[
\mathbf{x}_d \gets \begin{cases} \mathbf{x}_e  &\text{if } f_e < f_r \\
\mathbf{x}_r &\text{otherwise } \end{cases}, \qquad
f_d \gets \begin{cases} f_e  &\text{if } f_e < f_r \\
f_r &\text{otherwise } \end{cases}
\]

\textbf{else if } $f_0 \leq f_r < f_{d-1}$, use reflection
\[
\mathbf{x}_d \gets \mathbf{x}_r, \qquad f_d \gets f_r
\]

\textbf{else} ($f_r \geq f_{d-1}$) use contraction, outside if $f_r < f_d$ and inside otherwise:
\[
\mathbf{x}_{k} \gets \begin{cases}
\mathbf{x}_c + \psi \rho (\mathbf{x}_c - \mathbf{x}_d) & \text{if } f_r < f_d \\
\mathbf{x}_c - \psi (\mathbf{x}_c - \mathbf{x}_d) & \text{if } f_r \geq f_d,
\end{cases}
\qquad f_{k} \gets f(\mathbf{x}_{k})
\]
\quad \textbf{if } $f_{k} \leq f_r$ (outside) or $f_k < f_d$ (inside), accept:
\[\mathbf{x}_d \gets \mathbf{x}_k, \qquad f_d \gets f_k\]
\quad \textbf{else} perform shrink and re-evaluate:
\[
\mathbf{x}_i \gets \mathbf{x}_0 + \sigma (\mathbf{x}_i - \mathbf{x}_0), \qquad f_i \gets f(\mathbf{x}_i), \qquad i = 1, \ldots, d
\]
\textbf{end if }

\ENDFOR

\RETURN best point $\mathbf{x}_0$

\end{algorithmic}
\end{algorithm}

\begin{algorithm}[H]
\caption{L-BFGS-B with Central-Difference Gradients} \label{alg:lbfgs}
\begin{algorithmic}[1]
\REQUIRE Objective $L$ (Algorithm \ref{alg:loss}) as a function of the parameters $\mathbf{x} \in \mathbb{R}^d$ \\
Initial parameters $\mathbf{x}_0 \in \mathbb{R}^d$, \qquad set $\Lambda \subseteq [d]$ of log-parametrized coordinates \\
Finite-difference step $h > 0$ (default $10^{-4}$), \qquad first-step length $\rho_0 > 0$ (default $0.05$), \qquad bound radius $B > 0$ (default $5$)
\ENSURE Best parameters found, $\mathbf{x}^\ast \in \mathbb{R}^d$

\STATE Change variables:
\[
u_{0,i} \gets \begin{cases} \log x_{0,i} & i \in \Lambda \\ x_{0,i} & i \notin \Lambda \end{cases}, \qquad
x_i(\mathbf{u}) := \begin{cases} e^{u_i} & i \in \Lambda \\ u_i & i \notin \Lambda \end{cases}
\]
\[
\ell(\mathbf{u}) := L(\mathbf{x}(\mathbf{u}))
\]

\STATE Set finite-difference steps: $h_i \gets h$ for $i \in \Lambda$, and $h_i \gets h \, |x_{0,i}|$ otherwise

\STATE Define the gradient estimate, computing the $2d+1$ values of $\ell$ in parallel:
\[
\hat{g}_i(\mathbf{u}) \gets \frac{\ell(\mathbf{u} + h_i \mathbf{e}_i) - \ell(\mathbf{u} - h_i \mathbf{e}_i)}{2 h_i}, \qquad i \in [d]
\]
If a simulation fails at a point $\mathbf{u}$, use $\ell(\mathbf{u}) \gets \ell_{\text{best}} + 10^3 (1 + \|\mathbf{u} - \mathbf{u}_{\text{best}}\|^2)$, where $\mathbf{u}_{\text{best}}$ is the best point evaluated so far

\STATE Scale the objective so that the first step has length $\rho_0$:
\[
\kappa \gets \rho_0 / \|\hat{\mathbf{g}}(\mathbf{u}_0)\|
\]

\STATE Set bounds $|u_i - u_{0,i}| \leq B$ for $i \in \Lambda$ (and $\beta \geq 10^{-3}$ if $\beta$ is optimized)

\STATE Run L-BFGS-B on $\kappa \, \ell$ with gradient $\kappa \, \hat{\mathbf{g}}$ from $\mathbf{u}_0$, subject to the bounds, until its convergence criteria are met or the time budget is exhausted

\RETURN $\mathbf{x}^\ast \gets \mathbf{x}(\mathbf{u}_{\text{best}})$
\end{algorithmic}
\end{algorithm}

\section{Numerical Algorithms}\label{sec:numerical_algorithms}

This section describes in detail the numerical procedures used to model the evolution of contours.

Algorithm \ref{alg:weno3} describes a 3rd-order WENO interpolation procedure for functions defined on $\mathbb{T}^1$. It assumes that values of a function are known at evenly spaced points $v_j := -\pi + \tfrac{2\pi}{n}(j-1) \in \mathbb{T}^1, \quad j \in [n]$. Given $u \in \mathbb{T}^1$, the interpolation procedure uses a four-point stencil with indices $\{{\ell}', \ell, r, r'\} \subseteq [n]$ with $u \in [\ell, r)$. The procedure estimates the value of the function at $u$ using two quadratic Lagrange polynomials on the three-point stencils  $\{{\ell}', \ell, r\}$ and  $\{\ell, r, r'\}$, which are combined using Jiang--Shu smoothness measures \cite{jiang_efficient_1996}. A WENO procedure is used so that higher-order curvature information is preserved while not inducing spurious oscillations near discontinuities.

\begin{algorithm}[H]
\caption{WENO Interpolation on $\mathbb{T}^1$ using Quadratic Lagrange Polynomials} \label{alg:weno3}
\begin{algorithmic}[1]
\REQUIRE Interpolation locations $u_1, \ldots, u_m \in \mathbb{T}^1$ \\
Function values $f_1, \ldots, f_n \in V$ at evenly spaced points $v_j := -\pi + \tfrac{2\pi}{n}(j-1) \in \mathbb{T}^1, \quad j \in [n]$ 
\ENSURE Interpolated values $\tilde{f}_1, \ldots, \tilde{f}_m \in V$ at locations $u_1, \ldots, u_m$

\STATE Define interval length and machine epsilon:
\[\Delta \gets 2\pi / n, \qquad \varepsilon \gets \text{machine epsilon} \]

\STATE Compute stencil indices $\{(\ell_i', \ell_i, r_i, r_i')\}_{i\in[m]} \subseteq [n]$: 
\[
\ell_i \gets 1 + \lfloor (u_i + \pi)/\Delta \rfloor, \quad
r_i \gets (\ell_i + 1) \bmod n 
\]\[
\ell_i' \gets (\ell_i - 1) \bmod n, \quad
r_i' \gets (r_i + 1) \bmod n
\]

\STATE Map $u_i$ from $[v_{\ell_i}, v_{r_i})$ to $[0,1)$:
\[
\theta_i \gets \frac{u_i - (-\pi + (\ell_i -1) \, \Delta)}{\Delta}, \qquad  i \in [m]
\]

\STATE Initialize linear blending weight: 
\[
w_i \gets \frac{2}{3} - \frac{1}{3} \theta_i, \qquad i \in [m]
\]

\STATE Compute Jiang--Shu smoothness measures $(i \in [m])$:
\[
\begin{aligned}
(\beta_L)_i &\gets \varepsilon + (f_{r_i} - f_{\ell_i})^2 + \frac{13}{12}(f_{r_i} + f_{\ell_i'} - 2f_{\ell_i})^2,\\
(\beta_R)_i &\gets \varepsilon + (f_{r_i} - f_{\ell_i})^2 + \frac{13}{12}(f_{r_i'} + f_{\ell_i} - 2f_{r_i})^2
\end{aligned}
\]

\STATE Adjust weights using smoothness measures $(i \in [m])$:
\[
w_i \gets
\begin{cases}
\dfrac{w_i}{\,w_i + \left(\frac{(\beta_L)_i}{(\beta_R)_i}\right)^2 (1-w_i)\,}, & \text{if } (\beta_R)_i > (\beta_L)_i,\\[1.2em]
1 - \dfrac{(1-w_i)}{\, (1-w_i) + \left(\frac{(\beta_R)_i}{(\beta_L)_i}\right)^2  w_i \,}, & \text{otherwise.}
\end{cases}
\]

\STATE Evaluate left and right quadratic polynomials $(i \in [m])$:
\[
\begin{aligned}
(p_L)_i &\gets f_{\ell_i} + \tfrac{1}{2}(f_{r_i} - f_{\ell_i'})\theta_i + \tfrac{1}{2}(f_{r_i} + f_{\ell_i'} - 2f_{\ell_i})\theta_i^2,\\
(p_R)_i &\gets f_{r_i} + \tfrac{1}{2}(f_{r_i'} - f_{\ell_i})(\theta_i - 1) \\&\qquad\quad+ \tfrac{1}{2}(f_{r_i'} + f_{\ell_i} - 2f_{r_i})(\theta_i - 1)^2
\end{aligned}
\]

\STATE Compute final interpolated values $(i \in [m])$:
\[
\tilde{f}_i \gets w_i  (p_L)_i + (1 - w_i) (p_R)_i
\]
\end{algorithmic}
\end{algorithm}

Algorithms \ref{alg:derivative} and \ref{alg:primitive} show how derivatives and antiderivatives of functions defined on $\mathbb{T}^1$ can be computed using the Fast Fourier Transform (FFT).

\begin{algorithm}[H]
\caption{Computing Derivative of Discrete Functions on $\mathbb{T}^1$ using FFT}
\label{alg:derivative}
\begin{algorithmic}[1]
\REQUIRE Function values $f_1, \ldots, f_n \in \mathbb{R}$ at $n$ evenly spaced points in $\mathbb{T}^1$ 
\ENSURE Derivative values $f_1', \ldots, f_n' \in \mathbb{R}$ at $n$ evenly spaced points in $\mathbb{T}^1$
\STATE Vectorize:
\[\mathbf{f} \gets (f_1, \ldots, f_n) \in \mathbb{R}^n \]

\STATE Compute FFT:
\[
\hat{\mathbf{f}} \gets \texttt{FFT}(\mathbf{f})
\]

\STATE Define wavenumbers:
\[
\mathbf{k} \gets [0, 1, \ldots, \lfloor (n - 1) / 2 \rfloor , -\lfloor n / 2 \rfloor, \ldots, -1]
\] 

\STATE Multiply by $i \mathbf{k}$ and compute inverse FFT:
\[
\mathbf{f}' = \texttt{IFFT}( i \mathbf{k} \, \odot \, \hat{\mathbf{f}})
\]

\STATE Extract components:
\[(f_1', \ldots, f_n') \gets \mathbf{f}' \in \mathbb{R}^n\]

\end{algorithmic}
\end{algorithm}

A discrete function's antiderivative is periodic, and hence expressible via Fourier synthesis, if and only if the function has zero mean; a function with nonzero mean instead has an antiderivative containing a linear drift term, which is not periodic. To handle both cases, Algorithm \ref{alg:primitive} first separates out the mean $c$ of the input, computes a periodic (mean-zero) primitive of the remaining zero-mean part using the FFT, and then reintroduces the linear drift $c \cdot x_j$ corresponding to the removed mean. An optional final adjustment anchors the result at $x_1$ to produce a zero-Dirichlet primitive instead.

\begin{algorithm}[H]
\caption{Computing Primitive of Discrete Functions on $\mathbb{T}^1$ using FFT}
\label{alg:primitive}
\begin{algorithmic}[1]
\REQUIRE Function values $f_1, \ldots, f_n \in \mathbb{R}$ at $n$ evenly spaced points $x_j := -\pi + \tfrac{2\pi}{n}(j-1) \in \mathbb{T}^1$ \\
Type of primitive (mean-zero or zero-Dirichlet)
\ENSURE Primitive values $Pf_1, \ldots, Pf_n \in \mathbb{R}$ at $n$ evenly spaced points in $\mathbb{T}^1$

\STATE Compute the mean of $f$ and the mean-zero remainder:
\[
c \gets \frac{1}{n}\sum_{j=1}^n f_j, \qquad \tilde{f}_j \gets f_j - c, \quad j \in [n]
\]

\STATE Vectorize:
\[\tilde{\mathbf{f}} \gets (\tilde{f}_1, \ldots, \tilde{f}_n) \in \mathbb{R}^n \]

\STATE Compute FFT:
\[
\hat{\tilde{\mathbf{f}}} \gets \texttt{FFT}(\tilde{\mathbf{f}})
\]

\STATE Define reciprocal of wavenumbers:
\[
\bm{\xi} \gets \left[1, 1, \frac{1}{2}, \frac{1}{3}, \ldots, \frac{1}{\lfloor (n - 1) / 2 \rfloor} , -\frac{1}{\lfloor n / 2 \rfloor}, \ldots, -1 \right]
\]

\STATE Multiply by $-i \bm{\xi}$ and compute inverse FFT:
\[
P\tilde{\mathbf{f}} = \texttt{IFFT}( -i \bm{\xi} \, \odot \, \hat{\tilde{\mathbf{f}}})
\]

\STATE Reintroduce the linear drift corresponding to the removed mean, giving the mean-zero primitive:
\[
Pf_j \gets \left(P\tilde{\mathbf{f}}\right)_j + c \cdot x_j, \qquad j \in [n]
\]

\STATE For zero-Dirichlet primitive, additionally:
\[
Pf_j \gets Pf_j - Pf_1, \qquad j \in [n]
\]

\end{algorithmic}
\end{algorithm}

Algorithm \ref{alg:reparametrize} is a procedure for obtaining a constant-speed representation of a contour $\bm{\gamma}: \mathbb{T}^1 \to \mathbb{R}^2$, i.e. an alternate parametrization $\bm{\psi} : \mathbb{T}^1 \to \mathbb{R}^2$ defined so that $|\bm{\psi}'(u)|$ is constant for all $u \in \mathbb{T}^1$. This is done by computing a change of variables map $\xi: \mathbb{T}^1 \to \mathbb{T}^1$ which solves the initial value problem 
\begin{equation}\label{eq:reparametrize_ivp}
\xi'(u) = \frac{\sigma}{|\bm{\gamma}'(\xi(u))|}, \qquad \xi(-\pi) = -\pi
\end{equation}
where 
\[
\sigma := \frac{1}{2 \pi} \int_{\mathbb{T}^1} |\bm{\gamma}'(u)| \dd u
\]
is the average speed of the contour. Then $\bm{\psi}$ may be defined as the composition $\bm{\psi} := \bm{\gamma} \circ \xi$, and such a $\bm{\psi}$ will have the property that $|\bm{\psi}'(u)| = \sigma$ for all $u \in \mathbb{T}^1$ (i.e. $\bm{\psi}$ is a constant-speed representation of the same contour). Algorithm \ref{alg:reparametrize} demonstrates this procedure using discrete representations of the contours, by using the FFT to compute $|\bm{\gamma}'(u)|$ and then a 4th-order Runge--Kutta method to solve the IVP \eqref{eq:reparametrize_ivp}. Lastly, interpolation (see Algorithm \ref{alg:weno3}) is needed to compute the composition $\bm{\gamma} \circ \xi$.

\begin{algorithm}[H]
\caption{Resampling for Contours in $\mathbb{R}^2$}
\label{alg:reparametrize}
\begin{algorithmic}[1]
\REQUIRE Contour points $\bm{\gamma}_1, \ldots, \bm{\gamma}_n \in \mathbb{R}^2$
\ENSURE Contour points $\bm{\psi}_1, \ldots, \bm{\psi}_n \in \mathbb{R}^2$ describing  a constant-speed representation of the original contour

\STATE Compute directional derivatives of $\bm{\gamma}$ (see Algorithm \ref{alg:derivative}) $(i \in [2])$:
\[
(\bm{\gamma}_1)'_i, \ldots, (\bm{\gamma}_n)'_i \gets \texttt{Derivative}((\bm{\gamma}_1)_i, \ldots, (\bm{\gamma}_n)_i )
\]

\STATE Compute speed $(j \in [n])$ and average speed
\[
\sigma_j \gets \sqrt{ \sum_{i=1}^2 \left((\bm{\gamma}_j)'_i\right)^2 }, \qquad \sigma \gets \frac{1}{n}\sum_{j=1}^n \sigma_j
\]

\STATE Compute change of variables $\xi_1, \ldots, \xi_n \in \mathbb{T}^1$ to satisfy the IVP \eqref{eq:reparametrize_ivp}. See Algorithm \ref{alg:weno3}
\[
\xi_1 \gets -\pi,
\]
 \textbf{for} $j = 2$ to $n$ \textbf{do}
    \begin{align*}
    F_1 &\gets \texttt{Interpolate} (\xi_{j-1}, (\sigma/\sigma_1, \ldots, \sigma/\sigma_n)) \\
    F_2 &\gets \texttt{Interpolate} (\xi_{j-1} + \tfrac{1}{2} \tfrac{2\pi}{n}  F_1, (\sigma/\sigma_1, \ldots, \sigma/\sigma_n)) \\
    F_3 &\gets \texttt{Interpolate} (\xi_{j-1} + \tfrac{1}{2} \tfrac{2\pi}{n}  F_2, (\sigma/\sigma_1, \ldots, \sigma/\sigma_n)) \\
    F_4 &\gets \texttt{Interpolate} (\xi_{j-1} + \tfrac{2\pi}{n}  F_3, (\sigma/\sigma_1, \ldots, \sigma/\sigma_n)) \\
    \xi_j &\gets \xi_{j-1} + \tfrac{2\pi}{n} ( \tfrac{1}{6} F_1 + \tfrac{1}{3} F_2 + \tfrac{1}{3} F_3 + \tfrac{1}{6} F_4)
    \end{align*}
\textbf{endfor}

\STATE Compute constant-speed representation $\bm{\psi} : \mathbb{T}^1 \to \mathbb{R}^2$ : 
\[
\bm{\psi}_1, \ldots, \bm{\psi}_n \gets \texttt{Interpolate} ( (\xi_1, \ldots, \xi_n), (\bm{\gamma}_1, \ldots, \bm{\gamma}_n))
\]
\end{algorithmic}
\end{algorithm}

Algorithm \ref{alg:recalibrate} describes the procedure used to recalibrate contour points so that the first point is oriented in the $(1,0)$ direction from the centroid of the contour. This is done primarily because the training metric used to measure the discrepancy between a target contour $\tilde{\bm{\psi}}$ and a simulated contour $\bm{\psi}$ measures the distance between $\tilde{\bm{\psi}}(u)$ and $\bm{\psi}(u)$ for the same point $u \in \mathbb{T}^1$. Thus, rotating the parametrizations so that $u = -\pi$ points to the $(1,0)$ direction in each case ensures that similar parameter values match up with corresponding portions of each contour.

This procedure assumes that the interior of the contour $C \subseteq \mathbb{R}^2$ is star-shaped with respect to the centroid $\bm{\mu}$ of the contour, i.e. that for all $\mathbf{x} \in C$ the line segment between $\bm{\mu}$ and $\mathbf{x}$ is contained within $C$.

\begin{algorithm}[H]
\caption{Contour Recalibration}
\label{alg:recalibrate}
\begin{algorithmic}[1]

\REQUIRE Constant-speed contour points $\bm{\psi}_1, \ldots, \bm{\psi}_n \in \mathbb{R}^2$ 
\ENSURE Recalibrated constant-speed contour points $\bm{\psi}_1, \ldots, \bm{\psi}_n \in \mathbb{R}^2$ which describe the same contour, but with the first point oriented in the $(1,0)$ direction from the centroid

\STATE Compute centroid:
\[
\bm{\mu} \gets \frac{1}{n} \sum_{i=1}^n \bm{\psi}_i
\]

\STATE Compute angles of points from centroid $(j \in [n])$:
\[
\theta_j \gets \arctan( (\bm{\psi}_j)_2 - (\bm{\mu})_2, (\bm{\psi}_j)_1 - (\bm{\mu})_1 )
\]

\STATE Find indices near zero crossing (assuming counterclockwise orientation):
\[
k \gets \min \{ j: \theta_j > 0 \}, \qquad 
\ell \gets (k-1) \bmod n
\]

\STATE Use linear interpolation to find the approximate fractional index of zero crossing:
\[
r \gets \frac{k \theta_\ell - \ell \theta_k}{\theta_\ell - \theta_k}
\]

\STATE Get shifted angles in $\mathbb{T}^1$, $(i=1, \ldots, n)$:
\[
\theta_i' \gets  (\tfrac{2 \pi}{n} r + \theta_1) - 2\pi \lfloor (\tfrac{2 \pi}{n} r + \theta_1 + \pi)/(2\pi) \rfloor
\]

\STATE Interpolate points at shifted indices (see Algorithm \ref{alg:weno3}):
\[
\bm{\psi}_1, \ldots, \bm{\psi}_n 
\gets \texttt{Interpolate}( (\theta_1', \ldots, \theta_n'), (\bm{\psi}_1, \ldots, \bm{\psi}_n))
\]
\end{algorithmic}
\end{algorithm}

Algorithm \ref{alg:propagate2} shows a 2nd-order Exponential Time Differencing (ETD2) scheme for time-stepping the forced diffusion equation
\[
q_t(u, t) = q_{u u}(u, t) + G(u, t)
\]
where $q(\cdot, t): \mathbb{T}^1 \to \mathbb{R}$. In the Fourier domain, the equation reads
\[
\hat{q}_t(k, t) = - k^2 \hat{q}(k, t) + \hat{G}(k, t)
\]
which has the exact solution
\[
\hat{q}(k, t + \Delta t) 
= e^{-k^2 \Delta t} \hat{q}(k, t) + \int_0^{\Delta t} e^{-k^2(\Delta t - s)} \hat{G}(k, t+s) \dd s.
\]
If we approximate $\hat{G}(k, t+s)$ using linear interpolation 
\[
\hat{G}(\cdot, t+s) \approx \hat{G}(\cdot, t) + \frac{s}{\Delta t} \left(\hat{G}(\cdot, t+\Delta t) - \hat{G}(\cdot, t) \right)
\]
then substituting into the exact formula yields
\begin{align*}
\hat{q}(k, t + \Delta t) 
&=e^{-k^2 \Delta t} \hat{q}(k, t) + \Delta t \varphi_1(-k^2 \Delta t) \hat{G}(k, t) \\
&+  \Delta t \varphi_2(-k^2 \Delta t) \left(\hat{G}(k, t+\Delta t) - \hat{G}(k, t)  \right)
\end{align*}
where $\varphi_1(z) = \frac{e^z - 1}{z}$ and 
\[
\varphi_2(z) = \frac{\varphi_1(z) - 1}{z} = \frac{e^z - z - 1}{z^2}.
\]
This yields a second-order approximation for $q(u, t + \Delta t)$ after the inverse FFT is performed. Note that for small inputs, the functions $\varphi_1$ and $\varphi_2$ are computed using a series approximation to avoid small terms in the denominator. 

\begin{algorithm}[H]
\caption{Timestepping Forced Diffusion Equation on $\mathbb{T}^1$}
\label{alg:propagate2}
\begin{algorithmic}[1]
\REQUIRE Function values $q_{1, k}, \ldots, q_{n, k} \in \mathbb{R}$ at time $t_k$ \\
Time step $\Delta t > 0$ \\
Current-time forcing $G_{1, k}, \ldots, G_{n, k} \in \mathbb{R}$ and one-step-ahead forcing $G_{1, k+1}, \ldots, G_{n, k+1}$  at time $t_{k+1} := t_k + \Delta t$

\ENSURE Function values $q_{1, k+1}, \ldots, q_{n, k+1} \in \mathbb{R}$ at time $t_{k+1} := t_k + \Delta t$
\STATE Vectorize:
\[\mathbf{q}_k \gets (q_{1, k}, \ldots, q_{n, k} ) \in \mathbb{R}^n \]
\[\mathbf{G}_\ell \gets (G_{1, \ell}, \ldots, G_{n, \ell}) \in \mathbb{R}^n, \qquad \ell = k, k+1 \]

\STATE Define wavenumbers:
\[
\mathbf{k} \gets [0, 1, \ldots, \lfloor (n - 1) / 2 \rfloor , -\lfloor n / 2 \rfloor, \ldots, -1] \in \mathbb{Z}^n
\] 

\STATE Compute FFTs
\[
\hat{\mathbf{q}}_k \gets \texttt{FFT}(\mathbf{q}_k); \qquad 
\hat{\mathbf{G}}_\ell \gets \texttt{FFT}(\mathbf{G}_\ell), \quad \ell = k, k +1
\]

\STATE Compute exponential of Fourier-space diffusion operator (using vectorized operations):
\[
\mathbf{E}^{(0)} \gets \exp(- \mathbf{k}^2\Delta t) \in \mathbb{R}^n
\]

\STATE Compute coefficients for time-stepping, using vectorized operations (note: revert to series approximation for $|k_i^2\Delta t| \ll 1$):
\[
\mathbf{E}^{(1)} \gets \frac{\mathbf{E}^{(0)} - 1}{-\Delta t \mathbf{k}^2}  \in \mathbb{R}^n
\]
\[
\mathbf{E}^{(2)} \gets \frac{\mathbf{E}^{(1)} - 1}{-\Delta t \mathbf{k}^2}  \in \mathbb{R}^n
\]

\STATE Evaluate approximation to exact time step using vectorized operations:
\[
\hat{\mathbf{q}}_{k+1} \gets  \mathbf{E}^{(0)} \hat{\mathbf{q}}_k +
\Delta t \left( \mathbf{E}^{(1)} \hat{\mathbf{G}}_k + \mathbf{E}^{(2)} (\hat{\mathbf{G}}_{k+1} - \hat{\mathbf{G}}_k) \right)
\]

\STATE Compute inverse FFT:
\[
\mathbf{q}_{k+1} \gets  \texttt{IFFT}( \hat{\mathbf{q}}_{k+1})
\]

\STATE Extract components:
\[q_{1, k+1}, \ldots, q_{n, k+1}  \gets \mathbf{q}_{k+1} \in \mathbb{R}^n\]

\end{algorithmic}
\end{algorithm}

Algorithm \ref{alg:gamtotht} shows how to transform a contour from a representation as a parametrized function $\bm{\gamma}:\mathbb{T}^1 \to \mathbb{R}^2$ to the tangent-angle representation, consisting of a tangent-angle function $\theta :\mathbb{T}^1 \to \mathbb{R}$, a speed $\sigma$, and a centroid $\bm{\mu}$. The discrete values $\theta_1, \ldots, \theta_n$ of the tangent-angle function are only defined $\bmod 2\pi$, but because simulating the evolution of the contour requires simulating a diffusion equation for $\theta$, we must ensure that there are no unnecessarily large jumps in multiples of $2\pi$. To accomplish this, the discrete values are modified so that successive angles differ by less than $\pi$ (for example, if $\theta_{2} - \theta_1 > \pi$, then we subtract $2\pi$ from $\theta_2$). Note that this may mean that $\theta_n$ differs from $\theta_1$ by approximately an integer multiple of $2\pi$, which is the winding number of the contour.

By comparison, Algorithm \ref{alg:thttogam} (which converts back to the parametrized representation) is much more straightforward and utilizes mean-zero primitives (see Algorithm \ref{alg:primitive}).

\begin{algorithm}[H]
\caption{Transforming a Contour $\bm{\gamma}:\mathbb{T}^1 \to \mathbb{R}^2$ to Tangent-Angle Representation}
\label{alg:gamtotht}
\begin{algorithmic}[1]
\REQUIRE Contour points $\bm{\gamma}_1, \ldots, \bm{\gamma}_n \in \mathbb{R}^2$ 
\ENSURE Tangent-angle function values $\theta_1, \ldots, \theta_n \in \mathbb{R}$, \\
centroid $\bm{\mu} \in \mathbb{R}^2$, average speed $\sigma \in \mathbb{R}^+$

\STATE Compute directional derivatives of $\bm{\gamma}$ (see Algorithm \ref{alg:derivative}) $(i \in [2])$:
\[
(\bm{\gamma}_1)'_i, \ldots, (\bm{\gamma}_n)'_i \gets \texttt{Derivative}((\bm{\gamma}_1)_i, \ldots, (\bm{\gamma}_n)_i )
\]

\STATE Compute speed $(j \in [n])$, average speed, and centroid
\[
\sigma_j \gets \sqrt{ \sum_{i=1}^2 \left((\bm{\gamma}_j)'_i\right)^2 }, \qquad
\sigma \gets \frac{1}{n}\sum_{j=1}^n \sigma_j, \qquad
\bm{\mu} \gets \frac{1}{n}\sum_{j=1}^n \bm{\gamma}_j %
\]
\STATE Compute tangent-angle function values $(j \in [n])$:
\[
\theta_j \gets \arctan\left( - \frac{(\bm{\gamma}_j)'_1}{\sigma_j}, \frac{(\bm{\gamma}_j)'_2}{\sigma_j} \right)
\]

\STATE Clean tangent-angle function values so that $|\theta_j - \theta_{j+1}| \leq \pi$ for all $j\in[n-1]$:
\[
\delta_j \gets \theta_{j+1} - \theta_j, \qquad j \in [n-1]
\]
\[
J_j \gets 2 \pi \left( H(\delta_j - \pi) - H(-\delta_j - \pi) \right), \qquad j \in [n-1]
\]
\[
\theta_j \gets \theta_1 + \sum_{i=1}^{j-1} \left( \delta_i - J_i \right)
\]

\end{algorithmic}
\end{algorithm}

\begin{algorithm}[H]
\caption{Transforming a Contour from Tangent-Angle Representation to $\bm{\gamma}:\mathbb{T}^1 \to \mathbb{R}^2$}
\label{alg:thttogam}
\begin{algorithmic}[1]
\REQUIRE Tangent-angle function values $\theta_1, \ldots, \theta_n \in \mathbb{R}$ \\
centroid $\bm{\mu} \in \mathbb{R}^2$, average speed $\sigma \in \mathbb{R}^+$
\ENSURE Contour points $\bm{\gamma}_1, \ldots, \bm{\gamma}_n \in \mathbb{R}^2$

\STATE Compute mean-zero primitives componentwise $(j \in [n])$ (see Algorithm \ref{alg:primitive})
\[
(\bm{\gamma}_j)_1 \gets \sigma \cdot \texttt{Primitive}(-\sin(\theta_1), \ldots, -\sin(\theta_n)) + \mu_1
\]
\[
(\bm{\gamma}_j)_2 \gets \sigma \cdot  \texttt{Primitive}(\cos(\theta_1), \ldots, \cos(\theta_n)) + \mu_2
\]

\end{algorithmic}
\end{algorithm}

Due to numerical error, the discrete values $\theta_1, \ldots, \theta_n$ of  the tangent-angle function may drift so that they no longer represent a contour centered at $(0,0)$. This will cause the stored centroid value to be inaccurate. Instead of compounding the error by recalculating the centroid, Algorithm \ref{alg:normalize} demonstrates how to recenter the tangent-angle function by using several Newton steps. After these steps are performed, the discrete values are ``cleaned'' in a procedure similar to Algorithm \ref{alg:gamtotht} to ensure that $|\theta_{j+1} - \theta_j| \leq \pi$ for all $j \in [n-1]$. 

\begin{algorithm}[H]
\caption{Recenter Tangent-Angle Function}
\label{alg:normalize}
\begin{algorithmic}[1]
\REQUIRE Tangent-angle function values $\theta_1, \ldots, \theta_n \in \mathbb{T}^1$
\ENSURE New tangent-angle function values $\theta_1, \ldots, \theta_n \in \mathbb{T}^1$

\STATE Save first tangent angle: $\varphi \gets \theta_1$

\STATE Compute negative sines and cosines $(j \in [n])$:
\[
s_j  = -\sin(\theta_j), \qquad c_j = \cos(\theta_j)
\]

\STATE Newton iteration on $\mathbf{F}(x, y) := \sum_{j=1}^n \tfrac{(s_j+x, c_j+y)}{\sqrt{s_j^2 + c_j^2}}$:

\textbf{for} $i = 1$ to $4$ \textbf{do}
\[
\mathbf{F} \gets \sum_{j=1}^n \tfrac{(s_j, c_j)}{\sqrt{s_j^2 + c_j^2}} \]
\[D \mathbf{F} \gets 
\sum_{j=1}^n \frac{1}{(s_j^2 + c_j^2)^{3/2}}\begin{bmatrix} c_j^2 & - s_j c_j \\
-s_j c_j & s_j^2 \end{bmatrix} 
\]
\[
(s_j, c_j) \gets (s_j, c_j) -  (D\mathbf{F})^{-1} \mathbf{F}, \qquad j \in [n]
\]

\STATE Recompute tangent-angle function values $(j\in[n])$:
\[
\theta_j \gets \arctan\left( - \frac{s_j}{\sqrt{s_j^2 + c_j^2}}, \frac{c_j}{\sqrt{s_j^2 + c_j^2}} \right), \qquad j \in [n]
\]

\STATE Clean tangent-angle function values so that $|\theta_j - \theta_{j+1}| \leq \pi$ for all $j\in[n-1]$:
\[
\delta_j \gets \theta_{j+1} - \theta_j, \qquad j \in [n-1]
\]
\[
J_j \gets 2 \pi \left( H(\delta_j - \pi) - H(-\delta_j - \pi) \right), \qquad j \in [n-1]
\]
\[
\theta_j \gets \theta_1 + \sum_{i=1}^{j-1} \left( \delta_i - J_i \right)
\]

\STATE Re-anchor the tangent-angle values so that the first value is unchanged from its input value $\varphi$, preserving the reference direction of the contour:
\[
\theta_j \gets \theta_j - \theta_1 + \varphi, \qquad j \in [n]
\]

\end{algorithmic}
\end{algorithm}

Algorithm \ref{alg:nlforces} computes the nonlocal forces between each pair of colonies. For self-interactions $(i \in [m])$
\[
\texttt{NLF}^{(ii)}(u) 
 = \int_{\mathbb{T}^1} f\left(\frac{1}{2} \left\lvert \bm{\gamma}^{(i)}(u,t) - \bm{\gamma}^{(i)}(v,t) \right\rvert^2\right) |\partial_u \bm{\gamma}^{(i)}(v,t) | \dd v
\]
and for cross-interactions $(i,j \in [m]; i \neq j)$
\[
\texttt{NLF}^{(ij)}(u) = \int_{\mathbb{T}^1} g\left(\frac{1}{2} \left\lvert \bm{\gamma}^{(i)}(u,t) - \bm{\gamma}^{(j)}(v,t) \right\rvert^2\right) \left\lvert \partial_u \bm{\gamma}^{(j)}(v,t) \right\rvert \dd v. \]
Assuming that the parametrizations $\{\bm{\gamma}^{(i)}\}_{i \in [m]}$ each have constant speed $\sigma^{(i)}$, the computation of the integrals reduces to sums over the discrete contour points. Lastly, recall that the forces are multiplied by the time scaling $\tau > 0$, which is included as a model parameter.

\begin{algorithm}[H]
\caption{Compute Nonlocal Forces}
\label{alg:nlforces}
\begin{algorithmic}[1]
\REQUIRE Contour points $\{\bm{\gamma}_\ell^{(i)}: i \in [m], \ell \in [n] \} \subset \mathbb{R}^2$ at $n$ evenly spaced points in $\mathbb{T}^1$, for $m$ colonies \\
Average speeds $\{\sigma^{(i)}\}_{i=1}^m \subset \mathbb{R}$ \\
Interaction kernels $f, g:\mathbb{R}^+ \to \mathbb{R}^+$ \\
Time scaling factor $\tau \in \mathbb{R}^+$

\ENSURE Nonlocal forces $\{\texttt{NLF}^{(ij)}_\ell: i, j \in [m], \ell \in [n]\}$ on colony $i$ from colony $j$, at $n$ evenly spaced points in $\mathbb{T}^1$

\STATE Compute self-interactions $(i \in [m]; \ell \in [n])$:
\[
\texttt{NLF}^{(ii)}_{\ell} \gets \tau \sigma^{(i)} \frac{2 \pi}{n} \sum_{k=1}^n f \left( \frac{1}{2} \Big\|\bm{\gamma}_\ell^{(i)} - \bm{\gamma}_k^{(i)} \Big\|^2 \right)
\]
\STATE Compute cross-interactions $(i, j \in [m]; i \neq j; \ell \in [n])$:
\[
\texttt{NLF}^{(ij)}_{\ell} \gets \tau \sigma^{(j)} \frac{2 \pi}{n} \sum_{k=1}^n g \left( \frac{1}{2} \Big\|\bm{\gamma}_\ell^{(i)} - \bm{\gamma}_k^{(j)} \Big\|^2 \right)
\]

\end{algorithmic}
\end{algorithm} 

Algorithm \ref{alg:timestep} (continued in Algorithm \ref{alg:timestep2}) illustrates the procedure for computing a time step of the governing equations in the tangent-angle representation (see \eqref{eq:tangent-angle-governing-equations}, or \eqref{eq:tangent-angle-governing-equations-transport} when the transport mechanism is included). First, the curvature $\mathcal{K}^{(i)}$, the scaled forcing $\mathcal{F}^{(i)}$, and the bending parameter $\mathbf{a}^{(i)}$ are computed at the current time, together with the minimum distances between colonies, which come from the same pairwise distances used to compute the nonlocal forces. The time step restriction which guarantees that $\sigma^{(i)}$ stays positive (derived below) is then checked; with a fixed number of substeps per image it does not become active in practice (Section \ref{sec:optimization_problems}), but the check is retained as a safeguard. Estimates of $(\mathcal{K}^{(i)}, \mathcal{F}^{(i)}, \mathbf{a}^{(i)})$ at the next time are computed using linear extrapolation, and $\sigma^{(i)}$ is advanced to the next time step. The total diffusion force for the tangent-angle function is computed, and the tangent-angle function is advanced using a pseudo-exponential integrator. The centroids are then advanced by the trapezoidal rule. Note that the centroid velocity carries the factor $1/\sigma^{(i)}$, not the factor $1/{\sigma^{(i)}}^2$ which appears in the equation for $\theta^{(i)}$ (compare \eqref{eq:tangent-angle-governing-equations}). Finally, the tangent-angle function is recentered, the contour points are reconstructed, and, if the transport mechanism is included, each colony is translated rigidly by the transport displacement. The transport step is an explicit Euler step, which uses the minimum distances from the start of the step and the directions between the centroids after they have been advanced.

For the standard model, each stage of the update is second-order accurate in $\Delta t$: the exponential time-differencing update for $\theta^{(i)}$ (Algorithm \ref{alg:propagate2}), the second-order Adams--Bashforth update for ${\sigma^{(i)}}^2$, and the trapezoidal rule for the centroids, with the forces at the next time estimated by linear extrapolation. The convergence study in Section \ref{sec:optimization_problems} confirms that the loss converges at second order.

\textbf{Time Step Restriction.} The average speeds evolve according to $\partial_t [{\sigma^{(i)}}^2] = -2\mu_{a^{(i)}}$ (see \eqref{eq:tangent-angle-governing-equations}). Writing $\mathcal{M} = \mathcal{M}^{(i)}_k$ for the mean of $a^{(i)}$ at time $t_k$, $\dot{\mathcal{M}} = \partial_t \tilde{\mathcal{M}}^{(i)}_k$ for the estimate of its time derivative, and $\sigma = \sigma^{(i)}_k$, Algorithm \ref{alg:timestep} advances the squared speed by the second-order step
\[
\left(\sigma^{(i)}_{k+1}\right)^2 = \sigma^2 - 2 \Delta t\, \mathcal{M} - (\Delta t)^2 \dot{\mathcal{M}},
\]
whose right-hand side must be positive for $\sigma^{(i)}_{k+1}$ to be defined. This holds whenever $2 \Delta t |\mathcal{M}| + (\Delta t)^2 |\dot{\mathcal{M}}| < \sigma^2$, and a simple sufficient condition is
\begin{equation}\label{eq:dt_restriction}
\Delta t < \frac{1}{2} \frac{\sigma^2}{\sqrt{\mathcal{M}^2 + \tfrac{1}{2} \sigma^2 |\dot{\mathcal{M}}|}}.
\end{equation}
To see this, let $D = \mathcal{M}^2 + \tfrac{1}{2} \sigma^2 |\dot{\mathcal{M}}|$ and $x = |\mathcal{M}|/\sqrt{D} \in [0, 1]$, so that $\tfrac{1}{2} \sigma^2 |\dot{\mathcal{M}}| / D = 1 - x^2$. If \eqref{eq:dt_restriction} holds, then $2 \Delta t |\mathcal{M}| < \sigma^2 x$ and $(\Delta t)^2 |\dot{\mathcal{M}}| < \sigma^4 |\dot{\mathcal{M}}| / (4D) = \tfrac{1}{2} \sigma^2 (1 - x^2)$, so
\begin{align*}
2 \Delta t |\mathcal{M}| + (\Delta t)^2 |\dot{\mathcal{M}}| &< \sigma^2 \left( x + \tfrac{1}{2}(1 - x^2) \right) \\
&= \sigma^2 \left( 1 - \tfrac{1}{2}(1 - x)^2 \right) \leq \sigma^2.
\end{align*}
The bound \eqref{eq:dt_restriction} is never smaller than $1/\sqrt{2}$ times the largest step allowed by the condition $2 \Delta t |\mathcal{M}| + (\Delta t)^2 |\dot{\mathcal{M}}| < \sigma^2$, namely $\sigma^2 / \big(|\mathcal{M}| + \sqrt{\mathcal{M}^2 + \sigma^2 |\dot{\mathcal{M}}|}\big)$, and it equals that step when $\dot{\mathcal{M}} = 0$. At the optimized parameters of all $19$ per-colony fits (both kernel classes and both basins, with and without transport), the right-hand side of \eqref{eq:dt_restriction} exceeded the substep $\Delta t$ by a factor of at least $43$ throughout the simulation. 

\begin{algorithm}[H]
\caption{Advance to Next Time Step }
\label{alg:timestep}
\begin{algorithmic}[1]
\REQUIRE Tangent-angle values $\{\theta_{\ell, k}^{(i)}: i \in [m], \ell \in [n] \} \subset \mathbb{R}$ at time $t_k$, \qquad
Average speeds $\{\sigma^{(i)}_k\}_{i\in[m]}\subset \mathbb{R}$ at time $t_k$, \\
centroids $\{\bm{\mu}^{(i)}_k\}_{i\in[m]} \subset \mathbb{R}^2$  at time $t_{k}$, \qquad 
Interaction kernels $f, g:\mathbb{R}^+ \to \mathbb{R}^+$, \qquad
Time scaling factor $\tau \in \mathbb{R}^+$, \\
Model parameters $c^{(1)}, \ldots, c^{(m)}, \beta \in \mathbb{R}^+$, $\alpha \in \mathbb{R}$ ($\alpha = 0$ for the standard model), \qquad Gate width $w > 0$, \\
Time step $\Delta t \in \mathbb{R}^+$, \qquad Current time $t_k$, \qquad Past time $t_{k-1}$ \\
Past values of $\bm{\mathcal{K}}, \bm{\mathcal{F}}, \mathbf{a}$: $\{(\bm{\mathcal{K}}_{k-1}^{(i)}, \bm{\mathcal{F}}_{k-1}^{(i)}, \mathbf{a}_{k-1}^{(i)}), i \in [m]\}$

\ENSURE Time $t_{k+1} \in \mathbb{R}$, \qquad 
Tangent-angle values $\{\theta_{\ell, k+1}^{(i)}: i \in [m], \ell \in [n] \} \subset \mathbb{R}$ at time $t_{k+1}$, \\
Average speeds $\{\sigma^{(i)}_{k+1}\}_{i\in[m]} \subset \mathbb{R}$ at time $t_{k+1}$, \qquad
centroids $\{\bm{\mu}^{(i)}_{k+1}\}_{i\in[m]} \subset \mathbb{R}^2$ at time $t_{k+1}$, \qquad
contour points $\{\bm{\gamma}_{\ell, k+1}^{(i)}: i \in [m], \ell \in [n] \}$ at time $t_{k+1}$
\vspace{12pt}
\STATE Update forces (see Algorithms \ref{alg:derivative}, \ref{alg:nlforces}, \ref{alg:thttogam}), ($i\in[m]$):
\[
\bm{\mathcal{K}}^{(i)}_k\gets \texttt{Derivative}( \theta_{1, k}^{(i)}, \ldots, \theta_{n, k}^{(i)}) \in \mathbb{R}^n, \qquad 
\bm{\gamma}^{(i)}_{k} \gets \texttt{Contour}\left( \bm{\theta}^{(i)}_{k}, \sigma^{(i)}_{k}, \bm{\mu}^{(i)}_k \right) \]
\[
\bm{\mathcal{F}}^{(i)}_k \gets c^{(i)} \sigma^{(i)}_k \mathbf{1} - \texttt{NLForces}(\{\bm{\gamma}^{(i)}_k\}_{i\in[m]}, \{\sigma^{(i)}_k\}_{i\in[m]}, f, g, \tau) \in \mathbb{R}^n, \qquad
\mathbf{a}^{(i)}_k \gets \bm{\mathcal{K}}^{(i)}_k \odot \left( \bm{\mathcal{K}}^{(i)}_k - \bm{\mathcal{F}}^{(i)}_k \right)
\]
\STATE If the transport mechanism is included, record the minimum distances between colonies from the pairwise distances computed in Algorithm \ref{alg:nlforces}, $(i, j \in [m]; i \neq j)$:
\[
s^{(ij)}_{k} \gets \min_{\ell, \ell' \in [n]} \left \lvert \bm{\gamma}_{\ell, k}^{(i)} - \bm{\gamma}_{\ell', k}^{(j)} \right \rvert
\]

\STATE Compute mean of $\mathbf{a}$ and estimate its derivative over $[t_{k-1}, t_k]$, ($i \in [m]$)
\[
\mathcal{M}^{(i)}_k \gets \frac{1}{n}\sum_{\ell=1}^n a^{(i)}_{\ell,k}, \qquad 
\partial_t{\tilde{\mathcal{M}}^{(i)}_k} \gets  \frac{1}{n}\sum_{\ell=1}^n \frac{a^{(i)}_{\ell,k} - a^{(i)}_{\ell, k-1}}{t_k - t_{k-1}}
\]

\STATE Check the time step restriction \eqref{eq:dt_restriction}, which guarantees that every $\sigma^{(i)}_{k+1}$ is well defined and positive (a safeguard; with a fixed number of substeps per image it does not become active in practice, and Algorithm \ref{alg:timestepping} stops with an error if it does):
\[
\Delta t < \min_{i\in[m]} \frac{1}{2} \frac{\left(\sigma^{(i)}_k\right)^2}{\sqrt{\left(\mathcal{M}^{(i)}_k\right)^2 + \tfrac{1}{2} \left(\sigma^{(i)}_{k}\right)^2 \left\lvert \partial_t{\tilde{\mathcal{M}}^{(i)}_k} \right\rvert }}
\]

\STATE Set next time: $t_{k+1} \gets t_k + \Delta t$

\STATE Use linear extrapolation to estimate next values for $\mathbf{a}, \bm{\mathcal{F}}, \bm{\mathcal{K}}$, for $(i\in[m]; \ell \in [n])$:
\[
\tilde{a}^{(i)}_{\ell, k+1} \gets a^{(i)}_{\ell, k} + \Delta t \frac{a^{(i)}_{\ell, k} - a^{(i)}_{\ell, k-1}}{t_k - t_{k-1}}, \quad
\tilde{\mathcal{K}}^{(i)}_{\ell, k+1} \gets \mathcal{K}^{(i)}_{\ell, k} + \Delta t \frac{\mathcal{K}^{(i)}_{\ell, k} - \mathcal{K}^{(i)}_{\ell, k-1}}{t_k - t_{k-1}}, \quad 
\tilde{\mathcal{F}}^{(i)}_{\ell, k+1} \gets \mathcal{F}^{(i)}_{\ell, k} + \Delta t \frac{\mathcal{F}^{(i)}_{\ell, k} - \mathcal{F}^{(i)}_{\ell, k-1}}{t_k - t_{k-1}} 
\]

\STATE Advance average speeds using 2nd-order Adams--Bashforth explicit method $(i\in[m])$:
\[
\sigma^{(i)}_{k+1} \gets \sqrt{  \left(\sigma^{(i)}_k\right)^2 - 2(\Delta t)\left( \mathcal{M}^{(i)}_k + \tfrac{1}{2} \Delta t \cdot \partial_t{\tilde{\mathcal{M}}^{(i)}_k}\right) }
\]

\end{algorithmic}
\end{algorithm}

\makeatletter
\begin{algorithm}[H]
\caption{Advance to Next Time Step (continued from Algorithm \ref{alg:timestep})}
\label{alg:timestep1b}
\begin{algorithmic}[1]
\setcounter{ALC@line}{7}
\STATE Compute zero-Dirichlet primitives of $a$ (see Algorithm \ref{alg:primitive}), $(i \in [m])$:
\[
\mathbf{p}^{(i)}_k \gets \texttt{Primitive}(\mathbf{a}^{(i)}_k), \qquad
\tilde{\mathbf{p}}^{(i)}_{k+1} \gets \texttt{Primitive}(\tilde{\mathbf{a}}^{(i)}_{k+1})
\]

\STATE Compute diffusion force at time $t_k$ and estimate at time $t_{k+1}$ (see Algorithm \ref{alg:derivative}), $(i\in[m])$:
\[
\mathbf{G}^{(i)}_k \gets \mathbf{p}^{(i)}_k \odot  \bm{\mathcal{K}}^{(i)}_k - \texttt{Derivative}\left( \mathcal{F}^{(i)}_{1,k}, \ldots, \mathcal{F}^{(i)}_{n,k} \right)
\]
\[
\tilde{\mathbf{G}}^{(i)}_{k+1} \gets \tilde{\mathbf{p}}^{(i)}_{k+1} \odot  \tilde{\bm{\mathcal{K}}}^{(i)}_{k+1} - \texttt{Derivative}\left( \tilde{\mathcal{F}}^{(i)}_{1,k+1}, \ldots, \tilde{\mathcal{F}}^{(i)}_{n,k+1} \right)
\]

\STATE Advance tangent-angle function using pseudo-exponential integrator with effective time step  (see Algorithm \ref{alg:propagate2}), $(i\in[m])$:
\[
\omega^{(i)} \gets \left\lceil \tfrac{\theta_{n, k}^{(i)} - \theta_{1, k}^{(i)}}{2 \pi} \right\rceil, \qquad 
\eta_{\ell, k}^{(i)} \gets \theta_{\ell, k}^{(i)} - \omega^{(i)}\left( -\pi + \tfrac{2\pi}{n}(\ell - 1) \right), \qquad \ell \in [n]
\]
\[
\Delta t_\text{eff}^{(i)} = \Delta t \cdot \frac{1}{2} \left( \frac{1}{\left(\sigma^{(i)}_{k+1}\right)^2} + \frac{1}{\left(\sigma^{(i)}_{k}\right)^2} \right),
\qquad
\bm{\eta}^{(i)}_{k+1} \gets \texttt{ForcedDiffusionTimeStep}\left(\bm{\eta}^{(i)}_{k}, \Delta t_\text{eff}^{(i)}, \mathbf{G}^{(i)}_k, \tilde{\mathbf{G}}^{(i)}_{k+1}\right)
\]
\[
\theta_{\ell, k+1}^{(i)} \gets \eta_{\ell, k+1}^{(i)} + \omega^{(i)}\left( -\pi + \tfrac{2\pi}{n}(\ell - 1) \right), \qquad \ell \in [n]
\]

\end{algorithmic}
\end{algorithm}

\makeatletter
\begin{algorithm}[H]
\caption{Advance to Next Time Step (continued from Algorithm \ref{alg:timestep})}
\label{alg:timestep2}
\begin{algorithmic}[1]
\setcounter{ALC@line}{10}
\STATE Advance centroids using the trapezoidal rule, $(i\in[m])$:
\[
\mathbf{v}^{(i)}_k \gets \frac{1}{n} \sum_{\ell = 1}^n \left( p_{\ell ,k}^{(i)} \begin{bmatrix} -\sin(\theta^{(i)}_{\ell, k}) \\ \cos(\theta^{(i)}_{\ell, k}) \end{bmatrix} + \mathcal{F}^{(i)}_{\ell, k} \begin{bmatrix} \cos(\theta^{(i)}_{\ell, k}) \\ \sin(\theta^{(i)}_{\ell, k}) \end{bmatrix} \right)
\]
\[
\tilde{\mathbf{v}}^{(i)}_{k+1} \gets \frac{1}{n} \sum_{\ell = 1}^n \left( \tilde{p}_{\ell, k+1}^{(i)} \begin{bmatrix} -\sin(\theta^{(i)}_{\ell, k+1}) \\ \cos(\theta^{(i)}_{\ell, k+1}) \end{bmatrix} + \tilde{\mathcal{F}}^{(i)}_{\ell, k+1} \begin{bmatrix} \cos(\theta^{(i)}_{\ell, k+1}) \\ \sin(\theta^{(i)}_{\ell, k+1}) \end{bmatrix} \right)
\]
\[
\hat{\bm{\mu}}^{(i)}_{k+1} \gets \bm{\mu}^{(i)}_{k} + \Delta t \cdot \frac{1}{2} \left( \frac{\mathbf{v}^{(i)}_k }{\sigma^{(i)}_{k}}+ \frac{\tilde{\mathbf{v}}^{(i)}_{k+1}}{\sigma^{(i)}_{k+1}} \right)
\]

\STATE Recenter the tangent-angle function (Algorithm \ref{alg:normalize}) and reconstruct the contour points (Algorithm \ref{alg:thttogam}), $(i \in [m])$:
\[
\bm{\theta}^{(i)}_{k+1} \gets \texttt{Recenter}\left( \bm{\theta}^{(i)}_{k+1} \right), \qquad
\bm{\gamma}^{(i)}_{k+1} \gets \texttt{Contour}\left( \bm{\theta}^{(i)}_{k+1}, \sigma^{(i)}_{k+1}, \hat{\bm{\mu}}^{(i)}_{k+1} \right)
\]

\STATE Apply the nonlocal transport mechanism as a rigid translation of each colony (explicit Euler step; skip if $\alpha = 0$), $(i \in [m])$:
\[
\mathbf{d}^{(i)} \gets \Delta t \cdot \alpha \sum_{j \neq i} \frac{ \hat{\bm{\mu}}^{(i)}_{k+1} -  \hat{\bm{\mu}}^{(j)}_{k+1}}{\left \lvert \hat{\bm{\mu}}^{(i)}_{k+1} -  \hat{\bm{\mu}}^{(j)}_{k+1}\right \rvert} \, \chi_w\!\left( s^{(ij)}_{k} - \beta \right), \quad
\bm{\mu}^{(i)}_{k+1} \gets \hat{\bm{\mu}}^{(i)}_{k+1} + \mathbf{d}^{(i)}, \quad
\bm{\gamma}^{(i)}_{\ell, k+1} \gets \bm{\gamma}^{(i)}_{\ell, k+1} + \mathbf{d}^{(i)}, \quad \ell \in [n]
\]

\end{algorithmic}
\end{algorithm}
\makeatother

\end{document}